\documentclass[lettersize,journal]{IEEEtran}

\usepackage{graphicx}
\usepackage{multirow}
\usepackage{subcaption}
\usepackage{tcolorbox}
\usepackage{wrapfig}

\usepackage[ruled, vlined, linesnumbered, resetcount]{algorithm2e}
\SetKwInput{KwInput}{Input}                
\SetKwInput{KwOutput}{Output} 
\let\oldnl\nl
\newcommand{\nonl}{\renewcommand{\nl}{\let\nl\oldnl}}
\SetKwRepeat{Do}{do}{while}%

\usepackage{amsmath}
\usepackage{algpseudocode}
\usepackage{xcolor}
\def\HiLi{\leavevmode\rlap{\hbox to \hsize{\color{blue!35}\leaders\hrule height .8\baselineskip depth .5ex\hfill}}}
\def\HiLii{\leavevmode\rlap{\hbox to \hsize{\color{blue!10}\leaders\hrule height .8\baselineskip depth .5ex\hfill}}}

\usepackage{todonotes}
\newcommand{\Rev}[1]{\textcolor{black}{{#1}}}
\newcommand{\NewText}[1]{\textcolor{black}{{#1}}}
\newcommand{\TR}[2]{#2}

\newcommand{\MinorRev}[1]{\textcolor{black}{{#1}}}

\begin{document}

\title{TEASMA: A Practical Methodology for Test Adequacy Assessment of Deep Neural Networks}

\author{Amin Abbasishahkoo,
        Mahboubeh Dadkhah,
        Lionel Briand,~\IEEEmembership{Fellow,~IEEE,}
        Dayi Lin
\thanks{Amin Abbasishahkoo is with the School of EECS, University of Ottawa,
Ottawa, ON K1N 6N5, Canada (e-mail: aabba038@uottawa.ca).}
\thanks{Mahboubeh Dadkhah is with the School of EECS, University of Ottawa,
Ottawa, ON K1N 6N5, Canada (e-mail: mdadkhah@uottawa.ca).}
\thanks{Lionel Briand is with the School of EECS, University of Ottawa, Ottawa, ON K1N 6N5, Canada, and also with the Lero center, University of Limerick, Ireland (e-mail: lbriand@uottawa.ca).}
\thanks{Dayi Lin is with the Huawei Canada,
Kingston, ON K7K 1B7, Canada (e-mail: dayi.lin@huawei.com).}

\thanks{Manuscript received April 19, 2005; revised August 26, 2015.}}


\maketitle
\begin{abstract}
Successful deployment of Deep Neural Networks (DNNs), particularly in safety-critical systems, requires their validation with an adequate test set to ensure a sufficient degree of confidence in test outcomes. 
\Rev{Although well-established test adequacy assessment techniques from traditional software, such as mutation analysis and coverage criteria, have been adapted to DNNs in recent years, we still need to investigate their application within a comprehensive methodology for accurately predicting the fault detection ability of test sets and thus assessing their adequacy. In this paper, we propose and evaluate \textit{TEASMA}, a comprehensive and practical methodology designed to accurately assess the adequacy of test sets for DNNs. In practice, \textit{TEASMA} allows engineers to decide whether they can trust high-accuracy test results and thus validate the DNN before its deployment. Based on a DNN model's training set, \textit{TEASMA} provides a procedure to build accurate DNN-specific prediction models of the Fault Detection Rate (FDR) of a test set using an existing adequacy metric, thus enabling its assessment.  
We evaluated \textit{TEASMA} with four state-of-the-art test adequacy metrics: Distance-based Surprise Coverage (DSC), Likelihood-based Surprise Coverage (LSC), Input Distribution Coverage (IDC), and Mutation Score (MS). We calculated MS based on mutation operators that directly modify the trained DNN model (i.e., post-training operators) due to their significant computational advantage compared to the operators that modify the DNN's training set or program (i.e., pre-training operators).
Our extensive empirical evaluation, conducted across multiple DNN models and input sets, including large input sets such as ImageNet, 
reveals a strong linear correlation between the predicted and actual FDR values derived from MS, DSC, and IDC, with minimum $R^2$ values of 0.94 for MS and 0.90 for DSC and IDC.
Furthermore, a low average Root Mean Square Error (RMSE) of 9\% between actual and predicted FDR values across all subjects, when relying on regression analysis and MS, demonstrates the latter's superior accuracy when compared to DSC and IDC, with RMSE values of 0.17 and 0.18, respectively.
Overall, these results suggest that \textit{TEASMA} provides a reliable basis for confidently deciding whether to trust test results for DNN models.
}
\end{abstract}

\IEEEpeerreviewmaketitle

\section{Introduction}
Effective testing of Deep Neural Networks (DNN) is essential to ensure they can be trusted, particularly as they are increasingly deployed in safety-critical systems. However, assuming the testing outcome is satisfactory, to be able to trust it, it is crucial to assess the adequacy of the test set with which the DNN is being validated. 
%
\TR{C1.2, C3.4, C4.1, C4.2}{The goal of test adequacy assessment is therefore to ensure that a test set selected for evaluating a DNN model's performance is effective at identifying faults and can thus confidently validate the model's accuracy. In other words, when the DNN model achieves high accuracy on a test set, regardless of the test selection approach, the tester must determine if such results can be trusted and therefore whether to be confident about the model being deployed. Indeed, inadequate testing introduces the risk of unwarranted confidence in a model's accuracy. More precisely, high-accuracy test results, on a test set with high adequacy (predicted FDR), indicate that the model has been rigorously evaluated and is likely to perform as expected on unseen inputs from the same input distribution. In this paper, we propose \textit{TEASMA} (TEst ASsessment using Mutation Analysis), a practical and reliable methodology for accurately assessing the adequacy of unlabeled test sets. We thus expect that when a test set achieves high adequacy using \textit{TEASMA}, the high-accuracy test outcomes based on that test set are deemed trustworthy indicators of a high-performance model.
}

\TR{C2.1, C3.1}{
In the context of DNN testing, several techniques have been proposed for test adequacy assessment that are mainly relying on mutation analysis~\cite{shen2018munn, ma2018deepmutation, hu2019deepmutation++, humbatova2021deepcrime} or coverage criteria~\cite{dola2023input, kim2023evaluating, kim2019guiding, pei2017deepxplore, Ma2018DeepGaugeMT, tian2018deeptest, sun2019structural}.}
Mutation analysis, one of the most used techniques for test assessment in traditional software~\cite{papadakis2018mutation, papadakis2019mutation}, has been adapted to DNN models in recent years. DNN-specific Mutation Operators (MO) have been proposed~\cite{shen2018munn, ma2018deepmutation, hu2019deepmutation++, humbatova2021deepcrime} since traditional MOs are not directly applicable to DNNs~\cite{jahangirova2020empirical}. \Rev{Researchers} have proposed MOs that fall into two primary categories: post-training MOs~\cite{ma2018deepmutation, shen2018munn}, which mutate the DNN model obtained after the training process, and pre-training MOs~\cite{ma2018deepmutation, humbatova2021deepcrime}, which mutate either the training program or the training set and use these mutated artifacts in the training process to generate mutated models. 
Post-training MOs offer a practical advantage by not requiring retraining for mutant generation, resulting in vastly reduced computational costs compared to pre-training MOs. However, they directly modify a DNN model and thus are criticized for not realistically representing real faults~\cite{panichella2021we}.
The use of mutation analysis in traditional software is motivated by the strong association between mutants and faults reported in the literature~\cite{andrews2006using, just2014mutants, papadakis2018mutation}.  
However, the association between the mutants and DNN faults, and thus the ability to assess a test set based on its Mutation Score (MS), has never been studied. 
\TR{C2.1, C3.1}{
In this paper, we address this gap by investigating such an association for mutants generated by post-training MOs, motivated by their lower cost.}

\TR{C2.1, C3.1}{DNN testing has evolved to include a variety of coverage metrics inspired by coverage criteria for traditional software, 
including widely-used neuron and layer coverage criteria~\cite{pei2017deepxplore}. However, these criteria have shown no significant correlation with the number of mispredicted inputs or detected faults in a test set~\cite{aghababaeyan2021black, li2019structural, yang2022revisiting}.
In addition to these coverage metrics, distribution-based coverage metrics including Surprise Coverage (SC)~\cite{kim2019guiding, kim2023evaluating} and Input Distribution Coverage (IDC)~\cite{dola2023input} have been introduced. 
The SC metrics proposed by Kim \textit{et al.}~\cite{kim2019guiding, kim2023evaluating}, including Distance-based SC (DSC) and Likelihood-based SC (LSC), distinguish themselves from previous coverage metrics by measuring how surprising test inputs are for the DNN model relative to the training inputs. 
Unlike neuron and layer coverage, which can often be satisfied with a few test inputs exhibiting common input patterns, SC measures the novelty of test inputs.
IDC is a State-Of-The-Art (SOTA) coverage metric recently introduced by Dola \textit{et al.}~\cite{dola2023input}. This innovative framework offers a novel black-box test adequacy metric specifically designed for DNNs. Unlike white-box coverage metrics such as SC, which assess the internal behavior of a DNN, IDC focuses on the input domain by employing a Variational Autoencoder (VAE) to learn a low-dimensional latent representation of the input distribution.
}

\TR{C2.1, C3.1, C4.2}{
Though various test adequacy metrics have been proposed, such as MS, SC, and IDC, we need to devise and evaluate a comprehensive methodology for test adequacy assessment relying on such metrics. While these metrics have been validated to some extent, for example, the results of IDC demonstrating a correlation with misprediction,  a practical test adequacy assessment methodology is still lacking. Further, these metrics need to be compared in this particular application context. 
Such a methodology can provide engineers with an accurate assessment of the ability of a test set to detect faults, enabling them to confidently decide whether the test set is adequate for validating the DNN. 
}   

\TR{C2.1, C3.1, C4.2}{
In this paper, we address this gap by introducing and evaluating \textit{TEASMA}, a comprehensive and practical methodology for test adequacy assessment. \textit{TEASMA} relies on the training set and existing adequacy metrics to accurately predict the Fault Detection Rates (FDR) of a test set.}
Following \textit{TEASMA}, engineers can develop a specific FDR prediction model \NewText{using a selected adequacy metric}, along with Prediction Intervals (PI), for each DNN model based on its training set. This prediction model can then be used by engineers to predict a test set's FDR based on its \TR{C2.1, C3.1}{Adequacy Score (AS), calculated using the selected metric. Considering the prediction's PI, engineers can then decide} whether a test set is adequate for validation in their context. 
\TR{C2.1, C3.1}{
While \textit{TEASMA} provides the flexibility to leverage any adequacy metric, we apply \textit{TEASMA} using four SOTA metrics, including DSC, LSC, IDC, and MS and perform a comprehensive investigation to determine which ones can predict a test set's FDR accurately. }

\TR{C3.12}{Developing \textit{TEASMA} is further motivated by the costly and time-consuming process of preparing large test sets, as it allows for the assessment of test set adequacy without the need for human experts to label the test data, the latter being only required once the test set is considered adequate.
}

Using \textit{TEASMA}, a test set is assessed before labeling and execution, solely based on the training set used for training the DNN. Test inputs can then be labeled and executed once the test set is deemed adequate, thus preventing the waste of testing resources. 

We evaluated \textit{TEASMA} using \TR{C1.4, C2.7, C3.9, C4.10}{seven widely used DNN models and six image recognition input sets including ImageNet~\cite{ILSVRC15}}. 
\TR{C2.1, C3.1}{
We first show that there are strong relationships between FDR and AS based on all investigated metrics, i.e., MS, DSC, LSC, and IDC. Specifically for MS, this suggests that mutants generated by post-training MOs could be used as fault indicators.
However, despite these strong relationships, the shape of the relationships varies widely across subjects and is highly non-linear for most of them.  
Subsequently, we applied \textit{TEASMA}, 
using all four metrics,} 
to build DNN-specific FDR prediction models. We compared, for a large number of test subsets, their predicted FDR with the actual one. We could observe, 
\TR{C2.1, C3.1}{for all metrics except LSC}, a very high linear correlation between the two, with a slope close to 1 for the regression line. 
\TR{C2.1, C3.1}{
Consequently, we compared the accuracy of FDR predictions using MS, DSC, and IDC and concluded that MS is the most accurate metric for predicting FDR across all subjects.}

To summarize, the key contributions of this paper are as follows:

\begin{itemize}
    \item An empirical analysis of SOTA \TR{C2.1, C3.1}{test adequacy metrics, including DSC, LSC, IDC, and MS, showing that AS calculated using these metrics} has a strong relationship with FDR, though the shape of the relationship tends to be highly non-linear and varies widely across DNNs. Results thus confirm early suspicions that \TR{C2.1, C3.1}{these metrics} cannot directly be used as surrogates for FDR and thus for test assessments.

    \item \textit{TEASMA}, \TR{C2.1, C3.1}{an accurate methodology utilizing existing SOTA test adequacy metrics} for assessing a test set before the validation and deployment of a DNN. It is based on regression analysis using its training set, with the objective of building a predictor model for FDR, which can then be used as a basis for assessing the adequacy of test sets.
     
    \item A large empirical evaluation of \textit{TEASMA}, \TR{C2.1, C3.1}{relying on four existing SOTA test adequacy metrics} with \Rev{seven} widely used DNN models and \TR{C1.4, C2.7, C3.9, C4.10}{six} image input sets, \Rev{that took approximately 40 days of runtime}. Results suggest \textit{TEASMA} has the ability to build accurate predictors of a test set's FDR based on the DNN training set. 

    \item \TR{C2.1, C3.1}{ A comprehensive comparison between investigated metrics showing that MS calculated based on post-training MOs, is the most accurate metric for predicting FDR across all subjects, followed by DSC and IDC, whereas LSC proves to be ineffective for accurate FDR prediction.}

\end{itemize}

The remainder of this paper is structured as follows: Section \textrm{II} provides background and discussions of the test adequacy assessment techniques for DNNs involved in our experiments. Section \textrm{III} presents our proposed test adequacy assessment methodology. Section \textrm{IV} describes the experiments we performed to assess our methodology. Section \textrm{V} presents and discusses the results for each research question. Section \textrm{VI} discusses the related work, and Section \textrm{VII} concludes the paper.

%
\section{Background}
\label{sec:Background}
\TR{C2.1, C3.1}{This section is divided into four parts. The first three parts provide a brief overview of existing SOTA test adequacy assessment techniques for DNNs involved in our experiments. These include mutation analysis and two distribution-based coverage metrics: the widely used Surprise Coverage (SC) and the recently introduced Input Distribution Coverage (IDC). We also discuss in detail the existing methods for calculating the Adequacy Score (AS) of a test set based on these metrics. The last part describes a SOTA approach that can be leveraged to investigate the association between AS of a test set based on these metrics and its fault detection ability, enabling us to perform our experiments.
}

\subsection{\NewText{Mutation Analysis}}
\label{Sec:MutationTesting}

\Rev{In this section, we briefly overview mutation analysis for DNNs. We introduce two primary categories of existing Mutation Operators (MOs), namely pre- and post-training MOs, and discuss the cost of mutation analysis which is the key factor driving our choice for relying on post-training operators in \textit{TEASMA}. Finally, we summarize different ways of calculating MS for post-training MOs.
}
\subsubsection{Mutant Generation}
\label{Sec:MutantGeneration}

\begin{figure}[ht]
    \includegraphics[width = \columnwidth]{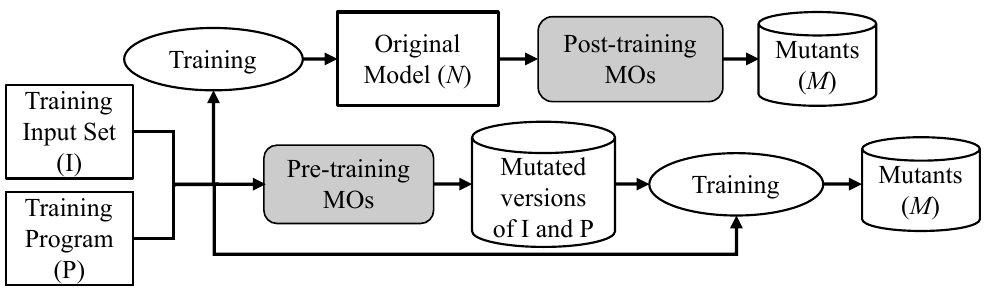}
    \centering
    \caption{Pre-training and post-training mutation operators}
    \label{fig:MutationOperators}
\end{figure}

Applying mutation testing to DNN-based systems requires defining specific MOs for DNN models~\cite{jahangirova2020empirical}. The proposed MOs can be broadly classified into two primary categories, namely pre-training~\cite{ma2018deepmutation, humbatova2021deepcrime} and post-training~\cite{ma2018deepmutation, shen2018munn, hu2019deepmutation++} operators, as illustrated in Figure~\ref{fig:MutationOperators}.
Pre-training MOs, also called source-level MOs, slightly modify the original version of either the program or the input set that has been used to train the original DNN model under test. These modified versions are then used by the training process to generate mutated models (i.e., mutants).  
Mutation testing using pre-training MOs is computationally expensive since it entails repeating the training process for generating each mutant. On the other hand, post-training MOs, also referred to as model-level MOs, modify the already trained original model, eliminating the need for additional training to generate mutants.

Ma \textit{et al.}~\cite{ma2018deepmutation} proposed DeepMutation, the first mutation testing framework for DNNs that includes both pre- and post-training MOs. Pre-training operators are classified into data-level and program-level operators whereas post-training operators are classified into weight-level, neuron-level, and layer-level operators, depending on the specific component of the DNN model that they modify.  
The DeepMutation++ tool~\cite{hu2019deepmutation++} was later introduced by the same authors implementing only the post-training operators.
Shen \textit{et al.}~\cite{shen2018munn} proposed the MuNN framework, including five post-training operators that modify neurons, activation function, bias values, and weight values of a DNN model. Since post-training MOs directly modify the DNN model, not all of them are applicable to any DNN model. For example, applying the Layer Deactivation ($LD$) operator of DeepMutation~\cite{ma2018deepmutation} on some layers may break the model's structure. Therefore, DeepMutation restricts the application of this operator to only those layers whose input and output shapes are consistent, if the model has such layers.

Humbatova \textit{et al.}~\cite{humbatova2021deepcrime} proposed DeepCrime, a mutation testing tool that includes 24 pre-training mutation operators inspired by real faults in deep learning systems~\cite{humbatova2020taxonomy, islam2019comprehensive, zhang2018empirical}.  The main goal was to obtain more realistic mutation operators.  
They also proposed a statistical approach~\cite{jahangirova2020empirical} for identifying killed mutants that requires multiple instances of each mutant. 
Therefore, DeepCrime repeats the training process on the original and modified versions of the training set and program many times, resulting in many instances of the original model and each mutant. This further increases the cost of mutation testing, requiring not only to repeat the training process and generate more mutants but also executing each test set on all mutant instances.   

Most of the MOs proposed for DNNs~\cite{ma2018deepmutation, humbatova2021deepcrime, shen2018munn} have parameters that define the scale of modifications by that MO and need to be configured before their application. For example, operators that modify training sets could be configured by defining the percentage of inputs to be modified. Therefore, each MO can be applied to the DNN with multiple configurations, some of them resulting in generating mutants with undesirable characteristics: equivalent, trivial, and redundant. It is essential to identify and filter such mutants since they not only increase the overall execution time of mutation analysis but may also affect the MS. Different approaches have been proposed to address this issue~\cite{ma2018deepmutation, hu2019deepmutation++, humbatova2021deepcrime}. 
In the DeepMutation++ tool, mutants with prediction accuracy lower than a specified threshold are eliminated. Then, for each input set, mutants with high error rates (i.e., killed by most of the inputs), are filtered out to prevent inflating the MS with easy-to-kill mutants. The DeepMutation++ sets the accuracy threshold to be 90\% of the original model's accuracy, and the error rate threshold to be 20\% of the input set size. 
Humbatova \textit{et al.}~\cite{humbatova2021deepcrime} defined killability and triviality scores for MOs in DeepCrime based on the training set. Redundant mutants are also identified and filtered out based on the subsumption relation between them~\cite{humbatova2021deepcrime}. Subsequently, to identify and exclude equivalent (non-killable), trivial, and redundant mutants, the entire training set is executed against all mutants generated by various configurations of each MO.

\subsubsection{Mutation Analysis Cost}
\label{Sec:MutationAnalysisCost}

The high computational cost of mutation analysis is its main drawback in practice. 
This cost is mainly due to performing two processes. Initially, one must utilize MOs to generate mutants. The second process involves executing a large number of test inputs against mutants, followed by the calculation of their MS. A precise calculation of MS usually necessitates filtering out the equivalent, trivial, or redundant mutants.
The cost issue is even more critical in the case of DNNs considering large input sets, which is usual in practice. In this context, the substantial computational overhead of mutant generation using pre-training MOs, due to the need to perform training on each mutant, is a significant drawback when compared with post-training MOs. 

\TR{C2.3, C3.3}{}
\Rev{In order to characterize the differences in computational cost between mutation analysis employing pre- and post-training MOs, we performed a preliminary study using pre-training mutants generated by DeepCrime with post-training mutants generated by DeepMutation++. Given the experiment's time-intensive nature, which entails consuming enormous amounts of computing resources, we limited it to just one simple DNN and input set, specifically LeNet5 and MNIST (subject S1 introduced later in Table~\ref{tab:Subjects}). 
Conducting mutation analysis on the same set of input subsets randomly sampled from the MNIST training set took less than 49 minutes using DeepMutation++, compared to over 123 hours with DeepCrime--more than 150 times the duration required by DeepMutation++. Consequently, post-training MOs in DeepMutation++ require two orders of magnitude less computational resources than pre-training MOs. While this experiment, due to its high computational requirements, is conducted on a single subject of modest size, the findings underscore the substantial difference in computational resource requirements for mutation analysis based on pre- and post-training MOs and justify our choice to employ the latter.}

\Rev{Given the arguments and preliminary results presented above, we rely on the more practical post-training MOs in our experiments. We investigate if they are able to generate mutants associated with faults that can support our objective, which is assessing the adequacy of test sets. In this paper, we deploy post-training MOs in \textit{TEASMA} for DNN test set adequacy assessment based on mutation analysis.
}

\subsubsection{Mutation Score Calculation}
\label{subSec:MutationTestingMetrics}
 
\NewText{Various methods have been proposed to identify killed mutants and calculate MS based on post-training MOs.} Similar to traditional software, killed DNN mutants can be identified by comparing the output of a mutant $M_i$ with the original model $N$ and the corresponding MS can be calculated as:  

\begin{equation} \label{Eq:DeepMutationStandardMS}
    StandardMS(T, M) = \frac{\sum_{M_i\in M} killed(T, M_i)}{\left| M \right|  }
\end{equation}

\noindent where $M$ is the set of mutants, $M_i \in M$ is $killed$ by test set $T$ ($killed(T, M_i) = 1$) if there is at least one input $t \in T$ that is correctly classified by $N$ and is not correctly classified by mutant $M_i$. DeepMutation~\cite{ma2018deepmutation}, however, calculates MS in a different way. For a classification problem with $k$ classes $C = \{C_1, ..., C_k \}$, a test input $t \in T$ kills class $C_j \in C$ of mutant $M_i \in M$ if $t$ is correctly classified as $C_j$ by the original model $N$ and is not classified as $C_j$ by mutant $M_i$. Accordingly, MS is calculated for a test set $T$ as:

\begin{equation} \label{Eq:DeepMutationMS}
    DeepMutationMS(T, M) = \frac{\sum_{M_i\in M} killedClasses(T, M_i)}{\left| M \right| \times \left| C \right| } 
\end{equation}

\noindent where $killedClasses(T, M_i)$ is the number of classes of mutant $M_i$ killed by inputs in $T$~\cite{ma2018deepmutation}. The difference when calculating MS for the same test set using $StandardMS$ based on Equation~(\ref{Eq:DeepMutationStandardMS}) and $DeepMutationMS$ based on Equation~(\ref{Eq:DeepMutationMS}) can be very large. 
For instance, consider a test set $TS$ from the input set of handwritten digits (MNIST). Each input $t \in TS$ is an image belonging to a class $C_j$. Let the test set $TS$ be represented by its correct input labels $label(TS) = \{3, 5, 3, 7\}$,  including two different images from the same class representing digit 3 but also two images from classes representing digits 5 and 7. Further, consider that all images are correctly predicted by $N$. If we have a set of mutants $M = \{M_1, M_2\}$, where the first two test inputs of $TS$ are mispredicted by $M_1$ and all test inputs are mispredicted by $M_2$, then $DeepMutationMS(TS, M) = \frac{2+3}{2 \times 10 }= 0.25$ while $StandardMS(TS, M) = 1$. 
The score calculated by $StandardMS$ for a test set $T$ is higher since killing a mutant $M_i$ with $T$ requires only one test input $t \in T$ to be mispredicted by $M_i$.   

Hu \textit{et al.}~\cite{hu2019deepmutation++} defined a killing score metric for each test input $t$ as the proportion of mutants in $M$ whose output is different from $N$.

\begin{equation} \label{Eq:KillingScore}
            KillingScore(t, M, N) = \frac{\left| M_i |  M_i \in M \wedge N(t) \neq M_i(t) \right|}{\left| M \right|}
\end{equation}

Humbatova \textit{et al.}~\cite{humbatova2021deepcrime} used this metric to calculate the MS of a test set $T$ on mutants generated by DeepMutation++~\cite{hu2019deepmutation++}:

\begin{equation} \label{Eq:KillingScoreBasedMS}
            KSBasedMS(T, M) =  \frac{\sum_{t\in T} KillingScore(t, M, N)}{\left| T \right|}
\end{equation}

However, calculating MS as the average $KillingScore$ of test inputs in a test set may not indicate the ability of the entire test set to kill mutants. A test set cannot achieve such a high MS unless many of its test inputs are able to kill most of the mutants. This is visible in results reported by Humbatova \textit{et al.}~\cite{humbatova2021deepcrime} where the MS calculated by $KSBasedMS$ for entire test sets ranges from 0.059 to 0.33.
Humbatova \textit{et al.}~\cite{humbatova2021deepcrime} further proposed a statistical approach for identifying killed mutants when applying pre-training MOs in DeepCrime where the training process is repeated $n$ times, leading to $n$ instances of the original model and mutants being generated (Section~\ref{Sec:MutantGeneration}). This further increases the cost of mutant generation using pre-training MOs and makes them computationally challenging for practical use in our context. 

Mutation analysis involves identifying test sets with high MS based on the key assumption that MS captures a test set's capability to detect faults. In DNNs, however, various MS formulas lead to different results and which test sets lead to high MS varies according to which formula is used. It is evident that more test sets can achieve a high MS using $StandardMS$ (Equation~(\ref{Eq:DeepMutationStandardMS})) than when using $DeepMutationMS$ (Equation~(\ref{Eq:DeepMutationMS})). We also expect that not many test sets can achieve a high MS using $KSBasedMS$ (Equation~(\ref{Eq:KillingScoreBasedMS})). Therefore, because we cannot a priori determine which MS formula satisfies our assumption to the best extent, it is essential to consider all of them when experimenting with mutation analysis for DNNs.

\subsubsection{Mutants and Faults}
\label{Sec:MutantsandFaults}

In traditional software, MOs are defined to introduce simple syntactical changes representing real mistakes often made by programmers into the source code of a program. Similarly, DeepCrime defined a set of pre-training MOs based on an analysis of real faults~\cite{humbatova2020taxonomy, islam2019comprehensive, zhang2018empirical} related to the training set or program used to train a DNN model. However, the high cost of generating mutants using pre-training MOs hinders their application in practice. Post-training MOs, in contrast, apply modifications directly to the original model but have been criticized for performing modifications that are not realistic~\cite{panichella2021we}. \TR{C3.5, C4.1, C4.6}{However, it is important to note that not all faults in DNN models stem from human error~\cite{ma2018deepmutation}. As emphasized by Ma et al.~\cite{ma2018deepmutation}, post-training operators are not intended to directly replicate human faults but rather to provide quantitative measurements of test set quality.}
Nevertheless, for mutation analysis to be effective in assessing the adequacy of test sets, regardless of how mutation operators are defined, it is important to generate mutants whose corresponding MS is, for a test set, a good predictor of fault detection. 
\TR{C3.5, C4.1, C4.6}{
Therefore, considering the strong practical advantages of post-training MOs, we decided to use these operators in our experiments. Although these operators are not derived from the analysis of human faults, we investigate their potential as predictors of fault detection and thus as a tool to assess the adequacy of test sets.}

The use of mutation analysis for assessing the quality of test sets in traditional software is supported by the observed association between mutants and real faults, as reported in the literature~\cite{andrews2006using, just2014mutants, papadakis2018mutation}. This association has been investigated by performing correlation \Rev{analysis }
between the MS of a test set and its capability to detect faults. Regression analysis has also been applied for modeling this relationship~\cite{andrews2006using, papadakis2018mutation}. In DNNs, however, the association between mutants and faults has never been investigated. The main challenge in such investigation is that, in contrast to traditional software where it is possible to identify real faults by isolating the faulty statement that caused a failure, identifying faults in DNNs is not a straightforward task due to the complexity and non-linear nature of the DNNs. However, a recent approach proposed by Aghababaeyan \textit{et al.}~\cite{aghababaeyan2021black} for estimating faults is a possible, practical, and automated solution. 
We describe \Rev{in Section~\ref{Sec:FaultEstimation}} how we rely on this approach to investigate the relationship between mutants and faults, and how MS can help predict FDR. 

\subsection{\NewText{Surprise Coverage}}
\label{Sec:SurpriseCoverage}

\TR{C2.1, C3.1}{
Surprise Coverage (SC) metrics have been introduced by Kim \textit{et al.}~\cite{kim2019guiding, kim2023evaluating} as innovative test adequacy metrics that quantify the surprise or novelty of test inputs with respect to the training data of a DNN. These metrics are derived from the Surprise Adequacy (SA) concept, which evaluates how surprising a test input is to a DNN based on its training set, thereby indicating the potential for revealing unexpected behavior or faults. SC evaluates the diversity of a test set by measuring the distribution of SA among its test inputs relative to the training input set. Kim \textit{et al.}~\cite{kim2023evaluating} define three SA metrics for each test input:
}

\TR{C2.1, C3.1}{
\textbf{Distance-based SA (DSA)} utilizes the Euclidean distance between the activation traces of test inputs and those observed during training. This metric is particularly useful in classification tasks, where inputs closer to decision boundaries are deemed more valuable for testing due to their higher potential to detect misclassifications.}

\TR{C2.1, C3.1}{
\textbf{Likelihood-based SA (LSA)} leverages Kernel Density Estimation (KDE)~\cite{wand1994kernel} to calculate the likelihood of the activation patterns of the test inputs given the distribution of training data activation patterns. Lower likelihoods suggest higher surprise, implying that the DNN is less familiar with such inputs, which might lead to higher fault revelation.}

\TR{C2.1, C3.1}{
\textbf{Mahalanobis Distance SA (MDSA)} employs the Mahalanobis distance to measure the difference between activation traces of the test input and the distribution of the activation traces derived from the training inputs. This method is beneficial due to its sensitivity to both the mean and covariance of the training data distribution, providing a robust indication of input novelty.}

\TR{C2.1, C3.1}{
Kim \textit{et al.}~\cite{kim2023evaluating} define three SC metrics: Likelihood-based Surprise Coverage (LSC), Distance-based Surprise Coverage (DSC), and Mahalanobis Distance Surprise Coverage (MDSC), which rely on LSA, DSA, and MDSA metrics, respectively. However, they only implemented LSC and DSC in their public repository~\cite{kim2023evaluating}.   
The SC computation involves dividing the observed SA values range into distinct intervals or buckets. SC is then defined as the proportion of these buckets containing at least one test input, measuring the test set's ability to cover a wide range of model behaviors:
}

\begin{equation}
    SC(T) = \frac{|\{b_i \mid \exists t \in T : SA(t) \in b_i\}|}{n}
\end{equation}

\TR{C2.1, C3.1}{
where $T$ is the set of test inputs, $b_i$ represents discrete buckets of SA values, and $n$ is the total number of buckets. In our experiments, we rely on the original implementation of the LSC and DSC metrics provided by Kim \textit{et al.}~\cite{kim2023evaluating}.
}

\subsection{\NewText{Input Distribution Coverage}}
\label{Sec:IDC}

\TR{C2.1, C3.1}{
Input Distribution Coverage (IDC), introduced by Dola \textit{et al.}~\cite{dola2023input}, is a black-box coverage metric that employs a novel approach by utilizing Variational Autoencoders (VAEs) to model the input distribution of a DNN, offering a framework to assess test adequacy based on the diversity and distribution of features within the test set. For each input set, a VAE is trained on its training input set, which is representative of the operational domain of any DNN trained with this input set. The IDC framework utilizes disentangled latent representations to learn a latent space composed of linear subspaces, where variation in each subspace corresponds to variation in a feature of the input set. However, methods for identifying the features present in a latent representation and accurately mapping these features to latent subspaces are not available. Consequently, the number of features representing an input space, i.e., the latent space dimensionality, is defined by a parameter in IDC.
}

\Rev{The VAE is trained to encode the high-dimensional input space into a lower-dimensional latent space, enabling the IDC framework to define a feasible coverage domain. 
Consequently, the latent space is divided into distinct partitions, meaning any latent linear subspace is encompassed by a subset of partitions.
These partitions represent unique regions that could potentially exhibit different behaviors or characteristics relevant to the DNN model’s performance. While testing the full combinatorial space is not practical, IDC relies on Combinatorial Interaction Testing (CIT)  to measure test coverage. IDC thus assesses the adequacy of a test set based on how thoroughly its test inputs cover the diverse partitions within the latent space. 
}

\TR{C2.1, C3.1}{
When assessing a test set, the encoder part of the VAE is used to obtain the latent representation of its test inputs. However, utilizing a VAE's latent representation is only possible for inputs that follow the same distribution as the VAE's training set, i.e., in-distribution inputs. Consequently, the IDC framework utilizes SOTA out-of-distribution (OOD) detectors to filter out OOD test inputs before processing them with the VAE. As a result, unlike all white-box DNN coverage metrics that measure test coverage for both in-distribution and OOD test inputs, IDC measures coverage solely based on in-distribution inputs. White-box DNN coverage metrics such as neuron coverage and extended neuron coverage cannot differentiate between valid and invalid test inputs and thus suffer from falsely inflated coverage by OOD inputs~\cite{dola2021distribution}. IDC overcomes the inclusion of OOD test inputs and enables a more accurate calculation of test coverage. 
}

\subsection{Fault Estimation}
\label{Sec:FaultEstimation}

The number of mispredicted test inputs is a well-known metric in evaluating DNN testing approaches~\cite{pei2017deepxplore, kim2019guiding, feng2020deepgini}. However, the number of mispredictions is not a good indicator of the number of faults and hence it is not an adequate metric to perform mutation analysis\TR{C4.7}{~\cite{aghababaeyan2021black}. }
Indeed, a single fault within a DNN model can make the model mispredict multiple test inputs. Subsequently, a test set that identifies a considerable number of mispredictions may actually detect only a few underlying faults, whereas another test set with the same number of mispredictions could detect a higher number of distinct faults. Therefore, assessing test sets based on mispredictions and thus investigating the relationship between the number of killed mutants and mispredictions can be misleading. 

Consequently, we rely on a recent approach introduced by Aghababaeyan \textit{et al.}~\cite{aghababaeyan2021black} for estimating faults in a DNN model, that are defined as distinct root causes of mispredictions~\cite{fahmy2021supporting}. Aghababaeyan \textit{et al.}~\cite{aghababaeyan2021black} cluster mispredicted inputs that exhibit similar features in three key steps: feature extraction, dimensionality reduction, and density-based clustering. Initially, VGG-16~\cite{simonyan2014very} is employed to extract the feature matrix from the mispredicted inputs, serving as the foundation for clustering. Subsequently, dimensionality reduction is applied to enhance clustering performance within the inherently high dimensional feature space. The HDBSCAN algorithm~\cite{campello2013density} is then utilized to cluster the mispredicted inputs based on the extracted features.

\TR{C2.4, C3.7}{
Aghababaeyan \textit{et al.}~\cite{aghababaeyan2021black} used both manual and metric-based evaluation to select the best hyperparameter configuration from several clustering configurations. For the metric-based evaluation, they used two established metrics:  Silhouette score~\cite{rousseeuw1987silhouettes} and Density-Based Clustering Validation (DBCV)~\cite{moulavi2014density}. 
To complement these quantitative assessments, they conducted a manual analysis to assess the quality of the final selected clusters. For this purpose, they generated heatmaps for each cluster, visualizing the distribution of features across inputs within a cluster. Their empirical analysis of the heatmaps showed that inputs in the same cluster are mispredicted due to the same fault in the model and, conversely, inputs in different clusters are mispredicted due to distinct faults.}
Therefore, if a test set contains only one of the test inputs in a cluster, then it is able to detect the underlying fault of that cluster.  

\TR{C3.7}{Similar failure clustering approaches have been proposed by Biagiola and Tonella~\cite{biagiola2024testing} and Attaoui \textit{et al.}~\cite{attaoui2023black} with different objectives. Biagiola and Tonella~\cite{biagiola2024testing} used the clustering approach to evaluate the diversity of the generated inputs while Attaoui \textit{et al.}~\cite{attaoui2023black} relied on the clusters for test selection aiming to efficiently improve the performance of the DNN model through retraining. Unlike Aghababaeyan \textit{et al.}~\cite{aghababaeyan2021black}, given their objectives, neither of these approaches proposed a fault analysis procedure to evaluate whether clusters capture faults. Consequently, in this study, we rely on the clustering approach introduced by Aghababaeyan  \textit{et al.}~\cite{aghababaeyan2021black} and adopt the same manual and metric-based approach to evaluate the identified fault clusters. 
}

Suppose we identify a set of $k$ Clusters of Mispredicted Inputs $CMI = \{CMI_1, ..., CMI_k\}$, we can then assume that each cluster $CMI_i$ contains inputs that are mispredicted due to the same root cause (i.e., the underlying fault $F_i$). 
Therefore, we say that test set $T$ $detects$ a fault $F_i \in F$ if there is at least one test input $t \in T$ such that  $t \in CMI_i$. Finally, we calculate the FDR for a test set $T$ as: 

\begin{equation}\label{Eq:FDR}
 FDR(T, F)= \frac{\sum_{F_i\in F} detect(T, F_i)}{ \left|F\right|  } 
\end{equation}

In the next section, we propose \textit{TEASMA}, a practical solution for test set adequacy assessment using mutation analysis. The objective is to predict the expected FDR for a given test set and DNN, based on mutation and regression analysis performed during training. We should note that while \textit{TEASMA} currently utilizes the fault estimation approach described in this section, it is flexible and can incorporate new fault identification approaches to improve its performance.

%

\begin{figure*}
    \includegraphics[width = \textwidth]{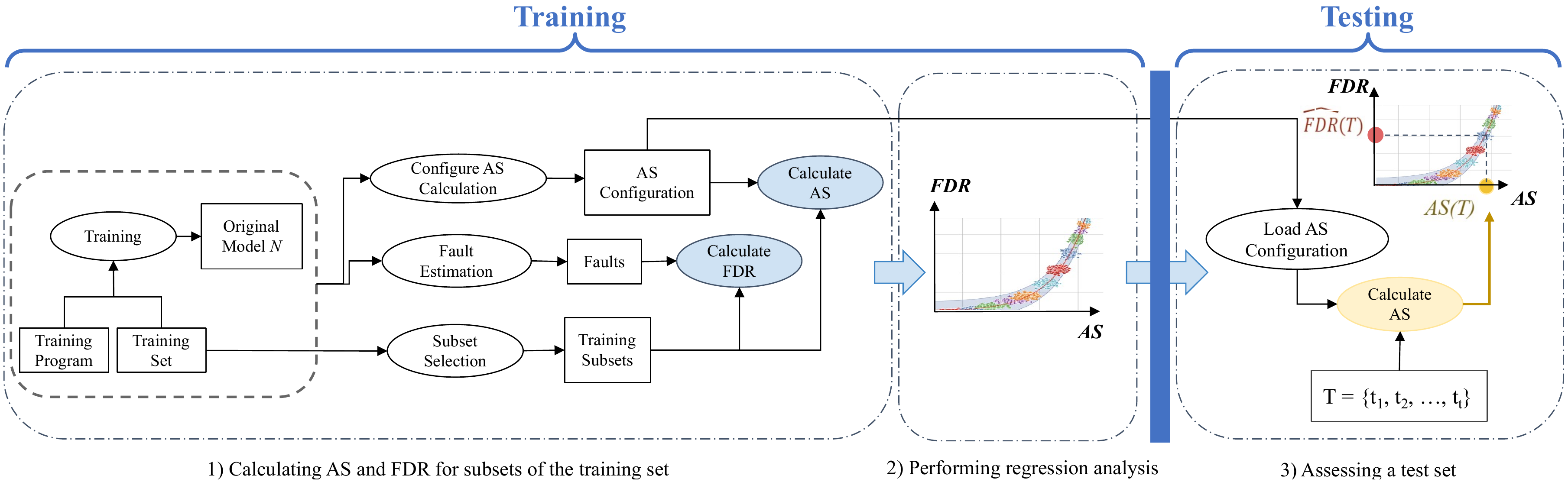}
    \centering
    \caption{The process of test adequacy assessment with \textit{TEASMA}}
    \label{fig:ApplicationProcess}
\end{figure*}

\section{The \textit{TEASMA} \Rev{Methodology} }
\label{Sec:Approach}

In this section, we present \textit{TEASMA}, a methodology for assessing the adequacy of test sets by predicting their fault detection capability, based on the training set \TR{C2.1, C3.1}{and utilizing existing adequacy metrics.}
The basic idea is to rely on regression analysis and the training set to accurately predict the FDR of test sets, which can then be used for evaluating their adequacy before proceeding with labeling and execution. 
\TR{C2.1, C3.1}{
\textit{TEASMA} provides the flexibility to perform adequacy assessment using any metric but one needs to determine which one yields the most accurate FDR predictions. In this study, we apply \textit{TEASMA} with four SOTA test adequacy metrics (MS, DSC, LSC, and IDC), which were described in Section~\ref{sec:Background}. We then investigate later in Section~\ref{sec:Results} which of these metrics can predict FDR more accurately.
}
We provide an overview of the main steps to apply \textit{TEASMA} in Figure~\ref{fig:ApplicationProcess} and a detailed description of these steps in \Rev{Algorithms~\ref{Alg:BuildPredictionModel}} and \ref{Alg:AssessingTestSets}. The process to be followed by engineers in practice is illustrated in Figure~\ref{fig:ApplicationProcess} and includes three high-level steps:

\noindent{\textbf{1) Calculating 
\TR{C2.1, C3.1}{Adequacy Score (AS)} and FDR for subsets of the training set.}} 

In this step, 
faults of the DNN model are estimated using the misprediction clustering approach described in Section~\ref{Sec:FaultEstimation} based on the training input set. 
Then, a large number of subsets of varying sizes are sampled from the training set, and their \Rev{AS} and FDR are calculated to be used in the next step. 
\TR{C2.1, C3.1}{
To calculate the AS of these training subsets, the selected adequacy metric needs to be configured. The configuration depends on the selected adequacy metric. As mentioned earlier, we repeat \textit{TEASMA} using MS, SC, and IDC for calculating AS in this study. Therefore, AS configuration in our experiments includes mutant generation, configuring SC parameters, and training a VAE using the training set. The AS configuration for the selected metric is saved to be reused by the last step of \textit{TEASMA}.}

\noindent{\textbf{2) Performing regression analysis.}}

In this step, regression analysis between FDR and \Rev{AS} is performed to obtain the best regression model for predicting FDR based on \NewText{AS}. This prediction model is then used in the next step to assess the adequacy of test sets. 
\TR{C2.1, C3.1}{
In this study, we investigate four different adequacy metrics, and thus perform regression analysis separately for FDR and each metric. The optimal regression model is then determined in terms of FDR prediction accuracy.}

\noindent{\textbf{3) Assessing test set adequacy.}}

\TR{C2.1, C3.1}{
In this step, given a test set, its AS is calculated based on the selected adequacy metric and the same configuration as in previous steps.
For instance, MS is calculated using the original DNN model and previously generated mutants. 
Similarly, SC is computed using the same configuration, including the DNN layer selected for computing SA, the number of buckets, and the lower and upper bounds of SA values, as in the first step. For IDC, the same pre-trained VAE is used.}
Then, using the computed \Rev{AS}, the prediction model obtained from the previous step is used to predict the test set's FDR ($\widehat{FDR}$) along with Prediction Intervals (PI).

Algorithm~\ref{Alg:BuildPredictionModel} describes the first two steps implemented in \textit{TEASMA} to create an accurate regression model for predicting FDR. As illustrated in Figure~\ref{fig:ApplicationProcess}, the first two steps of \textit{TEASMA} are applied entirely on the training set of the DNN. The algorithm starts by initializing an empty archive of subset samples $A$ (line 1). Then, to be able to perform \Rev{regression} analysis, \textit{TEASMA} is required to \Rev{configure the AS computation} (line 2) and identify faults in the DNN (line 3). 
\TR{C2.1, C3.1}{
The configuration process depends on the selected adequacy metric. In this study, it includes either generating mutants for computing MS, configuring the parameters for SC calculation, or training a VAE for IDC calculation.}

\begin{algorithm}[t!]
\DontPrintSemicolon
  \KwInput{\;
    \Indp $N$: the original DNN model under test \;    
    $TrSet$: the labeled training input set of $N$ \;
    $SN$: number of samples per each sampling size \;
    }
    
    \KwOutput{ \;
    \Indp $PM$: prediction model built for N} 
    
    \BlankLine
    $sample\ archive\ A \xleftarrow{} \emptyset$ \;

    \nonl
    \HiLii \Rev{Configure AS calculation} \;
    \Rev{$config \xleftarrow{} AdequacyConfig();$} \;

    \nonl
    \HiLii Estimate faults based on the training set\;
    $faults\ F \xleftarrow{} \Rev{EstimateFaults}(N, TrSet);$ \; 

    \BlankLine
    \nonl
    \HiLi 1) Calculate MS and FDR for training input subsets \;
    \Do{\Rev{$(1 - maxFDR \geq \theta)\: and \:(minFDR \geq \theta)$}}{
        $size \xleftarrow{} UpdateSamplingSize(A);$ \;
        $samples \xleftarrow{} Sampling(TrSet, size, SN);$ \;
        \ForEach{subset $sub_i$ in $samples$}{
        \Rev{$AS_i \xleftarrow{} AdequacyCalc(sub_i, N, config)$}  \;
        $FDR_i \xleftarrow{} FDRCalc(sub_i, F);$ \;
        }
        $A \xleftarrow{} AddtoArchive(samples, \Rev{AS}, FDR);$ \;
    }
    \BlankLine
    \nonl
    \HiLi 2) Perform regression analysis and select best model \;
    $regression\ models\ RM \xleftarrow{} \emptyset$ \;
    $RM \xleftarrow{} RegressionAnalysis(A);  $ \;
    $PM \xleftarrow{} argmax(RM, R^2);  $ \;
    \KwRet{$PM$}\; 
    
\caption{Building a specific prediction model for a DNN using its training set}
\label{Alg:BuildPredictionModel}
\end{algorithm}

The algorithm includes a main input sampling loop that selects a large number of input subsets with various sizes from the training set of the original DNN model $TrSet$, calculates their \Rev{AS} and FDR (lines 4-11), and adds them to the archive $A$. At the first iteration, when the archive is empty, the $UpdateSamplingSize$ function starts with a relatively small $size$. 
In each iteration, \textit{TEASMA} updates the sampling size (line 5) with a larger or smaller size according to the minimum and maximum FDR achieved by the samples currently present in the archive $A$. 
\TR{C4.8}{The sampling is performed until the entire FDR range ($0 \leq FDR \leq 1$) is covered. To achieve this, we define a stopping criterion based on the minimum ($minFDR$) and maximum ($maxFDR$) FDR values achieved by the samples in $A$ and a threshold parameter $\theta$. If $1 - maxFDR \geq \theta$, the
$UpdateSamplingSize$ function increases the sampling size to enable sampling subsets that are slightly larger than the largest subset in $A$ and can achieve higher FDR. Similarly, this function decreases the sampling size to enable sampling subsets slightly smaller than the smallest subset in $A$, if $minFDR \geq \theta$. This process is repeated with various subset sizes until $(1 - maxFDR < \theta)\: and \:(minFDR < \theta)$. 
We set $\theta = 0.05$ for our experiments in this study.
}

To be able to build a regression model, \textit{TEASMA} requires multiple input subsets of the same size. Therefore, for each subset $size$, \textit{TEASMA} samples $SN$ subsets from the training set (line 6). $SN$ is one of the input parameters of the algorithm and can be defined based on the size of the training set. 
For the reasons we describe later in Section~\ref{Sec:ExperimentTests}, input subsets can be sampled either randomly or uniformly. 
After sampling input subsets, \textit{TEASMA} calculates the \TR{C2.1, C3.1}{AS based on the selected metric and the FDR of all subsets (lines 7-9). When using MS, it can be computed based on any of the MS calculation approaches described in Section~\ref{subSec:MutationTestingMetrics}.}  
In the following sections, we investigate the relationship between FDR and \Rev{each AS }
to determine the best \Rev{metric} for building an accurate FDR prediction model. FDR is then computed based on the formula specified in Section~\ref{Sec:FaultEstimation}. Finally, the \Rev{AS} and FDR corresponding to all sampled subsets are added to the archive $A$ (line 10).

With a large number of input subsets of diverse sizes selected from the training set and their corresponding \Rev{AS} and FDR, \textit{TEASMA} builds a set of regression models (lines 12 and 13). 
For reasons we will explain in Section~\ref{Sec:ApproachRegression}, we consider non-linear regression models including quadratic, exponential, and regression trees in addition to linear regression.
To identify the best regression model, \textit{TEASMA} relies on common metrics for evaluating such models (Section~\ref{Sec:ApproachRegression}),  including the coefficient of determination ($R^2$) to measure the goodness of fit. 
Using these metrics, engineers can determine, for each DNN and training input set, the best regression model for predicting FDR. As a default, \textit{TEASMA} selects the regression model with the highest goodness of fit ($R^2$) (line 14).

\begin{algorithm}[t!]
\DontPrintSemicolon
  \KwInput{ \;
    \Indp $T$: an unlabeled test set to be assessed \;
    $N$: the original DNN model under test \;    
    }
    
    \KwOutput{ \;
    \Indp $\widehat{FDR}(T)$: predicted FDR of $T$ \;
    PI: prediction interval of $\widehat{FDR}(T)$ \; } 
    
    \BlankLine
    \nonl
    \HiLii Load \Rev{AS calculation configuration} for N \;
    \Rev{$config \xleftarrow{} LoadAdequacyConfig();$} \;
    
    \nonl
    \HiLii Load the prediction model built for N \;
    $PM \xleftarrow{} LoadPredictionModel(N)$ \;

    \BlankLine
    \nonl
    \HiLi 3) Assess test set T based on its \Rev{AS} \; 
    $\Rev{AS_T \xleftarrow{} AdequacyCalc(T, N, config);}$ \;
    $[\widehat{FDR}, PI] \xleftarrow{} Predict(PM, \Rev{AS_T})$ \;    
    
    \KwRet{[$\widehat{FDR}\ , \ PI$]}\; 
\caption{Assessing test sets}
\label{Alg:AssessingTestSets}
\end{algorithm}

Algorithm~\ref{Alg:AssessingTestSets} describes the third step of \textit{TEASMA}. As illustrated in Figure~\ref{fig:ApplicationProcess}, this step is performed on a set of unlabeled test inputs. 
\TR{C2.1, C3.1}{
Given a test set $T$ for the DNN $N$ to assess, \textit{TEASMA} starts by selecting the same configuration (e.g., generated mutants, MS calculation approach, or SC configurations)} as in the first step (line 1) and the prediction model $PM$ built in the second step (line 2). To predict the FDR of a test set $T$, \textit{TEASMA} first calculates 
\TR{C2.1, C3.1}{
AS (line 3), which is used in input of the prediction model $PM$ (line 4)}. \textit{TEASMA} provides engineers not only with the predicted FDR of $T$ but also with its prediction interval $PI$ to enable them to decide on the adequacy of a test set considering the lower or higher \Rev{confidence bounds of the prediction (Section~\ref{Sec:ApproachAssessment}). 
}

We argue that our proposed \Rev{methodology}---assuming an accurate FDR prediction model can be built specifically for each DNN using its training set---is practical.
Indeed, \Rev{even for the expensive MS calculation}, \textit{TEASMA} relies on post-training MOs, 
which are directly applied to the DNN model and do not entail any re-training, thus avoiding the main challenge with pre-training MOs.
Algorithm~\ref{Alg:BuildPredictionModel} is executed once for a specific DNN to generate mutants and create the prediction model. Subsequently, Algorithm~\ref{Alg:AssessingTestSets} can be employed multiple times to assess many test sets without incurring additional costs.
In practice, such assessment helps guide the selection of test inputs and significantly reduces test costs by only labeling a test set once it is considered adequate. However, since we build the FDR prediction model using the training set and use it to determine the adequacy of a test set, we implicitly assume that the two sets have similar distributions. This is, however, a common assumption for the training of DNNs to be considered adequate. 
\TR{C1.1, C2.8}{
In cases where this assumption would not hold such that test results would show poor model accuracy, accurate test adequacy assessment is not necessary or useful, as the DNN obviously needs to be retrained to achieve any confidence. }
 
Throughout the remainder of this section, we provide further details on key elements of \textit{TEASMA} and its application to image classification input sets.

\subsection{Calculating \Rev{AS}}
\label{Sec:ExperimentMutants}

\TR{C2.1, C3.1}{
It is essential to note that 
\textit{TEASMA} can employ any test set adequacy metric. However, the same metric must be used when calculating AS for both labeled training subsets (Step 1) and unlabeled test subsets (Step 3). 
In our study, the AS is calculated using four specific metrics: MS, DSC, LSC, and IDC. The MS calculation involves generating mutants and filtering out the trivial or equivalent ones. For SC calculation, parameters such as the lower and upper SA values are configured. To compute IDC, a VAE is trained using the training set. 
In Section~\ref{Sec:SACalculationSetup}, we describe the details of alternative AS calculations for each adequacy metric
}

\subsection{Estimating Faults}
\label{Sec:ExperimentFault}

\TR{C2.4, C3.7}{It is essential to note that \textit{TEASMA} can utilize any adequate fault identification approach. However, for the reasons outlined in Section~\ref{Sec:FaultEstimation}, this study utilizes the SOTA clustering approach introduced by Aghababaeyan  \textit{et al.}~\cite{aghababaeyan2021black}.}
\textit{TEASMA} estimates faults in the DNN model based on the mispredicted inputs of the training set. As we mentioned earlier, we implicitly assume that both the training set and the test set have a similar distribution. Except for transfer learning~\cite{niu2020decade} where there is an expected distribution shift between the training set and the test set, this default assumption is necessary for the training process to be considered adequate. Therefore, we can use the training set of the DNN model to identify faults in the DNN model and create a \Rev{model-specific FDR prediction model}. Note that, unlike test inputs, training inputs are labeled, and thus mispredicted inputs can be identified.   
To accomplish this, the entire training set is executed against the original DNN model, the mispredicted inputs are identified, and subsequently, a fault identification approach~\cite{aghababaeyan2021black} is applied to these mispredicted inputs. \textit{TEASMA} thus builds a specific fault prediction model for each individual DNN based on its training set. The performance of \textit{TEASMA} can potentially be enhanced by integrating new fault identification approaches and creating more precise prediction models.

\subsection{Sampling Input Subsets}
\label{Sec:ExperimentTests}

\textit{TEASMA} selects input subsets from the training set, calculates the \Rev{AS} and FDR of each subset, and builds a regression model. For this purpose, it requires a large number of input subsets that possess diverse \Rev{AS} and FDR values. Therefore, starting with sampling input subsets with a small size, \textit{TEASMA} updates the sampling size and continues with larger or smaller sizes where needed, until the \Rev{stopping criterion based on the minimum ($minFDR$) and maximum ($maxFDR$) FDR values achieved by the samples in $A$.} 
To build an accurate prediction model, \textit{TEASMA} requires the sampling of multiple subsets along the size range.  
The number of subsets sampled for each size is determined through an input parameter ($SN$) provided to \textit{TEASMA}, which we set to 300 in our experiments. Sampling is performed with replacement, allowing for the potential inclusion of the same test input in multiple subset samples. This approach serves the purpose of including diverse test inputs and \Rev{obtaining significant variance in AS and FDR values across samples.}

Since one of the MS formulas we use (Equation~(\ref{Eq:DeepMutationMS})) considers the number of images from each class $C_i$ in an input subset, \textit{TEASMA} complements the conventional approach of creating randomly sampled subsets by creating a set of uniformly sampled subsets. 
Random subsets may contain any number of input images from each class $C_i$ while uniform subsets contain an equal number for each class $C_i$. Uniform subsets may achieve higher MS values than random subsets based on Equation~(\ref{Eq:DeepMutationMS}). Therefore, we perform our experiments \Rev{using MS} with both types of input subsets and empirically determine which one serves our purpose better.

\subsection{Regression Analysis}
\label{Sec:ApproachRegression}

In this step, \textit{TEASMA} uses all the sampled input subsets from the training set to perform regression analysis between \Rev{AS} and FDR.  Given that post-training MOs may not be representative of real faults, we can expect non-linear relationships. In fact, as our subsequent results demonstrate, the shape of the relationship between \TR{C2.1, C3.1}{AS (calculated using MS, SC, and IDC)} and FDR is often highly non-linear and significantly varies across DNNs. Therefore, \textit{TEASMA} considers quadratic and exponential regression models, as well as regression trees when no regression function fits well. Regression trees enable the optimal partition of the \Rev{AS} range into optimal sub-ranges with similar FDR values while avoiding overfitting. Note that the best model shapes may differ across DNNs because of differences in their architecture and training sets. Consequently, for each DNN model, 
the best shape for the regression model is determined empirically, prioritizing simpler linear models when they fare well. Regardless of its shape, the regression model must be sufficiently accurate to enable practically useful FDR predictions. 
To measure the goodness of fit for each regression model, \textit{TEASMA} calculates the resulting coefficient of determination ($R^2$). \textit{TEASMA} also computes two common measures for evaluating the accuracy of predictions, namely the Mean Magnitude of Relative Error (MMRE) and the Root Mean Square Error (RMSE) to assess how accurately FDR can be predicted. To obtain more realistic results, \textit{TEASMA} performs a $K$-fold cross-validation procedure with $K=5$ and reports the average $R^2$, MMRE, and RMSE across all folds. Based on these metrics, engineers can determine for each DNN, the best regression model for predicting FDR based on \Rev{AS}.

\TR{C3.8, C4.1}{
It is important to note that according to \textit{TEASMA}'s methodology,  a specific prediction model needs to be built for each DNN model using a selected adequacy metric. This is, in practice, not an issue as one relies on the specific training set of each model. Thus, despite the generalizability of \textit{TEASMA} as an approach, it must be reapplied if the DNN model is updated or if another adequacy metric is employed.
}

\subsection{Assessing Test Sets}
\label{Sec:ApproachAssessment}
In the last step, when provided with a test set, \textit{TEASMA} computes its 
\TR{C2.1, C3.1}{AS using the original DNN and the same adequacy metric and configuration used in previous steps (e.g., previously generated mutants)}. Subsequently, it predicts the test set's FDR ($\widehat{FDR}$), including PIs based on the prediction model obtained from the previous steps.
Given a certain confidence level, such intervals specify the \Rev{probabilistic} range of actual FDR values for test sets achieving a specific \Rev{AS} value and thus provide engineers with valuable insights into the uncertainty of FDR predictions. 
\MinorRev{We compute PIs for linear, quadratic, and exponential regression models, using a standard method~\cite{montgomery2021introduction}.  In this method, the PIs are derived by adjusting the predicted values by a margin, which is determined by the standard error of the predictions. This adjustment accounts for the uncertainty in the estimates and includes a critical value from the t-distribution corresponding to the desired confidence level (95\% in our experiments). The PIs for quadratic and exponential are constructed similarly to linear regression. But in regression tree models, we compute PIs based on the bootstrapping method for non-parametric regression models~\cite{kumar2012bootstrap}, where the intervals are estimated non-parametrically using the percentile method, with bounds determined by the desired significance threshold (the 2.5th and 97.5th percentiles for a 95\% interval in our experiments) across the aggregated predictions from multiple trees.}
Depending on the context, one can then decide how conservative should decision-making be by considering the lower or higher bounds of the PI. The $\widehat{FDR}$ of a test set, if deemed sufficiently accurate based on its PI, can then be used to assess its capability to detect faults and therefore whether, in a given context, it is adequate to validate the DNN.

%
\section{Experimental Procedure}

In this section, we describe the empirical evaluation of \textit{TEASMA}, including the research questions we address, the subjects on which we conducted our evaluation, \TR{C2.1, C3.1}{the configurations for calculating AS,} and a comprehensive description of our experiments.

\subsection{Research Questions}
\label{Sec:RQs}
\TR{C2.1, C3.1}{
Our experiments are designed to answer the following research questions, relying on four test adequacy metrics for AS calculation: MS, DSC, LSC, and IDC}:

\textbf{RQ1: Can an accurate regression model be built to explain FDR as a function of \TR{C2.1, C3.1}{AS, based on any of the existing adequacy metrics?}}

In this question, we investigate the output of the second step of \textit{TEASMA} in Section~\ref{Sec:Approach}. We select a large number of input subsets from the training set and calculate their \TR{C2.1, C3.1}{AS using four existing adequacy metrics (i.e., MS, DSC, LSC, and IDC)} and their FDR using the faults estimated in the original model, as described in Section~\ref{Sec:ExperimentFault}. \NewText{We further break this question down into two sub-questions.}

\TR{C2.1, C3.1}{
\textbf{RQ1.1: Which adequacy metrics have a strong correlation with FDR?}}
In this question, we investigate, \Rev{for each adequacy metric, if the AS} of an input subset is significantly correlated with its FDR. 
\NewText{While a high correlation between AS and FDR demonstrates that there is a strong relationship between the two, it is also essential to further investigate the shape of their relationship. If the relationship between AS and FDR consistently follows a linear pattern with a slope of 1 for any of the adequacy metrics, then that metric is a surrogate for FDR and can be used directly to assess the adequacy of the test sets. However, our results, as explained in the next section, demonstrate that not only this is not the case for any of the investigated metrics, but the shape of the relationship greatly varies across subjects, thus justifying the next sub-question.}

\TR{C2.1, C3.1}{
\textbf{RQ1.2: Based on which adequacy metrics can an accurate regression model be built to predict FDR from AS based on the training set? }
For adequacy metrics with a strong correlation between AS and FDR,} 
we resort to regression analysis to model the relationship between \Rev{AS} and FDR. Consequently, we investigate if an accurate regression model can be built to predict the FDR of an input subset from its \Rev{AS}, based on the training set. If that is the case, the predicted FDR can serve as a good basis on which to decide whether a test set provides sufficient confidence about the validation of the DNN before its deployment. Though such questions have been investigated \Rev{using MS} for traditional software~\cite{andrews2006using, papadakis2018mutation}, that has not been the case for DNNs.

\TR{C2.1, C3.1}{
\textbf{RQ2: Which adequacy metrics lead to better FDR predictions on test sets?
}}
 
In this question, we evaluate the last step of \textit{TEASMA} in Section~\ref{Sec:Approach} and investigate \TR{C2.1, C3.1}{separately for each selected adequacy metric,} if we can accurately predict a test set's FDR using the regression model built in the previous step on the training set.

\TR{C2.1, C3.1}{
\textbf{RQ3: How accurately can we predict a test set's FDR?}
In this question, we compare the results obtained from the last step of \textit{TEASMA} using selected adequacy metrics and identify which one can be used to most accurately predict a test set's FDR, with a reasonably small confidence interval.}

\subsection{Subjects}
\label{Sec:Subjects}

\begin{table*}[htb]
    \centering   
    \small
    \caption{Information about experimental subjects}
    \resizebox{0.8\textwidth}{!}{
    \begin{tabular}{|c|      ccc|   cccc |     }
    \hline 
          ID  &Input   &Training   &Test           &Model   &Epochs   &Train      &Test     \\ 
              &set     &set        &set            &        &         &accuracy   &accuracy  \\ \hline 
          S1  &MNIST          &60,000  &10,000    &LeNet-5  &12         &96\%  &96\%     \\    \hline
          S2  &\multirow{2}{*}{Cifar-10}       &\multirow{2}{*}{50,000}  &\multirow{2}{*}{10,000}    &Conv-8  &20         &94\%  &85\%    \\  \cline{1-1}  \cline{5-8}
          S3  &  &    &      &ResNet20       &100     &97\%       &89\%      \\    \hline   
          S4  &Fashion-MNIST  &60,000  &10,000    &LeNet-4  &20         &94\%  &90\%   \\    \hline 
          S5  &\multirow{2}{*}{SVHN}           &\multirow{2}{*}{73257}  &\multirow{2}{*}{26032}    &LeNet-5  &20         &94\%  &87\%   \\    \cline{1-1}  \cline{5-8}
          S6  &  &    &      &VGG16       &10     &97\%       &93\%      \\    \hline 
          \NewText{S7}  &\NewText{Cifar-100}  &\NewText{50,000}    &\NewText{10,000}      &\NewText{ResNet152}       &\NewText{10}     &\NewText{89\%}       &\NewText{79\%}      \\    \hline
          \NewText{S8}  &\NewText{ImageNet}  &\NewText{1,281,167}    &\NewText{50,000}      &\NewText{InceptionV3}       &\NewText{pre-trained}     &\NewText{89\%}       &\NewText{77\%}      \\    \hline
          
    \end{tabular}
    }
    \label{tab:Subjects}
\end{table*}

We performed our study on different combinations of input sets and models. Since we rely on a SOTA fault estimation approach for DNNs that is tailored for image inputs, we selected a set of widely-used publicly available image input sets including MNIST~\cite{deng2012mnist}, Fashion-MNIST~\cite{deng2012mnist}, Cifar-10~\cite{Cifar10}, SVHN~\cite{netzer2011reading}, 
\TR{C1.4, C2.7, C3.9, C4.10}{}
\NewText{Cifar-100~\cite{krizhevsky2009learning}, and ImageNet~\cite{ILSVRC15}}. \NewText{Most of these input sets have also been used for evaluating SOTA mutation testing tools and operators~\cite{ma2018deepmutation, shen2018munn, hu2019deepmutation++, humbatova2021deepcrime}, as well as for evaluating adequacy metrics such as SC and IDC.}
MNIST is a set of images of handwritten digits and Fashion-MNIST contains images of clothes that are associated with fashion and clothing items. Each one of MNIST and Fashion-MNIST represents a 10-class classification problem and consists of 70,000 grayscale images of a standardized 28x28 size. Cifar-10 is another widely-used input set that includes a collection of images from 10 different classes (e.g., cats, dogs, airplanes, cars). SVHN~\cite{netzer2011reading}, a set of real-world house numbers collected by Google Street View, can be considered similar to MNIST since it includes images representing a sequence of digits from 10 classes of digits. The Cifar-10 and SVHN input sets contain 32x32 cropped colored images. 
\TR{C1.4, C2.7, C3.9, C4.10}{}
\NewText{Moreover, we selected two large-scale image input sets, namely Cifar-100~\cite{krizhevsky2009learning} and ImageNet~\cite{ILSVRC15}, that contain realistic and feature-rich images. The CIFAR-100 input set is a widely used benchmark in the field of computer vision, consisting of 100 classes of images, each containing 600 images. Among these images, 500 are allocated for training, while the remaining 100 are designated for testing, resulting in a total of 10,000 test images and 50,000 training images. Each image in the CIFAR-100 input set is sized 32×32 pixels and is in full color. ImageNet is a widely recognized large-scale input set designed for visual object recognition research in the field of computer vision. It includes over 14 million labeled images across 1,000 object categories~\cite{ILSVRC15}. We leverage its most commonly used subset, ImageNet-1k, from the ILSVRC2012 competition with a training set of 1,281,167 images and a validation set of 50,000 images for evaluation. 
The images in the ImageNet input set are sized 224x224 pixels, featuring a higher resolution compared to other input sets such as CIFAR-100.
}

In our experiments, we trained \Rev{seven} SOTA DNN models using the above input sets: LeNet-5~\cite{lecun1998gradient}, LeNet-4~\cite{lecun1998gradient}, ResNet20~\cite{he2016deep}, 
\TR{C1.4, C2.7, C3.9, C4.10}{ResNet152~\cite{he2016deep}, InceptionV3~\cite{szegedy2016rethinking}}, VGG16~\cite{simonyan2014very}, and an 8-layer Convolutional Neural Network (Conv-8). Since not all of the post-training MOs are applicable to every DNN model, as described in Section~\ref{Sec:MutantGeneration}, we selected a diverse range of DNN models with different internal architectures to ensure that all of DeepMutation's MOs are examined in our experiment as extensively as possible. 
\TR{C1.4, C2.7, C3.9, C4.10}{We should note that we use a pre-trained InceptionV3 model~\cite{szegedy2016rethinking} for the ImageNet input set. Additionally, following standard transfer learning practices, we employ a pre-trained ResNet152 model initially trained on the ImageNet input set, and then fine-tune it for the CIFAR-100 input set rather than training it from scratch~\cite{kornblith2019better}. This is motivated by the need for computational resource efficiency and improved performance. }

All the feasible combinations of DNN models and input sets we experiment with are referred to as subjects. Overall, we perform our experiment on \Rev{eight} subjects including a wide range of diverse image inputs and DNN architectures.
Table~\ref{tab:Subjects} lists the details of each subject including the size of test and training input sets, the number of epochs used to train the model, \Rev{and the accuracy of the DNN model on the training and test input sets.}

\TR{C1.1, C2.8}{
\NewText{It is important to note that our experiments were conducted using the original test sets of the subject models, as defined in their respective original sources~\cite{deng2012mnist, netzer2011reading, Cifar10, krizhevsky2009learning, ILSVRC15}, to eliminate potential biases, as we were not involved in their selection process. The subjects' DNN models cover a wide but high accuracy range, from 77\% to 96\%, as detailed in Table~\ref{tab:Subjects}.
As a result, scenarios where the model’s accuracy is poor, such as when a severe distribution shift occurs between the training and test set, are not considered in our experiments. Such scenarios reveal the need for retraining, and thus a test adequacy assessment becomes superfluous, as the goal of such assessment is to provide confidence in high-accuracy test results.} }

\subsection{\NewText{AS calculation Configuration}}
\label{Sec:SACalculationSetup}

\begin{figure}
    \includegraphics[width = \columnwidth]{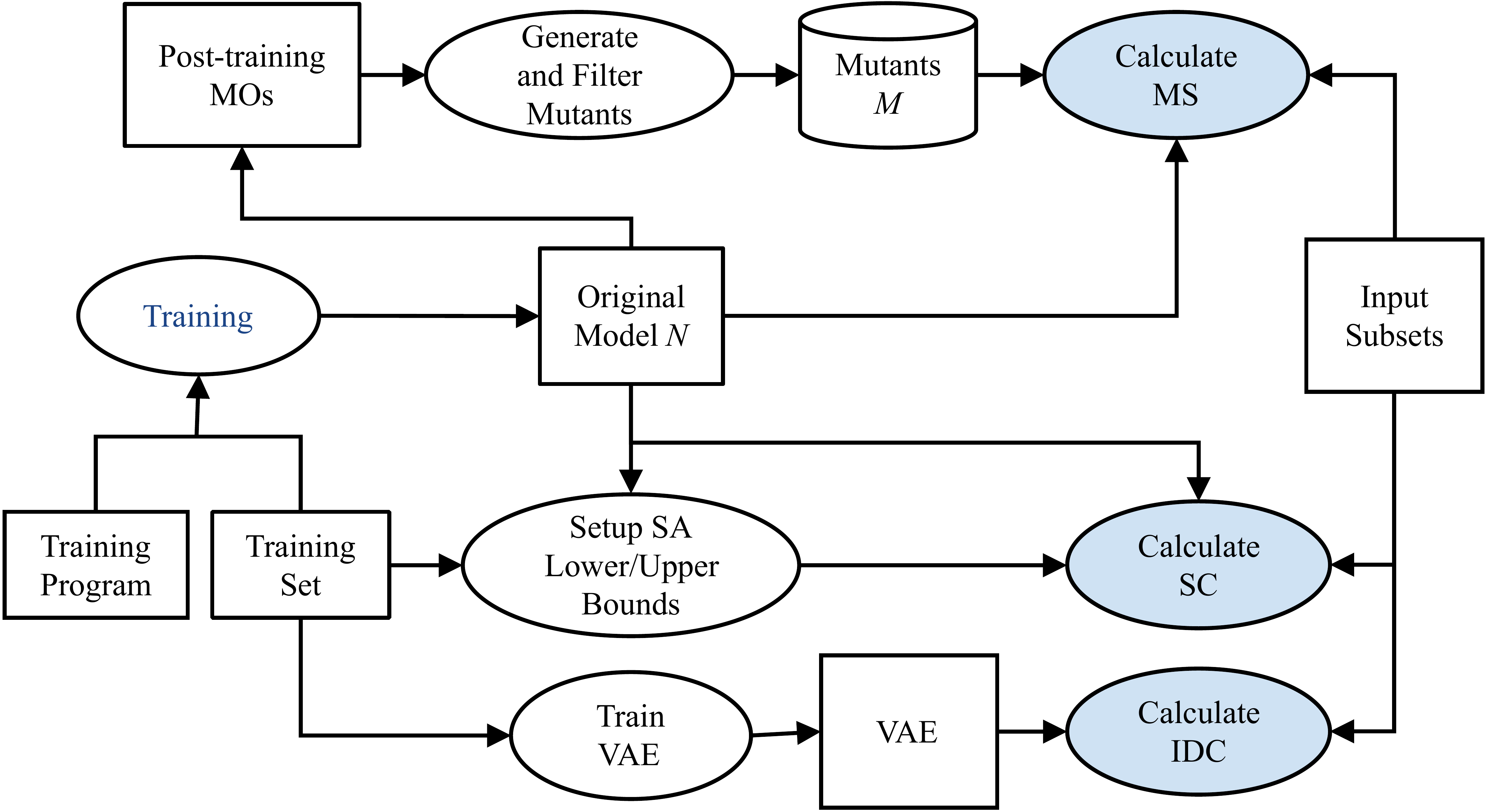}
    \centering
    \caption{Configuring MS, SC, and IDC calculation in our experiments }
    \label{fig:ASConfiguration}
\end{figure}

\TR{C2.1, C3.1}{In this section, we describe the configurations used in our experiments to calculate AS based on alternative adequacy metrics, as illustrated in Figure~\ref{fig:ASConfiguration}.} 
To calculate MS precisely, we need to complement the mutant generation process by identifying and excluding mutants that are trivial or equivalent. 
\Rev{In our experiments with MS, we use} DeepMutation++~\cite{hu2019deepmutation++}, the most recent SOTA mutation testing tool for DNNs that implements eight post-training MOs suitable for FNNs. Some of the implemented MOs need to be configured by setting values for their parameters. For example, the Neuron Effect Blocking (NEB) operator is configured by determining the number of neurons it affects. By default, the DeepMutation++ tool configures this operator to sample a predefined ratio of 1\% of the total neurons in a DNN model and limits the total number of generated mutants to 50. \Rev{We use} the default configuration for generating mutants.
Furthermore, \Rev{we perform the same procedure proposed by the DeepMutation++ tool and use} its default thresholds to filter out some of the generated mutants. 
Specifically, \Rev{we exclude} mutants when they yield less than 90\% of the original model's accuracy or mispredict more than 20\% of the inputs correctly predicted by the original model, as reported in the original papers~\cite{ma2018deepmutation, hu2019deepmutation++}. Additionally, \Rev{we perform} an extra step to filter out equivalent mutants.  
\TR{C3.14}{
For this purpose, the entire training set is executed against the mutants and those that cannot be killed by any training input are considered equivalent and therefore excluded.}
\Rev{Various approaches can be employed} for MS calculation as explained in Section~\ref{subSec:MutationTestingMetrics}. Consequently, in our experiment, we examine the correlation between FDR and MS, calculated using each approach, and empirically determine which approach results in the most accurate prediction model.
\TR{C1.3}{}
\NewText{The post-training MOs utilized in our experiments, as implemented in the DeepMutation++ tool~\cite{hu2019deepmutation++}, are as follows:} 

\begin{itemize}
\item \NewText{Gaussian Fuzzing (GF): Alters the weights of a DNN using a Gaussian distribution. This introduces variations in the connection strengths between neurons, subtly modifying the model's behavior.}

\item \NewText{Weight Shuffling ($WS$): Randomly shuffles the weights of connections for a selected neuron, affecting how inputs are processed by subsequent layers and potentially altering the output.}

\item \NewText{Neuron Effect Blocking ($NEB$): Temporarily disables a neuron's effect on subsequent layers by setting its outgoing weights to zero. This simulates the impact of a non-functional neuron on the model’s decision process.}

\item \NewText{Neuron Activation Inverse ($NAI$): Inverts the activation status of neurons. By changing the sign of output values before applying activation functions, this operator can alter the behavior of a model.}
\item \NewText{Neuron Switch ($NS$): Swaps two neurons within the same layer, changing their roles and potentially altering the layer's output pattern.}

\item \NewText{Layer Removal ($LR$): Randomly removes a layer of the DNNs, useful for assessing the redundancy and criticality of particular layers. The $LR$ operator is restricted to layers whose input and output shapes are consistent~\cite{ma2018deepmutation}.}

\item \NewText{Layer Addition ($LA$): Adds an additional layer to the model, which can introduce significant changes to the model's behavior and output. $LA_m$ operates under similar conditions as the LR operator to avoid breaking original DNNs.}

\item \NewText{Layer Duplication ($LD$): Inserts a copied layer immediately following its original layer. The LD operates under similar conditions as the LR operator to avoid breaking the original DNNs.}

\end{itemize}

\TR{C1.3, C4.11}{
It is important to note that not all MOs are applicable to every DNN architecture. Furthermore, the selection of MOs can affect the effectiveness of mutation analysis and thus, the results of applying \textit{TEASMA} with MS. Consequently, we investigate the potential impact of MO choices in our experiments. 
We measure the correlation for mutants generated by each MO separately and then attempt to find subsets of operators leading to higher correlation. We compare the highest correlation achieved by subsets of MOs with that obtained with all MOs. If the correlation of the former is higher than the latter, then we only consider mutants generated by the optimal subset with the highest correlation. 
}

\TR{C2.1, C3.1}{
To calculate SC, we use both the LSA and DSA metrics, consequently calculating DSC and LSC scores. 
When calculating DSA, first, the activation traces of training inputs for a chosen layer in the DNN are captured to establish a reference set. Then, for a given test input, the activation trace is computed at the same layer. Subsequently, the Euclidean distance between the activation trace of the test input and its nearest neighbor among the training traces is calculated, serving as the test input's DSA score.
When calculating DSA for a training input, as required in Algorithm~\ref{Alg:BuildPredictionModel}, it is necessary to exclude this particular input from the training set; otherwise, the DSA value would be zero. 
Similarly, to accurately compute LSA for a training input, it is removed from the training set, and the probability density is recalculated based on the remaining inputs. Subsequently, LSA is calculated using the updated density estimation. This process is repeated for each training input to calculate its DSA and LSA values.
In our experiments, following the approach of the original paper~\cite{kim2023evaluating}, we calculated DSA and LSA using the deepest (i.e., the last hidden) layer of the model. Additionally, we set the number of buckets for calculating LSC and DSC to 1,000, consistent with the value employed in the original paper~\cite{kim2023evaluating}. The lower bound and upper bound for calculating the SC of both the training and test subsets are determined using the SA values observed during the training. 
}

\TR{C2.1, C3.1}{
The IDC framework instantiation involves training a VAE component, selecting an existing OOD component, and configuring IDC parameters~\cite{dola2023input}. In the IDC workflow, a VAE is trained with the training set of each input set to model the input distribution in a latent space. 
In our experiments, we employ the publicly available VAEs from Dola \textit{et al.}~\cite{dola2023input} for our first four subjects, which share identical input sets with those used in the original evaluation of the IDC framework. For the remaining input sets, we trained new VAEs. Consequently, in our experiments, six VAEs are employed for the MNIST, Cifar-10, Fashion-MNIST, SVHN, Cifar-100, and ImageNet input sets. All these VAEs use $\beta$-TCVAE~\cite{chen2018isolating} model based on the architecture proposed by Burgess \textit{et al.}~\cite{burgess2018understanding}, with latent sizes of 
6, 32, 6, 64, 32, and 64, respectively.
}

\TR{C2.1, C3.1}{
To calculate the IDC of a test set, its test inputs are encoded into the latent space using the VAE's encoder component. Then test set's coverage is measured based on the adequacy of its test inputs in covering the partitions of the latent space. Crucially, an OOD detector is employed in the IDC framework to filter OOD test inputs and ensure that only realistic, in-distribution inputs are processed by the VAE's encoder, making this the most computationally expensive step of the workflow. However, in our experiments, where IDC is computed for test inputs directly selected from the original test set of an input set, the OOD filter step becomes redundant as these inputs are by default in-distribution. Consequently, we exclude the OOD detection step of the IDC workflow in our experiments. Additionally, we use the default IDC parameter configurations from the original implementation of the IDC framework~\cite{dola2023input}.
As shown in Figure~\ref{fig:ASConfiguration}, calculating IDC for input subsets does not require the presence of the DNN model, making it independent of the model and relying solely on the feature interactions of its inputs.
}

\subsection{Experiments}
\label{Sec:ExperimentVariations}

This section explains how we conduct our experiments to evaluate \textit{TEASMA} \TR{C2.1, C3.1}{individually for each selected metric} and answer the research questions described in Section~\ref{Sec:RQs}. 
To answer RQ1, we perform the first step of the \textit{TEASMA} approach (Section~\ref{Sec:Approach}) as detailed in Algorithm~\ref{Alg:BuildPredictionModel} using the training input set of our subjects. 
As a result, we calculate FDR and \Rev{AS} for a large number of diverse input subsets sampled from the training set. 
Subsequently, we measure the correlation between FDR and \Rev{AS} using the non-parametric Spearman coefficient based on all subsets, since we cannot assume linearity.  
Next, using all the training subsets, we proceed to the second step of \textit{TEASMA}, regression analysis, relying on both linear and non-linear models. \textit{TEASMA} performs a $K$-fold cross-validation procedure with $K=5$ and reports the average of $R^2$, MMRE, and RMSE across all folds for each regression model separately. The best regression model can be selected based on these metrics.

To summarize, the experiment that we conducted \TR{C2.1, C3.1}{for each adequacy metric} using the training set to answer RQ1 included the following steps:
\begin{enumerate} 
    \item Perform the first step of \textit{TEASMA} and calculate \Rev{AS} and FDR for each input subset 
    \item Measure the Spearman correlation between FDR and \Rev{AS}  
    \item Perform the second step of \textit{TEASMA} and conduct regression analysis between FDR and \Rev{AS} using linear and non-linear models (quadratic, exponential, and regression trees)
    \item Evaluate regression models to identify the best one based on their $R^2$, MMRE, and RMSE calculated based on $K$-fold cross-validation 
\end{enumerate}

To address RQ2, we sample a large number of input subsets from the test set, calculate their \TR{C2.1, C3.1}{AS using the same adequacy metric and configuration from the previous steps}, and leverage the previously built and selected regression model to predict FDR ($\widehat{FDR}(T)$). We should note that \Rev{for calculating MS,} we determine killed mutants by comparing their output with that of the original model and thus the actual labels of test inputs are not required. \TR{C2.1, C3.1}{Similarly, calculating AS based on SC and IDC does not necessitate knowledge of the inputs' labels.
It is important to remark that we consistently use the same set of randomly sampled test subsets in our experiments with all adequacy metrics.}

To estimate the actual FDR ($ActualFDR$) of each test subset, we use the same fault clusters identified based on the training set as described in Section~\ref{Sec:ExperimentFault} and we assign each mispredicted test input to one of the clusters. Since the fault estimation approach (Section~\ref{Sec:FaultEstimation}) employs the HDBSCAN algorithm~\cite{campello2013density}, a density-based hierarchical clustering method where each cluster is characterized by a number of core points, we assign each mispredicted test input to the cluster featuring a core point that is closest to the input.
To ensure an accurate calculation of $ActualFDR$ and allow the test set to reach the maximum FDR value of 1, we exclusively take into account fault clusters that can be detected by the test set. To achieve this, we limit our denominator in Equation~(\ref{Eq:FDR}) to the number of fault clusters with at least one mispredicted test input assigned to them. Consequently, we conduct our experiments based on the faults detectable by the test set, so that our results are not affected by the varying quality of the provided test sets across subjects.

Subsequently, we are able to compare for each test subset, the predicted FDR with the actual one and investigate if the former is closely aligned with the latter and can thus be trusted to make decisions about the test set.
Therefore, to answer RQ2 \TR{C2.1, C3.1}{for each metric}, we went through the following steps:
\begin{enumerate}
    \item Randomly sample a large number of input subsets of diverse sizes from the test set
    \item Calculate \TR{C2.1, C3.1}{AS, using the same metric and configuration from the previous steps,} and FDR ($ActualFDR$) for each test subset
    \item Use previously selected prediction model to compute $\widehat{FDR}$ based on \Rev{AS} for each test subset   
    \item Based on all test subsets, measure the linear correlation ($R^2$) between $\widehat{FDR}$ and $ActualFDR$ and check the slope of the regression line 
\end{enumerate}

\TR{C2.1, C3.1}{
To address RQ3, we investigate the accuracy of selected metrics within \textit{TEASMA}, in in terms of FDR predictions. 
Specifically, we focus on adequacy metrics identified in RQ2 that exhibit better FDR predictions characterized by higher linear correlation ($R^2$) with the actual FDR and slopes closer to 1. We compute RMSE between $\widehat{FDR}$ and $ActualFDR$ and compare such prediction accuracy for each metric. Subsequently, we identify the metric with the lowest average RMSE across subjects, to determine the best metric.
}

As described in Section~\ref{subSec:MutationTestingMetrics}, different methods have been proposed to determine whether a mutant is killed and therefore to calculate MS. To fully address our research questions, we ran a comprehensive analysis by considering all the MS calculation variants described in Section~\ref{subSec:MutationTestingMetrics}. As a result, we ran three variations of our experiments that only differ in the way they calculate MS: (E1) $DeepMutationMS$, (E2) $StandardMS$, and (E3) $KSBasedMS$.

\NewText{The total execution time to conduct our entire experiments, on all subjects using the four investigated adequacy metrics, on one Nvidia P100 GPU with 12GB memory, would have taken approximately five months. Through extensive parallelization and using two separate computing environments, we reduced it to 40 days.
We conducted our experiments with the IDC metric on a machine with two Nvidia Quadro RTX 6000 GPUs (each with 24GB GPU memory), an Intel Xeon Gold 6234 16-Core CPU, and 187GB of RAM. For the remaining metrics, we used a cloud computing environment provided by the Digital Research Alliance of Canada~\cite{computecanada}, utilizing the Cedar cluster with 1352 various Nvidia GPUs (P100 Pascal and V100 Volta) with memory ranging from 8 to 32 GB.
}

\subsection{Data Availability}

\MinorRev{
We have released the replication package of this paper online including the code, mutant models, detailed information on the MO subset analysis, and comprehensive results of all examined regression models for all metrics across all subjects~\cite{replicationpackage}.
}

\section{Results}
\label{sec:Results}

In this section, we report results pertaining to our research questions and discuss their practical implications. \TR{C2.4, C3.6}{Furthermore, to enable a full validity analysis of our results, we report the details of the fault clusters used in our experiments.}

\subsection{
\textbf{RQ1: Can an accurate regression model be built to explain FDR as a function of \TR{C2.1, C3.1}{AS, based on any of the existing adequacy metrics?}}}

\subsubsection{\NewText{\textbf{RQ1.1}}}  
\TR{C2.1, C3.1}{Which adequacy metrics have a strong correlation with FDR?}

To investigate the extent of the correlation between \Rev{AS} and FDR, \Rev{for each adequacy metric}, we randomly sample a large number of input subsets from the training set. We then compute the non-parametric Spearman correlation coefficient, for all adequacy metrics in Table~\ref{tab:CorrelationMSFDR}, between \Rev{AS} and FDR for all these subsets.

\begin{table}[b]
    \centering   
    \small
    \caption{Spearman correlation coefficients computed between \Rev{AS and FDR using each adequacy metric,} for random input subsets of the training set for each subject}
     \resizebox{0.99\columnwidth}{!}{
    \begin{tabular}{  |c|  c|  c|  c|  c|  c|  c|  c|  c|  c|}
    \cline{1-10} 
             \multicolumn{2}{|c|}{Adequacy metric}       &\multicolumn{8}{c|}{Subjects}     \\ \cline{1-10}
               &Variants         &S1     &S2  &S3     &S4  &S5  &S6   &\NewText{S7}  &\NewText{S8}\\ \cline{2-10} 
            \multirow{2}{*}{MS}  &E1  &\textbf{0.99}   &\textbf{0.99}   &\textbf{0.99}  &\textbf{0.99}  &\textbf{0.99} &\textbf{0.98}   &\NewText{\textbf{1}} &\NewText{\textbf{1}} \\\cline{2-10}  
                                    &E2          &0.92       &0.73       &0.66       &0.91       &0.86      &0.91   &\NewText{0.53} &\NewText{0.45}  \\  \cline{1-10}           
            \multirow{2}{*}{\NewText{SC}}    &\NewText{DSC}      &\NewText{0.99}       &\NewText{0.99}       &\NewText{0.99}       &\NewText{0.99}       &\NewText{0.99}      &\NewText{0.99} &\NewText{0.99}   &\NewText{1}  \\  \cline{2-10}
            &\NewText{LSC}      &\NewText{0.99}       &\NewText{0.99}       &\NewText{0.99}       &\NewText{0.99}       &\NewText{0.99}      &\NewText{0.99}   &\NewText{0.99} &\NewText{1}  \\  \cline{1-10}

            \multicolumn{2}{|c|}{\NewText{IDC}}  &\NewText{0.99}       &\NewText{0.99}       &\NewText{0.99}       &\NewText{0.99}       &\NewText{0.99}      &\NewText{1}   &\NewText{1} &\NewText{1}  \\  \cline{1-10}
            
    \end{tabular}
    }
    \label{tab:CorrelationMSFDR}
\end{table}

\Rev{When using MS for calculating AS, we consider different MS calculation variants.} 
The correlation analysis results using Spearman correlation coefficients for experiment variants E1 and E2 are shown in Table~\ref{tab:CorrelationMSFDR}. \Rev{We further extended our experiment to include a large number of uniformly sampled input subsets from the training set since the MS calculation proposed by DeepMutation~\cite{ma2018deepmutation} in Equation~(\ref{Eq:DeepMutationMS}) considers the number of classes of each mutant killed by inputs in a test set. However,} we only report results for random subsets as they yielded slightly higher correlations and reporting results from uniform subsets does not provide new insights. 

We did not report the results for experiment variant E3 since all of the correlation values for both random and uniform subsets across all subjects are below 0.1, indicating an absence of correlation. This can be explained by the fact that MS is determined by computing the average $KillingScore$ (Equation~(\ref{Eq:KillingScore})) of all inputs within a subset. Consequently, for subsets of different sizes, MS calculated based on $KSBasedMS$ (Equation~(\ref{Eq:KillingScoreBasedMS})) falls within a similar range (0 to 0.1). As discussed in Section~\ref{subSec:MutationTestingMetrics}, this confirms that calculating the MS of a subset as the average capability of its inputs to kill mutants cannot be used as an indicator of the capability of the entire subset to kill mutants. 

E1 shows the strongest correlation across all subjects between MS and FDR (above 0.98 across all subjects). In this experiment, we calculate MS as proposed originally by DeepMutation (Equation~(\ref{Eq:DeepMutationMS})). In contrast, for experiment E2, $R^2$ ranges from 0.45 to 0.92, which is still a strong correlation but not nearly as high as for E1.  
The method for identifying killed mutants and calculating MS therefore has a substantial impact on the results of mutation analysis. Given the strong correlations achieved in E1 and E2 across all subjects, we conclude that mutants generated by DeepMutation's post-training MOs are highly associated with faults.
Given the above results, \Rev{in our experiments using MS for AS calculation,} we focus solely on $DeepMutationMS$ and random subsets from now on. 

\TR{C2.1, C3.1}{
As reported in Table~\ref{tab:CorrelationMSFDR}, the correlation analysis using Spearman correlation coefficients for DSC, LSC, and IDC showed results similar to MS based on variant E1, indicating a strong correlation (above 0.99) across all subjects. These results suggest that all utilized metrics, DSC, LSC, IDC, and MS calculated using $DeepMutationMS$ are all promising metrics for test assessment. However, beyond demonstrating a strong correlation, it is essential to investigate the shape of the relationship between AS and FDR for these metrics, which we explore in the following section.} 

\TR{C1.3, C4.11}{}
Before proceeding to the next steps with MS, we investigated the potential impact of MO choices when calculating MS. Ideally, MOs capable of generating mutants associated with faults are preferred in mutation analysis~\cite{just2014mutants}. Therefore, it is imperative to select mutants based on which MS exhibits a stronger correlation with FDR. 
We calculated the correlation between FDR and MS individually for each MO as well as for different subsets of operators. 
Our results show that none of the MOs or MO subsets achieved a significantly higher correlation than the correlation obtained with all MOs. Consequently, we proceeded to the next step considering, for each subject, all of the applicable MOs and performed regression analysis to model the relationship between MS and FDR on training set. 

\begin{tcolorbox}   
    \Rev{\textbf{Answer to RQ1.1}: Distribution-based coverage metrics including DSC, LSC, and IDC have a very strong Spearman correlation with FDR. Post-training MOs generate mutants that are highly associated with faults; thus, based on these mutants, MS also has a very strong correlation with FDR, especially when computed with $DeepMutationMS$.  }
\end{tcolorbox}

\begin{figure*} 
     \centering
     \begin{subfigure}[b]{0.23\textwidth}
         \centering
         \includegraphics[width=\textwidth]{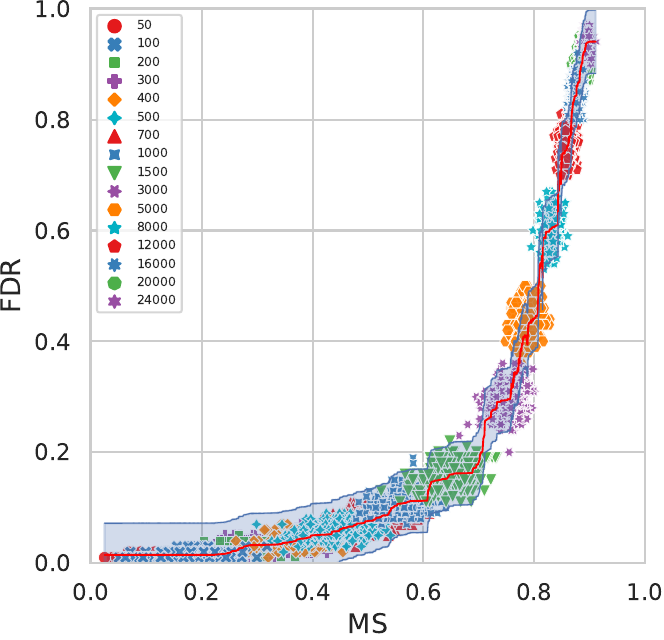}
         \caption{$Subject\ S1$}
         \label{fig:TrainS1}
     \end{subfigure}
     \hfill
     \begin{subfigure}[b]{0.23\textwidth}
         \centering
         \includegraphics[width=\textwidth]{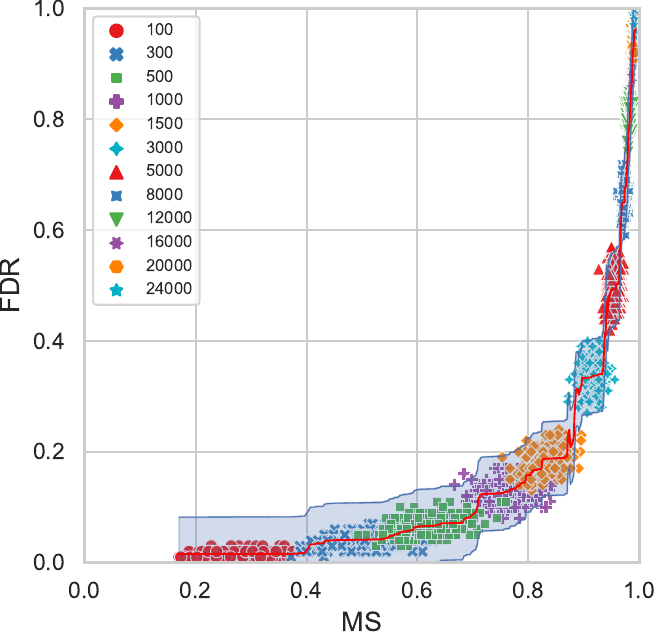}
         \caption{$Subject\ S2$}
         \label{fig:TrainS2}
     \end{subfigure}
     \hfill
     \begin{subfigure}[b]{0.23\textwidth}
         \centering
         \includegraphics[width=\textwidth]{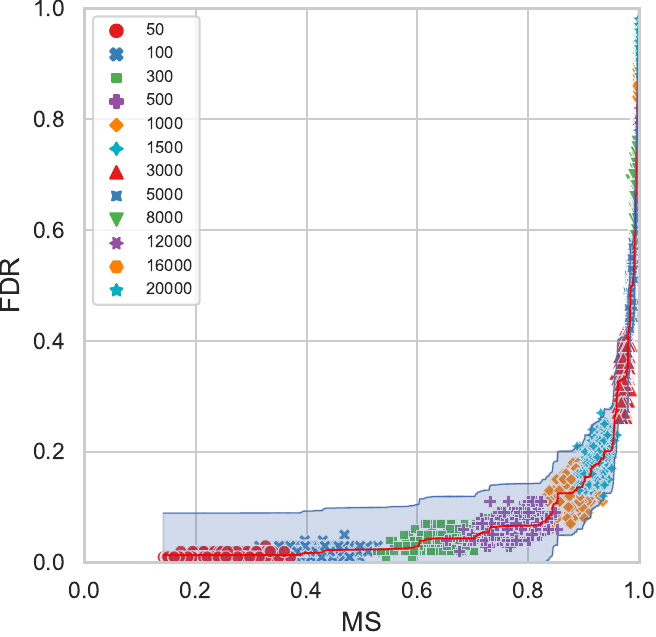}
         \caption{$Subject\ S3$}
         \label{fig:TrainS3}
     \end{subfigure}
     \hfill
     \begin{subfigure}[b]{0.23\textwidth}
         \centering
         \includegraphics[width=\textwidth]{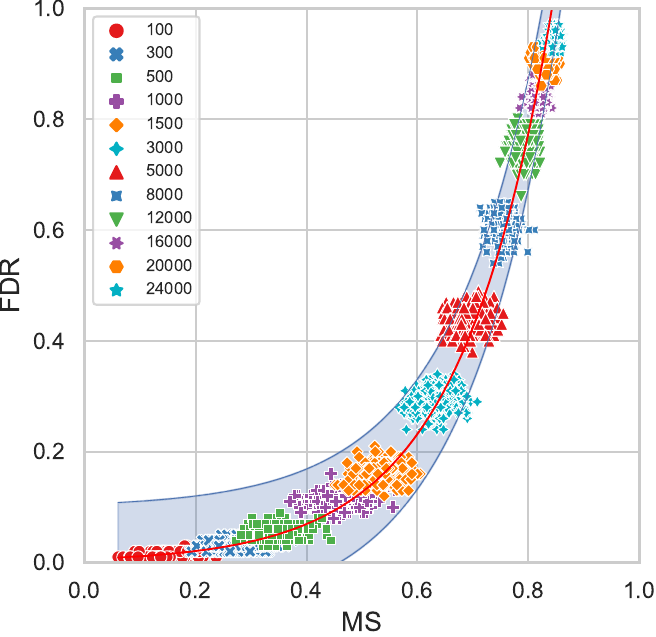}
         \caption{$Subject\ S4$}
         \label{fig:TrainS4}
     \end{subfigure}
     \hfill
     \begin{subfigure}[b]{0.23\textwidth}
         \centering
         \includegraphics[width=\textwidth]{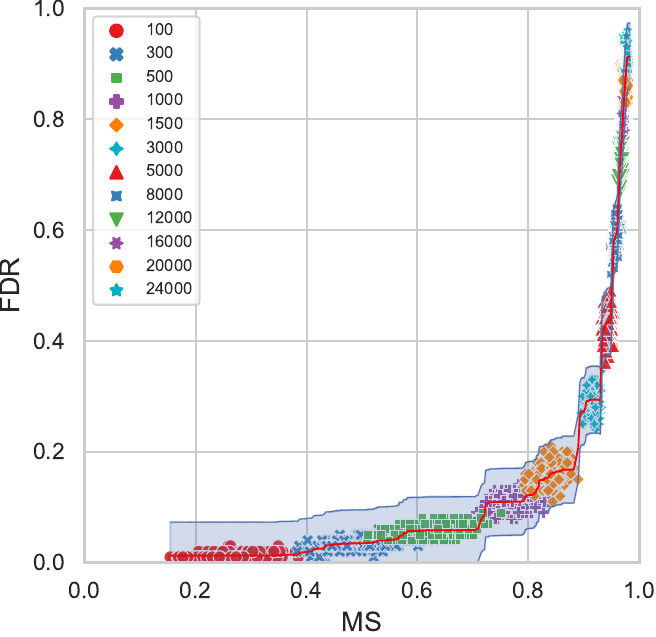}
         \caption{$Subject\ S5$}
         \label{fig:TrainS5}
     \end{subfigure}
     \hfill
     \begin{subfigure}[b]{0.23\textwidth}
         \centering
         \includegraphics[width=\textwidth]{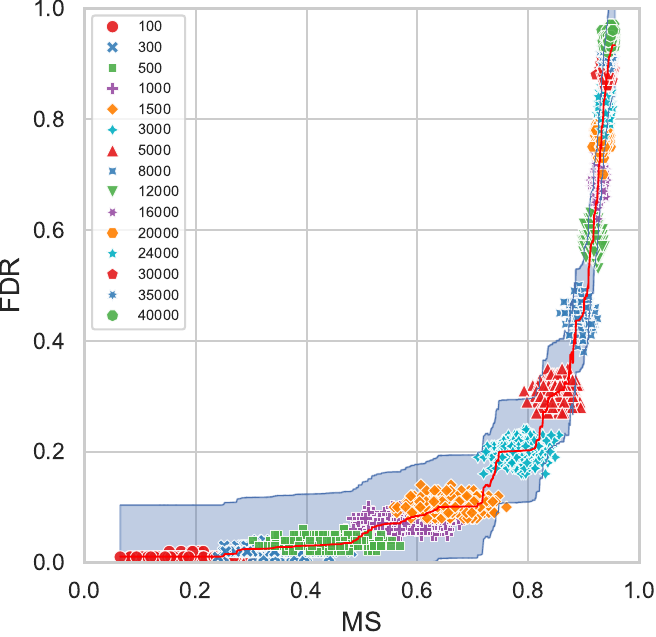}
         \caption{$Subject\ S6$}
         \label{fig:TrainS6}
     \end{subfigure}
     \hfill
     \begin{subfigure}[b]{0.23\textwidth}
         \centering
         \includegraphics[width=\textwidth]{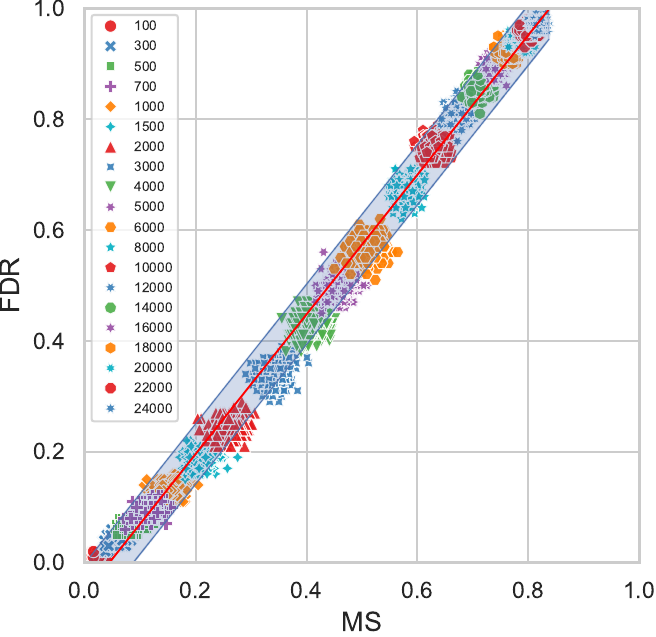}
         \caption{$\NewText{Subject\ S7}$}
         \label{fig:TrainS7}
     \end{subfigure}
     \hfill
     \begin{subfigure}[b]{0.23\textwidth}
         \centering
         \includegraphics[width=\textwidth]{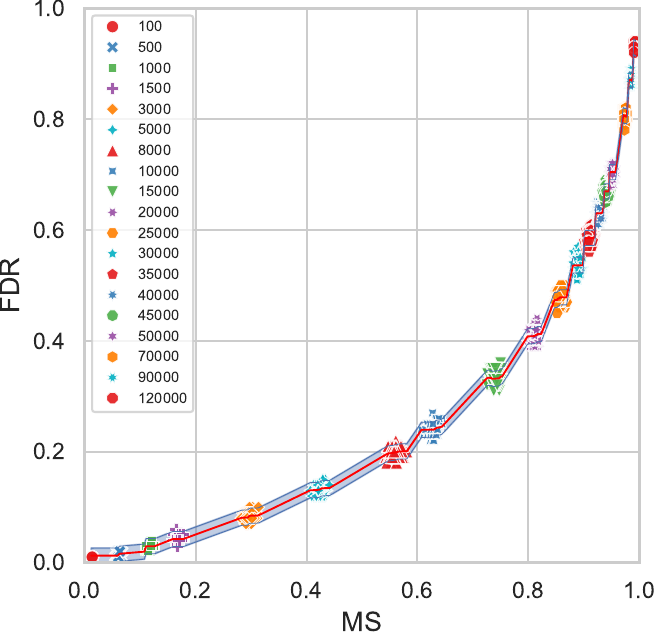}
         \caption{$\NewText{Subject\ S8}$}
         \label{fig:TrainS8}
     \end{subfigure}
        \caption{Selected regression models across subjects using MS based on the training set}
        \label{fig:MSTrainSetRegressionAll}
\end{figure*}

\begin{table}[t]
    \centering   
    \caption{Accuracy and shape of the selected regression models, \NewText{built using each adequacy metric}, when predicting FDR based on the training set and using a 5-fold cross-validation procedure}
     \resizebox{0.99\columnwidth}{!}{
    \begin{tabular}{  |c|  c|    c| c| c| c| c| c|  c| c|}
    \cline{1-10} 
    Adequacy  &Accuracy      &\multicolumn{8}{c|}{Subjects}     \\ \cline{3-10}
    metric &metric       &S1   &S2     &S3     &S4     &S5  &S6  &\NewText{S7}  &\NewText{S8} \\ \cline{1-10}
    \multirow{4}{*}{MS (E1)} &$R^2$       &0.99   &0.99   &0.99      &0.98   &0.99   &0.98  &\NewText{0.99}  &\NewText{1} \\ \cline{2-10}    
    &MMRE        &0.18   &0.13   &0.16      &0.14   &0.1   &0.13  &\NewText{0.25}  &\NewText{0.04}    \\ \cline{2-10}
    &RMSE        &0.03   &0.04   &0.04      &0.05   &0.03   &0.05  &\NewText{0.03}   &\NewText{0.01} \\ \cline{2-10}
    &Shape*      &RT     &RT     &RT        &E      &RT     &RT  &\NewText{L}  &\NewText{RT}    \\ \cline{1-10}
    
    \multirow{4}{*}{DSC} &$R^2$       &\NewText{1}   &\NewText{1}   &\NewText{0.99}      &\NewText{1}   &\NewText{1}   &\NewText{0.99}  &\NewText{0.99}   &\NewText{0.99} \\ \cline{2-10}    
    &MMRE        &\NewText{0.17}   &\NewText{0.11}   &\NewText{0.14}      &\NewText{0.09}   &\NewText{0.09}   &\NewText{0.09}  &\NewText{0.09}   &\NewText{0.06}  \\ \cline{2-10}
    &RMSE        &\NewText{0.02}   &\NewText{0.02}   &\NewText{0.03}      &\NewText{0.02}   &\NewText{0.02}   &\NewText{0.02}  &\NewText{0.03}  &\NewText{0.02}  \\ \cline{2-10}
    &Shape*      &\NewText{RT}     &\NewText{RT}     &\NewText{RT}        &\NewText{RT}      &\NewText{RT}     &\NewText{RT}  &\NewText{RT}  &\NewText{RT}     \\ \cline{1-10}

    \multirow{4}{*}{LSC} &$R^2$       &\NewText{1}   &\NewText{1}   &\NewText{1}      &\NewText{1}   &\NewText{1}   &\NewText{0.99}  &\NewText{1}  &\NewText{0.99}   \\ \cline{2-10}    
    &MMRE        &\NewText{0.18}   &\NewText{0.10}   &\NewText{0.13}      &\NewText{0.11}   &\NewText{0.11}   &\NewText{0.09}  &\NewText{0.08}   &\NewText{0.09}   \\ \cline{2-10}
    &RMSE        &\NewText{0.02}   &\NewText{0.02}   &\NewText{0.02}      &\NewText{0.03}   &\NewText{0.02}   &\NewText{0.02}  &\NewText{0.02}  &\NewText{0.02}    \\ \cline{2-10}
    &Shape*      &\NewText{Q}     &\NewText{RT}     &\NewText{RT}        &\NewText{E}      &\NewText{Q}     &\NewText{RT}  &\NewText{RT} &\NewText{E} \\ \cline{1-10}

    \multirow{4}{*}{IDC} &$R^2$       &\NewText{1}   &\NewText{1}   &\NewText{1}      &\NewText{1}   &\NewText{1}   &\NewText{1}  &\NewText{1}  &\NewText{1}    \\ \cline{2-10}    
    &MMRE        &\NewText{0.02}   &\NewText{0.08}   &\NewText{0.11}      &\NewText{0.21}   &\NewText{0.14}   &\NewText{0.13}  &\NewText{0.10}  &\NewText{0.04}
    \\ \cline{2-10}
    &RMSE        &\NewText{0.02}   &\NewText{0.02}   &\NewText{0.02}      &\NewText{0.02}   &\NewText{0.02}   &\NewText{0.02}  &\NewText{0.02}  &\NewText{0.01} 
  \\ \cline{2-10}
    &Shape*      &\NewText{L}     &\NewText{RT}     &\NewText{RT}        &\NewText{L}      &\NewText{L}     &\NewText{Q}  &\NewText{Q}  &\NewText{RT}  \\ \cline{1-10}
    
    \multicolumn{10}{l}{
            *Selected regression model shapes include \NewText{Linear (L)}, Exponential (E), \NewText{Quadratic (Q)},
        }    \\
    \multicolumn{10}{l}{
              and Regression Tree (RT)
        }    \\
    \end{tabular}     
   }
     \label{tab:TrainingSetRegressionAccuracy}
\end{table}

\subsubsection{\NewText{\textbf{RQ1.2}}} 
\TR{C2.1, C3.1, C4.15}{Based on which adequacy metrics can an accurate regression model be built to predict FDR from AS based on the training set? 
}

\Rev{As described in the \textit{TEASMA}'s procedure in Section~\ref{Sec:Approach}, for each subject, multiple regression models are examined starting with a simple linear regression model and then using more complex non-linear models (quadratic, exponential, and regression trees) when needed. As outlined in Algorithm~\ref{Alg:BuildPredictionModel}, \textit{TEASMA} selects by default the regression model with the highest $R^2$ as the best model.} 
\Rev{In Table~\ref{tab:TrainingSetRegressionAccuracy}, we report detailed results for the selected regression model including $R^2$, average RMSE, and MMRE across all $5$-folds, along with the model shape for all subjects.}  


\TR{C2.1, C3.1}{The results for MS show a minimum $R^2$ value of 0.98, while the other three metrics show a minimum $R^2$ value of 0.99 across all subjects, indicating that all these regression models explain most of the variance in FDR. The average RMSE, across all $K$-folds for all subjects, ranges from 0.01 to 0.05 for MS, from 0.02 to 0.03 for DSC and LSC, and from 0.01 to 0.02 for IDC, indicating high prediction accuracy. Additionally, the average MMRE ranges from 0.04 to 0.25 for MS, from 0.06 to 0.17 for DSC, from 0.08 to 0.18 for LSC, and from 0.02 to 0.21 for IDC, indicating a low relative error. These results suggest that all four investigated metrics can be used to very accurately predict FDR in the training set through (non-linear) regression analysis.}

\TR{C2.1, C3.1}{Table~\ref{tab:TrainingSetRegressionAccuracy} also reports the shape of the regression model using all four adequacy metrics across all subjects.}  
\TR{C1.4, C2.7, C3.9, C4.10}{In the case of MS, only S7 exhibits linearity,} the shape of S4 is exponential, and regression trees are the best for all other subjects. 
\Rev{
However, when using SC metrics, the selected regression model consistently has a non-linear shape, with regression trees emerging as the best for all subjects when using DSC. For LSC, the regression models have an exponential shape for two subjects, quadratic for another two, and for the remaining subjects, regression trees are the best.
Regarding IDC, four subjects exhibit a linear shape while regression trees are the best for the remaining subjects. As we can see, there is a wide variation of shapes across metrics and subjects. 
}

This suggests that \Rev{across all metrics,} standard regression functions are often not a good fit and regression trees offer more flexibility by optimally partitioning the \Rev{AS} range and computing averages within partitions. To prevent overfitting of the regression trees, which would yield poor results on test sets, we limited their maximum depth to 5.

\TR{C2.1, C3.1}{
To visualize our results, we present scatterplots and fitted regression curves between FDR and AS with 95\% prediction intervals (PI) for each subject, based on all input subsets sampled from the training set.
}
\TR{C4.9}{We compute PIs in linear, quadratic, and exponential regressions using a standard method based on the fitted regression models, and for the regression tree models, we compute PIs based on a bootstrapping method for non-parametric regression models~\cite{kumar2012bootstrap}.}
PIs depict, for each specific \Rev{AS} value, the likely range of potential FDR values for a 95\% confidence level. 
\TR{C2.1, C3.1}{
The regression models built using all metrics display narrow PIs across all subjects, suggesting that these models can be confidently used to predict the FDR of a test set with limited uncertainty.}

The fitted regression curves for MS across all subjects are displayed in Figure~\ref{fig:MSTrainSetRegressionAll}.
\TR{C1.4}{For all subjects, except S7, where the shape is highly non-linear,} as subset size increases, MS initially shows rapid growth compared to FDR, indicating that the first mutants are relatively easy to kill when compared to actual faults. But then the rate of increase in MS starts to quickly slow down over time, suggesting that mutants are rapidly becoming more difficult to kill. As illustrated in Figure~\ref{fig:MSTrainSetRegressionAll}, identical MS values can result in very different levels of fault detection abilities across subjects. For instance, in subject S6, an MS of 0.8 is associated with an FDR of 0.2 while in subject S4, the same MS is associated with an FDR of 0.8. \TR{C1.4}{The only exception is subject S7, where the relationship between FDR and MS is linear. However, the slope of the regression line is slightly greater than one, suggesting that killing mutants is harder than detecting faults.} These observations confirm that, though mutants may not be representative of faults, it does not prevent us from accurately modeling their relationship. 

\MinorRev{
Our results with other investigated metrics indicate that} when using both SC metrics, DSC, and LSC, the shape of the selected regression model is non-linear across all subjects.
However, when using DSC, while the maximum FDR is consistently achieved across all subjects, the maximum DSC coverage (close to 1) is only achieved for subject S8. Moreover, the coverage of subsets with FDR close to 1 varies across subjects and is less than 0.4 for S6 and S7, less than 0.6 for S1, S2, and S5, and less than 0.7 for S3 and S4. For LSC, the maximum coverage is achieved on three subjects (S2, S4, and S6) but the coverage of subsets with FDR close to 1 varies across subjects. 
This indicates that achieving maximum coverage based on DSC and LSC is harder than detecting all faults. As a result, these metrics cannot provide a stopping criterion for selecting an adequate test set other than achieving the maximum coverage, which often requires selecting a large number of inputs. Consequently, this further justifies the importance of a test assessment methodology that can accurately predict a test set's FDR.
In general, the highly non-linear relationship between AS and FDR when using MS, DSC, and LSC, on most subjects, indicates that none of these metrics can be directly used to assess test sets and decide on their adequacy. 

The shape of the selected regression model between IDC and FDR using the training subsets is linear for three subjects, with a slope that varies across these subjects and is consistently higher than 1. For the remaining subjects, the selected regression models are non-linear. We can therefore draw similar conclusions as above for other metrics. 
To summarize, the shape of the relationship between AS and FDR, using the investigated metrics, greatly varies across subjects, which is likely due to the characteristics of the DNN models and associated training sets. For example, differences in internal architectures across DNN models determine the distribution of mutants generated by post-training MOs and contribute to such variation in relationships. Though this is not an issue in practice, this highlights the need to determine the best regression model empirically for each DNN model based on its training set, as specified by \textit{TEASMA} (Section~\ref{Sec:Approach}).

\begin{tcolorbox}   
    \textbf{Answer to RQ1.2}: \Rev{An accurate regression model with a minimum $R^2=0.98$ (based on cross-validation) can be fitted for each subject based on its training set to accurately capture the relationship between FDR and AS,  whether calculated using MS, DSC, LSC, or IDC.} The shape of the regression model is often non-linear and greatly varies across subjects. 
\end{tcolorbox}


\begin{table}[b]
    \centering   
    \small
    \caption{Linear correlation between $\widehat{FDR}$ and $ActualFDR$, for all random test subsets, \NewText{obtained using each adequacy metric} across all subjects}
     \resizebox{0.99\columnwidth}{!}{
    \begin{tabular}{  |c|  c|    c| c| c| c| c| c|  c| c|   }
    \cline{1-10} 
    Adequacy  &Accuracy      &\multicolumn{8}{c|}{Subjects}    \\ \cline{3-10}
    metric    &metric       &S1   &S2     &S3     &S4     &S5  &S6  &\NewText{S7}  &\NewText{S8}    \\ \cline{1-10}
    \multirow{2}{*}{MS (E1)} &$R^2$       &0.97   &0.98   &0.94   &0.97   &0.98   &0.98  &\NewText{0.99}  &\NewText{0.99}  \\ \cline{2-10}    
    &Slope        &1.21   &1.19   &1.16   &1.23   &0.95   &1.08  &\NewText{1.18}  &\NewText{1.21}   \\ \cline{1-10}
    
    \multirow{2}{*}{DSC} &$R^2$       &\NewText{0.99}   &\NewText{0.95}   &\NewText{0.90}      &\NewText{0.98}   &\NewText{0.90}   &\NewText{0.95}  &\NewText{0.97}   &\NewText{0.98}   \\ \cline{2-10}    
    &Slope        &\NewText{1.46}   &\NewText{1.61}   &\NewText{1.67}      &\NewText{0.93}   &\NewText{0.81}   &\NewText{1.43}  &\NewText{0.93}   &\NewText{1.23}  \\ \cline{1-10}

    \multirow{2}{*}{LSC} &$R^2$       &\NewText{0.79}   &\NewText{0.97}   &\NewText{0.90}      &\NewText{0.89}   &\NewText{0.99}   &\NewText{0.94}  &\NewText{0.93}  &\NewText{0.99}   \\ \cline{2-10}    
    &Slope        &\NewText{3.56}   &\NewText{2.01}   &\NewText{1.64}      &\NewText{8.09}   &\NewText{1.90}   &\NewText{8.32}  &\NewText{1.48}   &\NewText{1.48}   \\ \cline{1-10}

    \multirow{2}{*}{IDC} &$R^2$       &\NewText{0.99}   &\NewText{0.94}   &\NewText{0.90}      &\NewText{0.98}   &\NewText{0.99}   &\NewText{0.90}  &\NewText{0.97}  &\NewText{0.99}    \\ \cline{2-10} 
    &Slope        &\NewText{1.45}   &\NewText{1.53}   &\NewText{1.62}      &\NewText{1.48}   &\NewText{1.10}   &\NewText{1.52}  &\NewText{1.45}  &\NewText{1.41}  
    \\ \cline{1-10} 
    \end{tabular}    
   }
     \label{tab:TestSubsetAccuracy}
\end{table}

\begin{figure*}[h]
     \centering
     \begin{subfigure}[b]{0.23\textwidth}
         \centering
         \includegraphics[width=\textwidth]{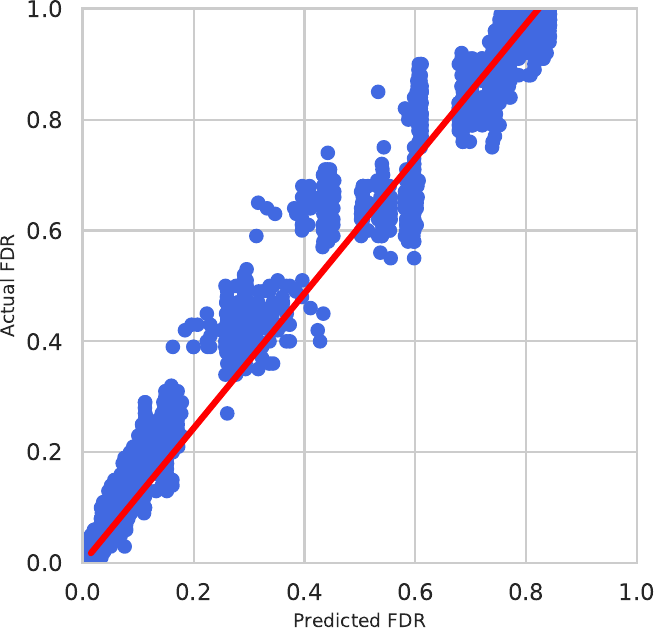}
         \caption{$Subject\ S1$}
         \label{fig:TestS1}
     \end{subfigure}
     \hfill
     \begin{subfigure}[b]{0.23\textwidth}
         \centering
         \includegraphics[width=\textwidth]{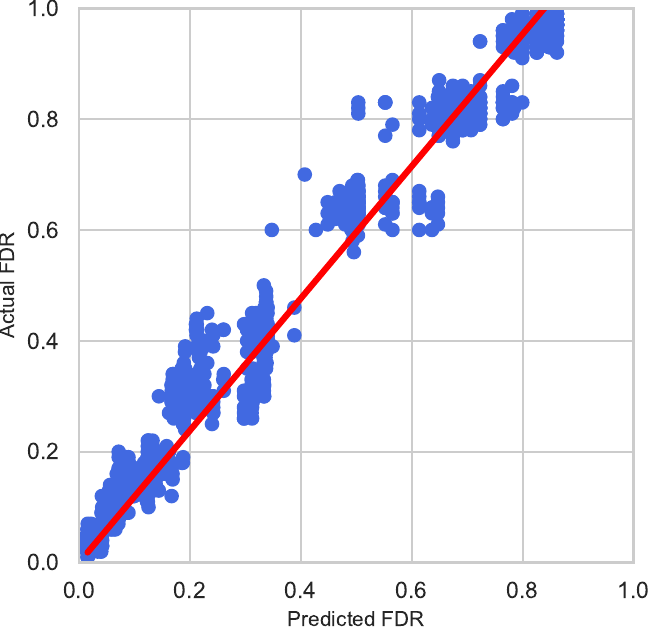}
         \caption{$Subject\ S2$}
         \label{fig:TestS2}
     \end{subfigure}
     \hfill
     \begin{subfigure}[b]{0.23\textwidth}
         \centering
         \includegraphics[width=\textwidth]{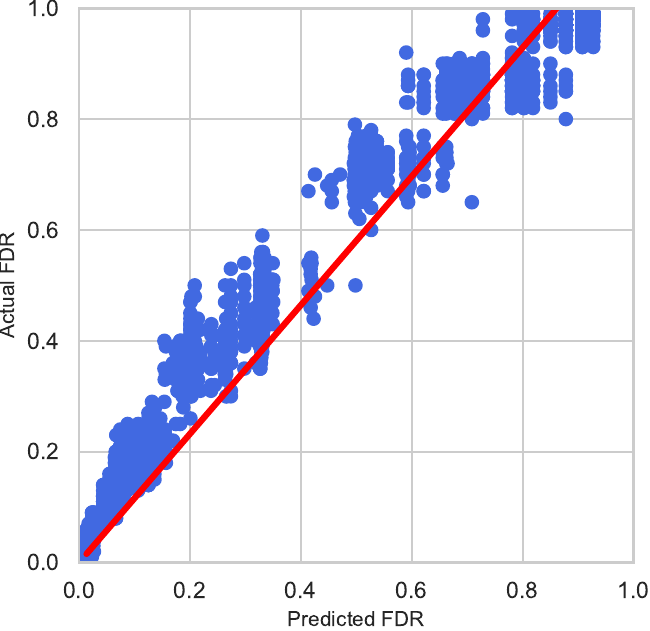}
         \caption{$Subject\ S3$}
         \label{fig:TestS3}
     \end{subfigure}
     \hfill
     \begin{subfigure}[b]{0.23\textwidth}
         \centering
         \includegraphics[width=\textwidth]{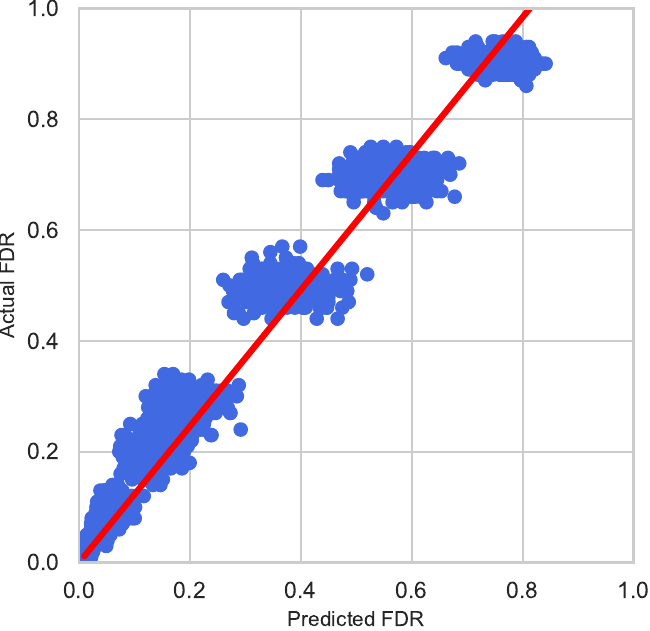}
         \caption{$Subject\ S4$}
         \label{fig:TestS4}
     \end{subfigure}
     \hfill
     \begin{subfigure}[b]{0.23\textwidth}
         \centering
         \includegraphics[width=\textwidth]{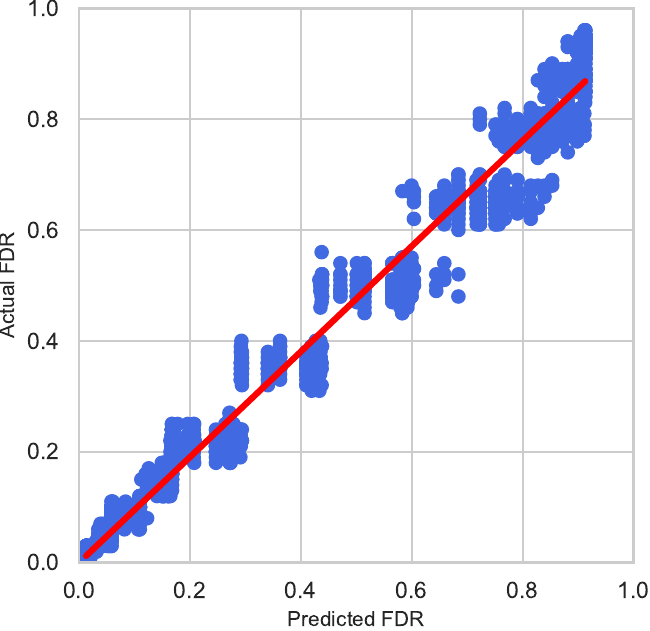}
         \caption{$Subject\ S5$}
         \label{fig:TestS5}
     \end{subfigure}
     \hfill
     \begin{subfigure}[b]{0.23\textwidth}
         \centering
         \includegraphics[width=\textwidth]{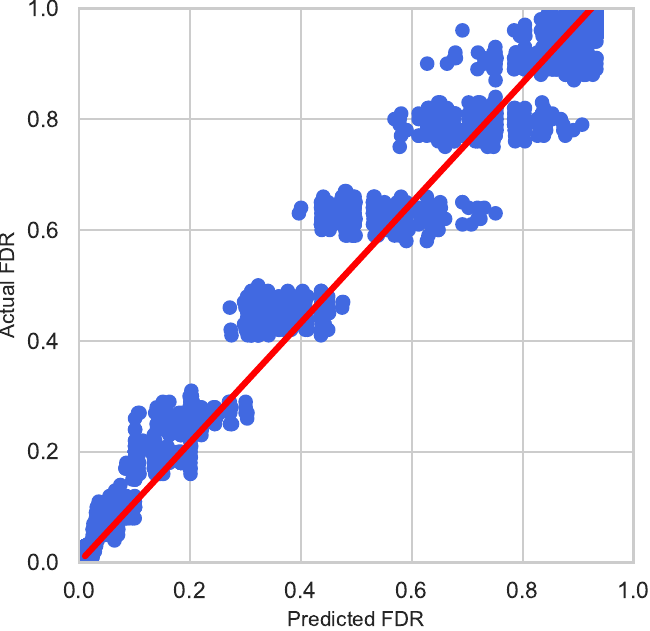}
         \caption{$Subject\ S6$}
         \label{fig:TestS6}
     \end{subfigure}
     \hfill
     \begin{subfigure}[b]{0.23\textwidth}
         \centering
         \includegraphics[width=\textwidth]{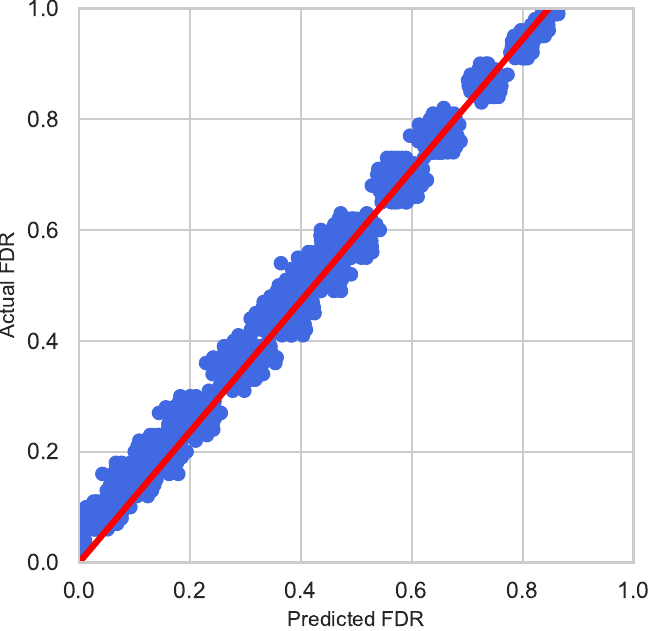}
         \caption{$\NewText{Subject\ S7}$}
         \label{fig:TestS7}
     \end{subfigure}
    \hfill
    \begin{subfigure}[b]{0.23\textwidth}
         \centering
         \includegraphics[width=\textwidth]{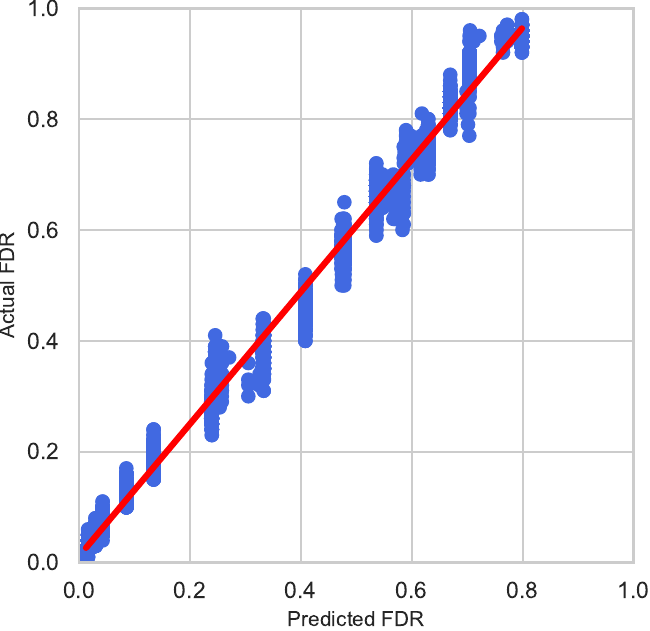}
         \caption{$\NewText{Subject\ S8}$}
         \label{fig:TestS8}
     \end{subfigure}
        \caption{Linear regression line between $\widehat{FDR}$ and $ActualFDR$ \NewText{using MS}}
        \label{fig:MSLinearRegressionFDRandFDRhat}
\end{figure*}

\Rev{\subsection{\textbf{RQ2: Which adequacy metrics lead to better FDR predictions on test sets?
}}}

To address this question, we analyze the relationship of the predicted FDR ($\widehat{FDR}$) with the actual one ($ActualFDR$) in the test set. To do so, for each subject, we selected a large number of random test subsets and measured the fit ($R^2$) and slope of a linear regression model between $\widehat{FDR}$ and $ActualFDR$. If the former can be trusted as an indicator of the latter, we should have a high $R^2$ and a slope close to 1. 
\TR{C2.1, C3.1}{
We should note that we repeat this process with each adequacy metric, consistently using the same set of test subsets.
Table~\ref {tab:TestSubsetAccuracy} reports the results using all investigated test adequacy metrics.} 
We can see a minimum $R^2$ of 0.94 and a slope that is either slightly below or above 1 \Rev{when using MS}, across all subjects. 
\TR{C2.1, C3.1}{
The results for DSC and IDC indicate a minimum $R^2$ value of 0.90. However, slopes tend to deviate more from 1 when compared to those observed with MS. Notably, for IDC, the slope is consistently greater than 1, suggesting that the prediction models built using this metric tend to underpredict FDR. LSC shows the lowest performance among investigated metrics, with a minimum $R^2$ value of 0.79 and slopes that consistently and significantly exceed 1. In particular, LSC shows substantial deviations, with slopes exceeding 2 for subject S2, 3 for subject S1, and 8 for subjects S4 and S6, indicating significant inaccuracies in FDR prediction for these cases.
}

\TR{C2.1, C3.1}{
For each metric, we calculate the average absolute deviation of the slopes of the linear regression lines between $\widehat{FDR}$ and $ActualFDR$ from 1 across all subjects. 
MS resulted in the lowest average deviation of 0.16, indicating that the slopes of the linear regression lines obtained using this metric across all subjects are closest to 1, on average. 
DSC has an average deviation of 0.34, significantly higher than MS, indicating it is less accurate than MS. IDC, with an average deviation of 0.44, is slightly higher than DSC but still relatively close. In contrast, LSC shows the largest average deviation of 2.56, indicating that its linear regression lines deviate the most from 1. 
}

\TR{C2.1, C3.1}{
In summary, results indicate that the regression models developed on the training set using all investigated metrics, except for LSC, yield FDR predictions that are closely aligned with the actual FDR of the test sets. Therefore, FDR predictions obtained with \textit{TEASMA} using MS, DSC, and IDC metrics} provide a good basis on which to assess the adequacy of test sets and whether good accuracy results can be trusted to decide about deploying a DNN model.

\TR{C2.1, C3.1}{
The linear regression lines between $\widehat{FDR}$ derived from MS and $ActualFDR$, where the intercept is set to zero, are shown for all subjects in Figure~\ref{fig:MSLinearRegressionFDRandFDRhat}.
}
These regression lines clearly show a strong alignment between the two. The slopes of the lines fall within the range [0.95 1.23], thus indicating no strong bias in the FDR predictions.  
For most subjects, the prediction models tend to slightly underestimate the actual FDR, with the exception of subject S5, where a slight overestimation is observed.
\MinorRev{Our linear regression results for the other adequacy metrics show that} DSC exhibits the widest range of slope variations, from 0.81 for subject S5 to 1.67 for subject S3. While the prediction models tend to underestimate the actual FDR for most subjects, they overestimate it for three subjects. In contrast, IDC is the only metric where the regression slopes consistently exceed 1, ranging from 1.10 to 1.62. This suggests that for test subsets, detecting all faults is easier than achieving maximum IDC coverage.

\begin{tcolorbox} 
    \textbf{Answer to RQ2}: 
    \NewText{A linear regression model between $\widehat{FDR}$ and $ActualFDR$ on the test set can be fitted with a minimum $R^2$ value of 0.94 using MS, and 0.90 using DSC, and IDC, with slopes slightly below or above 1. 
    This indicates that, for each subject, the selected regression model built on the training set using these metrics, accurately predicts the FDR of a test set based on its AS.} 
    Consequently, \textit{TEASMA} provides a reliable \Rev{methodology} for assessing test sets and whether to trust high accuracy results when validating a DNN model.
\end{tcolorbox}


\begin{table}[b]
    \centering   
    \small
    \caption{\NewText{Accuracy of the FDR predictions using accurate adequacy metrics}}
     \resizebox{0.99\columnwidth}{!}{
    \begin{tabular}{  |c|  c|    c| c| c| c| c| c|  c| c|    c|}
    \cline{1-11} 
    Accuracy  & Adequacy     &\multicolumn{8}{c|}{Subjects}  &\multirow{2}{*}{Avg}    \\ \cline{3-10}
    metric &metric       &S1   &S2     &S3     &S4     &S5  &S6  &\NewText{S7}  &\NewText{S8}  & \\ \cline{1-11}
    \multirow{3}{*}{RMSE}  
    &MS (E1)       &\NewText{\textbf{0.12}}   &\NewText{\textbf{0.11}}   &\NewText{\textbf{0.12}}      &\NewText{0.09}   &\NewText{\textbf{0.05}}   &\NewText{\textbf{0.07}}  &\NewText{0.10}   &\NewText{\textbf{0.09}}  &\NewText{\textbf{0.09}}  \\ \cline{2-11}
    &DSC        &\NewText{0.21}   &\NewText{0.24}   &\NewText{0.26}      &\NewText{\textbf{0.06}}   &\NewText{0.16}   &\NewText{0.16}  &\NewText{\textbf{0.08}}   &\NewText{0.10} &\NewText{0.17} \\ \cline{2-11}
    &IDC        &\NewText{0.20}   &\NewText{0.23}   &\NewText{0.25}      &\NewText{0.15}   &\NewText{0.06}   &\NewText{0.23}  &\NewText{0.20}   &\NewText{0.15} &\NewText{0.18} \\ \cline{1-11} 
    \end{tabular}    
   }
     \label{tab:TestFDRFredictionAccuracy}
\end{table}

\subsection{\textbf{\NewText{RQ3: How accurately can we predict a test set's FDR?}}}

\TR{C2.1, C3.1}{
To address this question, we compare the adequacy metrics identified in the previous question as accurate predictors of FDR: MS, DSC, and IDC. To evaluate the prediction accuracy of each metric, we employ RMSE, a standard and easy to interpret measure for assessing prediction accuracy. RMSE measures the average magnitude of the errors between predicted and actual values. It computes the square root of the average squared differences between predicted and actual values, making it sensitive to larger errors and penalizing them more than smaller ones. This characteristic of RMSE is particularly valuable in our context, where large errors between the predicted FDR and the actual FDR are especially undesirable.
Table~\ref{tab:TestFDRFredictionAccuracy} presents the RMSE values, between $\widehat{FDR}$ and $ActualFDR$, for each metric. MS achieves the lowest RMSE among all metrics for most subjects (five out of seven) and has the second-lowest RMSE, with only a small difference from the lowest value, for the remaining two subjects, indicating that MS is consistently accurate across different subjects. The average RMSE for MS is 0.09, indicating that, on average, the difference between the predicted FDR and the actual FDR is 0.09. This low value suggests that predictions using MS are quite accurate. In contrast, the average RMSE values for DSC and IDC are 0.17 and 0.18, respectively, nearly double that of MS. These higher RMSE values imply that DSC and IDC are significantly less accurate at predicting FDR compared to MS.
}

\TR{C2.1, C3.1}{
While the shape of the relationship between IDC and FDR is linear for some subjects and close to linearity for others, indicating it could be an accurate metric, these results show that IDC is not the most accurate for predicting FDR. 
Our investigation reveals that the shape of the relationship between IDC and FDR for test subsets is not similar to that of training subsets. 
One explanation for this can be that, unlike other metrics, calculating IDC is entirely independent of the DNN model and is solely based on input features. Consequently, inputs that are correctly predicted by the DNN model--and thus do not contribute to fault detection--can still increase the IDC coverage. 
Given that the accuracy of the DNN model on test sets is typically lower than on its training set, the proportion of correctly predicted inputs that can increase coverage without contributing to FDR is higher in the training subsets than in the test subsets.
This discrepancy affects the shape of the relationship between IDC and FDR on the training subsets, making it different from the relationship on the test subsets. It is important to note that the relationship between IDC and FDR on the test subsets is mostly non-linear. For subjects where a linear relationship exists for both training and test subsets, the slope of the linear regression line is higher for the test subsets. This indicates that, for training and test subsets with identical IDC coverage, the test subsets consistently achieve higher FDR. 
}

\begin{tcolorbox} 
    \NewText{\textbf{Answer to RQ3}:  Its relatively low RMSE (9\% on average)  across all subjects, indicates that MS calculated based on DeepMutation's post-training MOs is a good solution for accurately predicting FDR, following the \textit{TEASMA} methodology, and fares better than solutions relying on DSC and IDC. }
\end{tcolorbox}

\subsection{Fault clusters}

\TR{C2.4, C3.6}{
In this section, we report on the fault clusters we used in our experiments since they play a central role in our experiments. These clusters are obtained by applying the approach described in Section~\ref{Sec:FaultEstimation} on the DNN model's training set as outlined in Section~\ref{Sec:ExperimentFault}. 
This clustering approach, like many others, relies on several hyperparameters such as the number of neighbors considered locally connected in UMAP and the minimum size of clusters in HDBSCAN. These hyperparameters should be effectively fine-tuned to achieve the best clustering results. Therefore, following the original implementation~\cite{aghababaeyan2021black}, we applied this approach to each subject using several hyperparameter configurations. We then selected the optimal configuration through the same manual and metric-based evaluations introduced in the original paper~\cite{aghababaeyan2021black}. 
}

\TR{C2.4, C3.6}{As described in Section~\ref{Sec:FaultEstimation}, the clusters obtained for each subject were evaluated based on the Silhouette score. For each subject, the configuration that achieved the highest Silhouette score was selected to obtain the final clusters. Table~\ref{tab:FaultClusters} provides detailed information on the fault clusters obtained for each subject in our experiments, including the total number of fault clusters as well as their Silhouette score and DBCV value. 
Achieving a Silhouette score close to 1 is rarely feasible in practice but, with a minimum Silhouette score of 0.60 across all subjects, we seem to achieve good clustering based on what is reported in the literature~\cite{rousseeuw1987silhouettes, aghababaeyan2021black}. This score suggests that the clusters are well-defined and distinct, reflecting a good balance between intra-cluster cohesion and inter-cluster separation. The clusters' DBCV scores, which exceed 0.5 for most subjects, provide additional evidence supporting the quality of the clustering results. These scores further suggest that the final clusters are not only well-separated but also dense, which is indicative of high-quality clustering. 
}

\begin{table}[t]
    \centering   
    \caption{\NewText{Information about fault clusters}}
    \resizebox{0.99\columnwidth}{!}{
    \begin{tabular}{|c|       c|  c|  c| c| c| c| c| c|    }
    \hline 
        \NewText{Subjects}     &\NewText{S1}  &\NewText{S2}  &\NewText{S3}  &\NewText{S4}  &\NewText{S5}  &\NewText{S6}  &\NewText{S7}  &\NewText{S8}  \\  \hline 
        \NewText{\#Faults}   &\NewText{307}  &\NewText{354}  &\NewText{191}  &\NewText{512}  &\NewText{607}  &\NewText{590}  &\NewText{718}  &\NewText{3118}  \\  \hline 
          
        \NewText{Silh.}    &\NewText{0.88}  &\NewText{0.70}  &\NewText{0.69}  &\NewText{0.79}  &\NewText{0.66}  &\NewText{0.78}  &\NewText{0.78}  &\NewText{0.60}  \\  \hline
        \NewText{DBCV}    &\NewText{0.81}  &\NewText{0.51}  &\NewText{0.52}  &\NewText{0.66}  &\NewText{0.51}  &\NewText{0.66}  &\NewText{0.66}  &\NewText{0.47} \\    \hline   
       
    \end{tabular}
    }
    \label{tab:FaultClusters}
\end{table}

\begin{figure}[b] 
     \centering
     \begin{subfigure}[b]{0.49\columnwidth}
         \centering
         \includegraphics[width=\columnwidth]{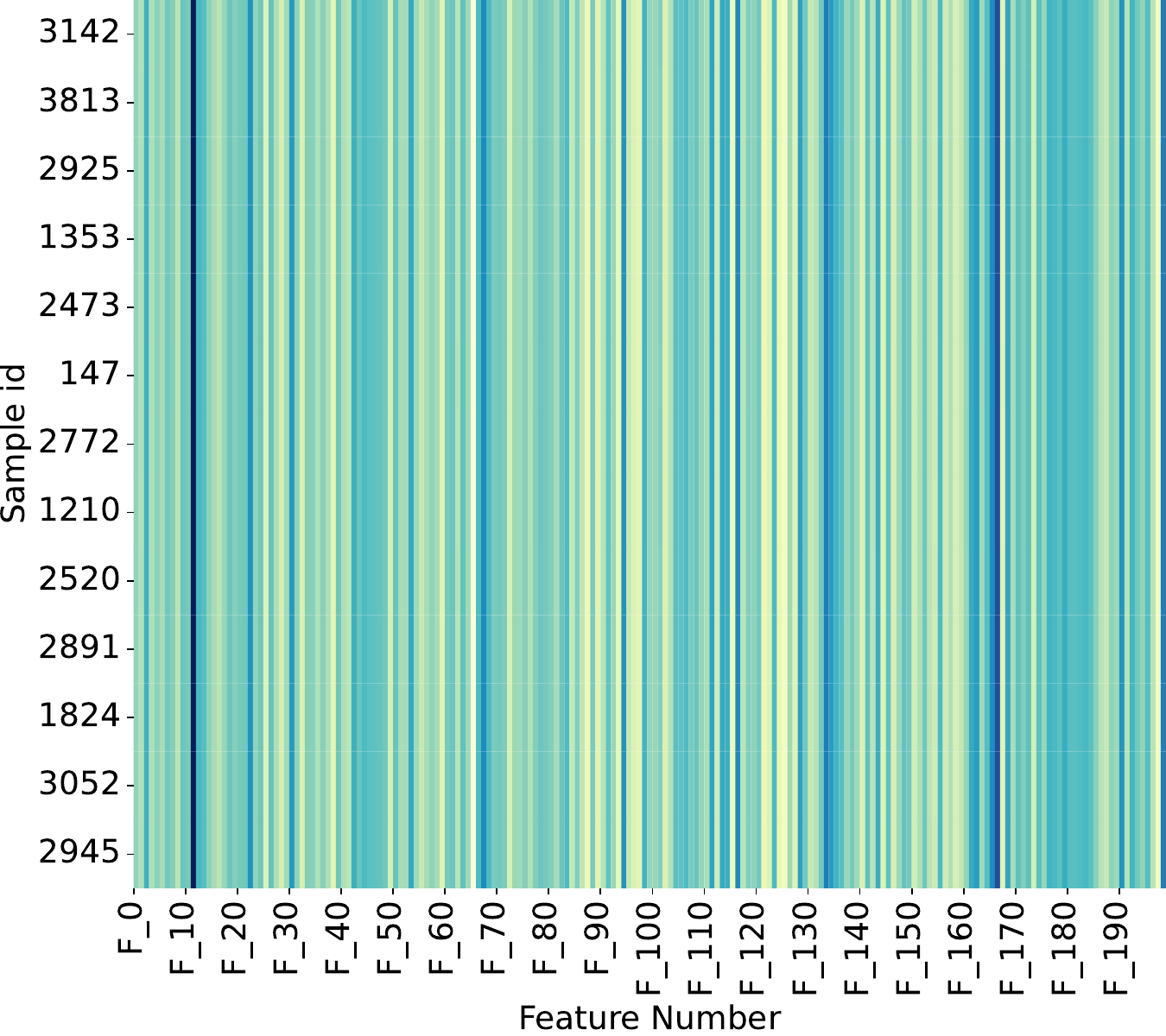}
         \caption{$A\: final\: cluster$}
         \label{fig:Heatmap1}
     \end{subfigure}
     \hfill
     \begin{subfigure}[b]{0.49\columnwidth}
         \centering
         \includegraphics[width=\columnwidth]{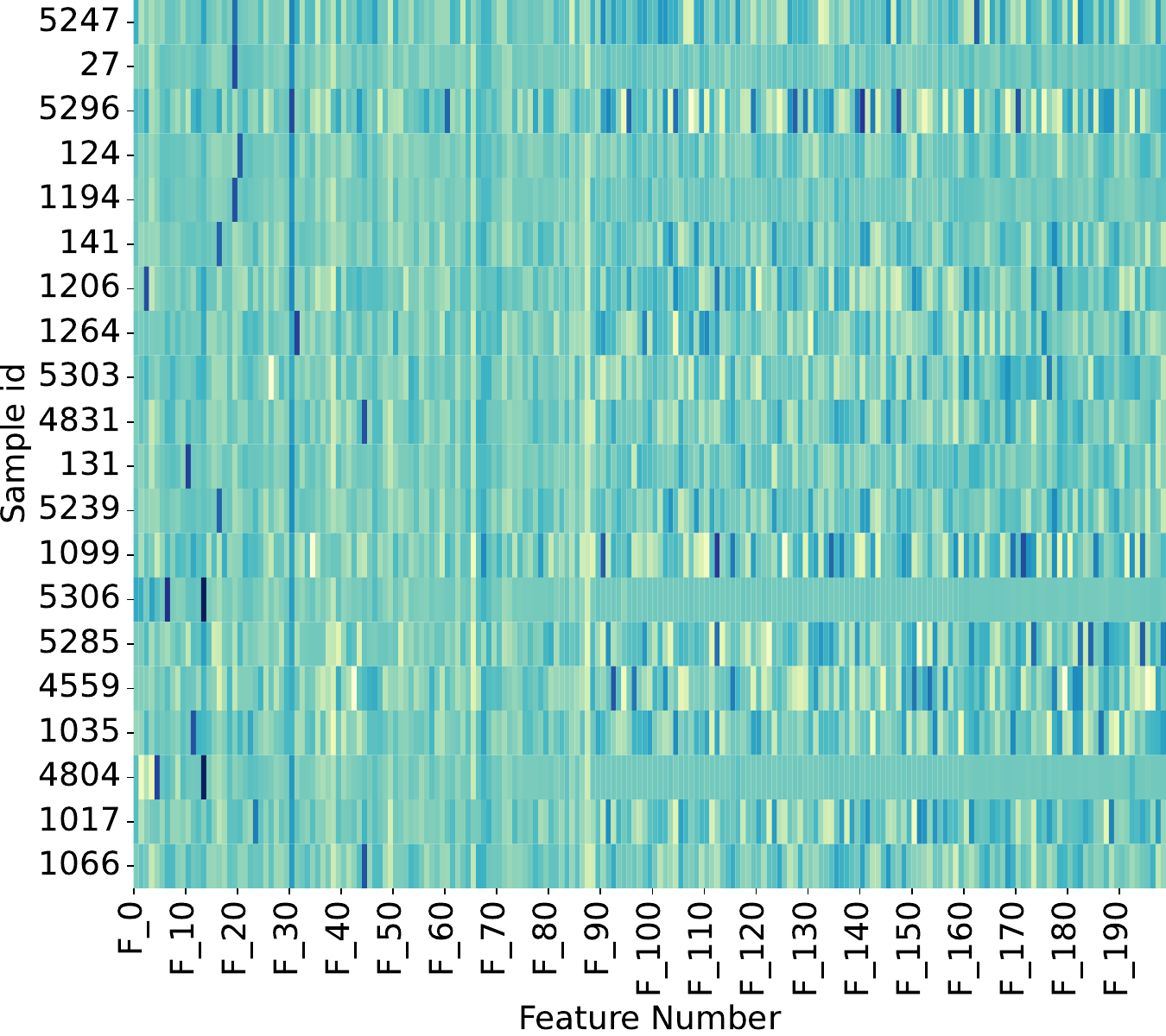}
         \caption{$A\: noisy\: cluster$}
         \label{fig:Heatmap2}
     \end{subfigure}
     \hfill
        \caption{Heatmap examples of the fault clusters}
        \label{fig:Heatmaps}
\end{figure}

\TR{C2.4, C3.6}{Furthermore, we conducted an exhaustive manual evaluation of the features for each cluster as described in the original paper~\cite{aghababaeyan2021black}. For this purpose, we evaluate features' heatmaps related to final clusters.
Two representative examples of these heatmaps are depicted in Figure~\ref{fig:Heatmaps}. Each heatmap's rows correspond to the input IDs within a cluster, while columns represent their respective features, with colors indicating feature values. The first heatmap, in Figure~\ref{fig:Heatmap1}, is related to a quality cluster characterized by well-clustered inputs sharing common feature distribution patterns. Conversely, the second heatmap, in Figure~\ref{fig:Heatmap2}, illustrates a cluster containing poorly-clustered inputs. We observed that most of the clusters look like Figure~\ref{fig:Heatmap1}, demonstrating coherent feature distribution patterns and indicating effective clustering.
}

\subsection{Threats to Validity} \label{sec:Threats}

We discuss in this section the different threats to the validity of our results and describe how we mitigated them. 

\noindent{\textbf{Internal threats to validity.}} 
One of the internal threats to validity pertains to our sampling approach, as our results are based on sampled input subsets. \Rev{To address this, we selected a large number of randomly sampled input subsets of varying sizes. Furthermore, when experimenting with MS, we employed uniform sampling in addition to random sampling, since one of the MS calculation formulas used in our experiments considers the number of images from each input class in a subset.} 
Another internal threat to validity stems from selected parameters and configurations used throughout the experiments. This encompasses parameters for clustering mispredicted inputs to estimate faults and for configuring the calculation of \Rev{AS}. To mitigate this concern, regarding clustering parameters, we explored different settings for each subject and opted for the best configuration based on the evaluation of the Silhouette index~\cite{rousseeuw1987silhouettes} and DBCV scores~\cite{moulavi2014density}, as suggested by Aghababaeyan \textit{et al.}~\cite{aghababaeyan2021black}. For MO configuration, we used the default settings of the DeepMutation++ tool, which yielded accurate results in terms of FDR prediction. 
\TR{C2.1, C3.1}{
For calculating DSC and LSC, following the original study~\cite{kim2023evaluating}, we used the deepest layer of the model and set the number of buckets to 1000. The lower and upper bounds were set using the SA values observed during training. For calculating IDC, we used the publicly available VAEs for subjects sharing identical input sets with those used in the original study~\cite{dola2023input} and trained new VAEs only for the remaining input sets.}

\noindent{\textbf{Construct threats to validity.}}   
A potential threat is related to the calculation of MS, SC, and FDR. To mitigate this concern and ensure the robustness of our results, we considered multiple ways to compute MS, as described in Section~\ref{subSec:MutationTestingMetrics}.
\TR{C2.1, C3.1}{Regarding the SC calculation, it is important to note the implementations we employed. We primarily utilized the original implementations of DSC and LSC provided by Kim \textit{et al.}~\cite{kim2023evaluating}. However, due to the substantial computational cost associated with calculating LSC on our largest input set, ImageNet, we opted for an alternative implementation provided by Weiss \textit{et al.}~\cite{weiss2021review} for this specific dataset. It has been thoroughly evaluated in the corresponding study by Weiss \textit{et al.} and has shown comparable outcomes to the original implementation. We thus ensure the feasibility of our experiments while maintaining the integrity of our results.}
For FDR, we leveraged a SOTA fault estimation approach~\cite{aghababaeyan2021black} based on clustering mispredicted inputs. This approach has been thoroughly validated and the resulting clusters have been carefully analyzed by the authors. We reused their publicly available approach to achieve accurate fault identification.

\noindent{\textbf{External threats to validity.}} 
To address potential concerns regarding the generalizability of \textit{TEASMA}, we performed our experiment on multiple subjects including various combinations of \Rev{seven} DNN models with different internal architectures and \Rev{six} input sets containing diverse types of images. \NewText{The selected input sets include large input sets, such as Cifar-100 and ImageNet, to assess the practicality and scalability of the proposed methodology.} 
\TR{C3.8, C4.1}{Our results show great consistency in terms of FDR prediction accuracy across all these models and input sets, further validating the generalizability of \textit{TEASMA}.} 
\TR{C2.1, C3.1}{Another potential threat may arise from the specific adequacy metrics selected for evaluating \textit{TEASMA}. To mitigate this threat, we investigated four SOTA metrics, including the widely-used DSC and LSC metrics and the recent input feature coverage IDC metric, as well as mutation analysis.} 
Another threat might be related to the specific set of MOs that we considered in our experiment \Rev{with MS}. For practical reasons, we focused on post-training MOs and investigated all the MOs implemented by the DeepMutation++ tool, also considering whether subsets of MOs might yield better results. Other MOs may yield different results in the future.

\MinorRev{
\subsection{Practical Guidelines for Using TEASMA}}

\MinorRev{}
In practice, when applying \textit{TEASMA}, one must first build an accurate regression model, relying on post-training MOs, or one of the alternative metrics. Then, for each test set being considered, one must compute AS and then use the regression model to predict FDR. When relying on MS, previously generated mutants are used to calculate the test set's MS. Therefore, the main cost is to build the regression model using the training set. 


While \textit{TEASMA} can be employed with various metrics, the choice of adequacy metric directly impacts both the accuracy of the FDR predictions and the overall computation time taken by the process, particularly for large input sets or complex DNN architectures. 
Our empirical results show that MS, DSC, and IDC are all reliable metrics for predicting FDR, though MS is the most accurate. However, their computational requirements vary significantly since, in our experiments,  running \textit{TEASMA} with MS took roughly five times longer than with DSC and about eight times longer than with IDC. Therefore, whether MS is applicable depends on the complexity of the DNN model under test and the size of its training set, as well as available computational resources and time constraints. 

However, for high-criticality DNNs, such as those used in autonomous driving or healthcare, where accurate FDR predictions are crucial, MS is the recommended choice when possible. Conversely, for less critical systems or when faster test evaluations are required, IDC provides a good balance between accuracy and performance.


%

\NewText{
\section{Related Work}
\TR{C2.2}{}
In this section, we review prior work related to our proposed methodology from two perspectives: test selection and generation, and test adequacy assessment, within the context of DNN models.
}

\NewText{
\subsection{Test Selection and Generation}
Testing and enhancing a DNN model requires large labeled input sets. In many contexts, however, the scarcity of labeled inputs poses a significant challenge, especially when these models are applied in and fined-tuned to different deployment contexts. 
Consequently, recent research has focused on developing innovative methods for selecting unlabeled test inputs to not only test (after labeling) but also improve the model via retraining with the selected set~\cite{hu2024test, hu2022empirical}.
}

\NewText{
Several selection methods assign a score to each test input that captures its priority for getting labeled and tested, enabling testers to select test inputs with higher scores based on their labeling budget. 
} 
\TR{C1.5}{}
\NewText{Unlike surprise adequacy metrics introduced by Kim \textit{et al.}~\cite{kim2019guiding}, such as DSA, LSA, and MDSA, which measure how surprising or unexpected a new test input is to a DNN model given its training input set (as detailed in Section~\ref{sec:Background}), uncertainty-based metrics, such as DeepGini~\cite{feng2020deepgini}, Margin~\cite{ma2021test}, and MaxP~\cite{ma2021test}, estimate the model's confidence in predicting each test input. They aim to prioritize inputs on which the model is least confident in its predictions and that are thus most informative or challenging for a model. Similarly, entropy-based methods~\cite{byun2019input} compute the entropy of the predicted class probabilities for each test input and prioritize inputs with higher entropy, indicating greater uncertainty in the model's predictions.
}

\NewText{
Other test selection approaches, rather than scoring individual inputs, select a test subset with a predefined labeling budget. For instance, Aghababaeyan \textit{et al.}~\cite{aghababaeyan2023deepgd} proposed DeepGD, a black-box multi-objective search-based approach. DeepGD selects a fixed-size subset of inputs guided by the uncertainty of model output probabilities and the diversity of selected inputs' features. It has been shown to be able to select test sets capable of effectively revealing diverse faults in the DNN model and guiding the enhancement of the model through retraining with the selected inputs. 
Ma \textit{et al.}~\cite{hu2022empirical} conducted an empirical study with various selection metrics (including DSA, DeepGini, and the entropy-based metric) and random selection to investigate the impact of various ranges of distribution shift on their effectiveness for model enhancement. Their observations revealed that no individual metric consistently outperforms the others across all distribution shift ranges. Building on these insights, they introduced a new approach for selecting a fixed-size subset of inputs, which mitigates the effects of distribution shifts and exhibits up to a 30\% improvement in model accuracy.
Evaluating the performance of test selection methods on test sets with a distribution shift from the training set is particularly important for methods that aim to enhance the model in new deployment contexts through retraining. 
In this paper, however, our focus is on assessing the adequacy of test sets. In scenarios where a distribution shift occurs between the training set and the deployment environment and leads to poor model accuracy, model retraining is necessary and test adequacy assessment is not useful as one cannot have confidence in the model regardless of the test set used. After proper retraining, test adequacy assessment can be performed on the retrained model to validate its high-accuracy test results. In our experiments, to prevent any potential biases, we intentionally relied on the subject models' original test sets since we had no involvement in their selection. 
}

\TR{C3.1}{
Zohdinasab \textit{et al.}~\cite{zohdinasab2023efficient} introduced DeepHyperion-CS, a tool for generating inputs that explores the feature space to cover unseen feature combinations or improve performance compared to similar existing inputs. However, the study did not investigate the potential of using feature space coverage as a metric for test adequacy or its correlation with a test set's fault detection capability. 
Selecting appropriate features is crucial when employing feature-based coverage metrics, as their effectiveness hinges on choosing features that accurately represent significant system properties. Zohdinasab \textit{et al.}~\cite{zohdinasab2023efficient} aim to provide an interpretable characterization of DL systems by identifying which features of the test inputs influence the system's incorrect behavior.
Nonetheless, identifying such features is challenging and involves a systematic approach that requires domain experts' assistance. This manual process is time-consuming and labor-intensive, requiring human assessors to analyze existing inputs and select relevant features for a given domain.
In contrast, Attaoui \textit{et al.}~\cite{attaoui2023black} proposed a feature-based test selection approach for the image domain that relies on a pre-trained model to automatically extract features of the error-inducing inputs. Consequently, they can detect the root causes of mispredictions by clustering these inputs based on their features. Finally, they use these clusters to effectively retrain and enhance the performance of the DNN model.   
}

\NewText{
Regardless of the selection or generation approach, assessing the adequacy of a selected test set is essential to mitigate the risk of insufficient testing. Based on our findings, \textit{TEASMA} can accurately assess the adequacy of a test set and thus confidently verify whether high-accuracy test outcomes can be trusted. 
}

\NewText{
\subsection{Test Adequacy Assessment}
\TR{C2.2}{}
Test adequacy assessment refers to the process of evaluating the quality of a test set and its capacity to validate test outcomes~\cite{jia2010analysis}. 
In traditional software testing, coverage metrics~\cite{kochhar2015code} and mutation testing~\cite{jia2010analysis} stand out as pivotal methods for assessing test adequacy. Coverage metrics, including statement, branch, condition, and path coverage, measure the comprehensiveness of a test suite and its ability to execute different parts of the system code. 
\MinorRev{Similarly, researchers have proposed various neuron coverage metrics in the context of DNN models for test assessment~\cite{pei2017deepxplore, Ma2018DeepGaugeMT, tian2018deeptest, sun2019structural, yang2022revisiting, gerasimou2020importance}. }
They aim to measure how well the test set covers the model's internal neuron activations, ensuring that a wide range of the DNN model's behaviors are examined during testing.
However, these metrics have shown no significant correlation with the number of mispredicted inputs or detected faults in a test set~\cite{li2019structural, chen2020deep}. 
Consequently, in our experiments to evaluate \textit{TEASMA}'s methodology with coverage metrics, we relied on SOTA surprise coverage metrics, DSC and LSC, alongside the recently introduced black-box coverage metric IDC, as described in Section~\ref{sec:Background}. 
}

\NewText{
Mutation testing has been established as a widely-used technique for assessing test suites in traditional software~\cite{jia2010analysis, andrews2006using}. 
As described in detail in Section~\ref{sec:Background}, this technique has been adapted for DNNs and specific MOs have been proposed. 
The potential of these operators to address the high cost of labeling test inputs in DNNs by guiding test input generation and prioritization has been investigated in the literature~\cite{riccio2021deepmetis, wang2021prioritizing}. 
}
\TR{C3.2}{
Riccio \textit{et al.}~\cite{riccio2021deepmetis} developed DeepMetis, a search-based input generation tool that utilizes the mutants generated by DeepCrime's pre-training operators to guide the process of test input generation. Their goal is to generate inputs that can augment an existing test set and increase its capability to kill pre-training mutants. Given that DeepCrime's operators are based on real faults~\cite{jahangirova2020empirical}, test sets that kill more mutants are assumed to be able to detect more faults.
}
\TR{C1.5, C3.2}{
Wang \textit{et al.}~\cite{wang2021prioritizing} proposed PRIMA, a test input prioritization approach based on mutation analysis. In addition to four post-training MOs similar to DeepMutation's operators, they define input MOs for some specific domains including text and image inputs. PRIMA prioritizes inputs that are able to kill many mutated models based on the assumption that such inputs are more likely to reveal faults. Additionally, it prioritizes inputs that produce different prediction results with many mutated versions. The rationale is that if a slight mutation on a test input alters the prediction results, it indicates that the mutated part is actively utilized by the model, making it sensitive to capturing errors in the model. Both DeepMetis~\cite{riccio2021deepmetis} and PRIMA~\cite{wang2021prioritizing} aim to achieve test sets with higher MS. However, neither of these approaches investigate the correlation between MS and FDR, nor do they intend to help assess the adequacy of selected or generated test sets. 
}

\NewText{The application of post-training operators for adversarial sample detection in DNNs has also been investigated by Wang \textit{et al.}~\cite{wang2019adversarial}. They introduced a new approach relying on DeepMutation's MOs, based on the observation that adversarial samples are more sensitive to mutations in the DNN model compared to normal samples. This sensitivity is leveraged to distinguish between normal and adversarial samples at runtime.
}

\NewText{
Despite the availability of various test adequacy metrics such as MS, DSC, LSC, and IDC, a critical gap remains in the development and evaluation of a comprehensive methodology for test adequacy assessment. 
\textit{TEASMA} addresses this gap by providing a reliable and validated methodology to evaluate test set adequacy, regardless of the selection or generation technique employed, utilizing and comparing SOTA test adequacy metrics. 
Our results show that \textit{TEASMA} when deployed with DSC, IDC, and MS calculated based on post-training MOs, provides an accurate prediction of the fault detection ability of test sets and therefore their adequacy. MS seems to yield the most accurate results. 
}

\section{Conclusion}
\Rev{In this paper, we propose \textit{TEASMA}, a practical methodology for assessing the adequacy of unlabeled test sets for DNNs. Such assessments are important in practice, regardless of the test set selection strategy, in order to determine whether high-accuracy test results can be trusted to validate DNN models. Once a test set is deemed adequate, it can then be labeled and executed.  
}

\Rev{Our empirical analysis investigated the relationship between the Adequacy Scores (AS) of test sets, calculated using four SOTA test adequacy metrics, including Distance-based Surprise Coverage (DSC), Likelihood-based Surprise Coverage (LSC), Input Distribution Coverage (IDC), and Mutation Score (MS) calculated based on post-training mutation operators. Results show a strong correlation between AS and the Fault Detection Rate (FDR) for all these metrics. However, the shape of the relationships is typically not linear and greatly varies across DNN models. Consequently, these metrics are not accurate surrogate measures for FDR and cannot be used directly for test adequacy assessment. 
}
\Rev{\textit{TEASMA}'s methodology provides a procedure to accurately predict the FDR of an unlabeled test set from its AS, relying on a selected adequacy metric, so as to confidently decide whether a test set can provide reliable test results.
\textit{TEASMA} is based on regression analysis using the training set and can be deployed with any test adequacy assessment metric for DNNs.
}

\Rev{We conducted an extensive empirical evaluation using multiple DNN models and input sets, including large input sets such as ImageNet, with four SOTA adequacy metrics: MS, DSC, LSC, and IDC.  
Our results indicate that, unlike LSC, prediction models built using MS, DSC, and IDC can accurately predict a test set's FDR--- with  $R^2$ values of 0.94 for MS and 0.90
for DSC and IDC, and a slope close to 1---which can then reliably be used to decide about the adequacy of a test set. 
Additionally, the average Root Mean Square Error (RMSE) between predicted and actual FDR values when using MS is 0.09, indicating a higher prediction accuracy than that of DSC and IDC, with RMSE values of 0.17 and 0.18, respectively. Therefore, MS based on post-training operators shows to be the most accurate adequacy metric when deployed within \textit{TEASMA}.
Overall, \textit{TEASMA} is thus a reliable methodology to ensure the proper validation of a DNN before deployment, a particularly important task in critical contexts. 
}

\TR{C3.13}{Given the practicality of \textit{TEASMA} and its ability to accurately predict FDR, we intend to rely on it, in future work, to guide the automated generation of adequate test sets for DNN models.}

\section*{Acknowledgements}
This work was supported by a Research Grant from Huawei Technologies Canada Company, Ltd., the Science Foundation Ireland under Grant 13/RC/2094- 2, and the Canada Research Chair and Discovery Grant programs of the Natural Sciences and Engineering Research Council of Canada (NSERC).

\bibliographystyle{IEEEtran}
\bibliography{main.bib}

\begin{thebibliography}{10}
\providecommand{\url}[1]{#1}
\csname url@samestyle\endcsname
\providecommand{\newblock}{\relax}
\providecommand{\bibinfo}[2]{#2}
\providecommand{\BIBentrySTDinterwordspacing}{\spaceskip=0pt\relax}
\providecommand{\BIBentryALTinterwordstretchfactor}{4}
\providecommand{\BIBentryALTinterwordspacing}{\spaceskip=\fontdimen2\font plus
\BIBentryALTinterwordstretchfactor\fontdimen3\font minus \fontdimen4\font\relax}
\providecommand{\BIBforeignlanguage}[2]{{%
\expandafter\ifx\csname l@#1\endcsname\relax
\typeout{** WARNING: IEEEtran.bst: No hyphenation pattern has been}%
\typeout{** loaded for the language `#1'. Using the pattern for}%
\typeout{** the default language instead.}%
\else
\language=\csname l@#1\endcsname
\fi
#2}}
\providecommand{\BIBdecl}{\relax}
\BIBdecl

\bibitem{shen2018munn}
W.~Shen, J.~Wan, and Z.~Chen, ``Munn: Mutation analysis of neural networks,'' in \emph{2018 IEEE International Conference on Software Quality, Reliability and Security Companion (QRS-C)}.\hskip 1em plus 0.5em minus 0.4em\relax IEEE, 2018, pp. 108--115.

\bibitem{ma2018deepmutation}
L.~Ma, F.~Zhang, J.~Sun, M.~Xue, B.~Li, F.~Juefei-Xu, C.~Xie, L.~Li, Y.~Liu, J.~Zhao \emph{et~al.}, ``Deepmutation: Mutation testing of deep learning systems,'' in \emph{2018 IEEE 29th international symposium on software reliability engineering (ISSRE)}.\hskip 1em plus 0.5em minus 0.4em\relax IEEE, 2018, pp. 100--111.

\bibitem{hu2019deepmutation++}
Q.~Hu, L.~Ma, X.~Xie, B.~Yu, Y.~Liu, and J.~Zhao, ``Deepmutation++: A mutation testing framework for deep learning systems,'' in \emph{2019 34th IEEE/ACM International Conference on Automated Software Engineering (ASE)}.\hskip 1em plus 0.5em minus 0.4em\relax IEEE, 2019, pp. 1158--1161.

\bibitem{humbatova2021deepcrime}
N.~Humbatova, G.~Jahangirova, and P.~Tonella, ``Deepcrime: mutation testing of deep learning systems based on real faults,'' in \emph{Proceedings of the 30th ACM SIGSOFT International Symposium on Software Testing and Analysis}, 2021, pp. 67--78.

\bibitem{dola2023input}
S.~Dola, M.~B. Dwyer, and M.~L. Soffa, ``Input distribution coverage: Measuring feature interaction adequacy in neural network testing,'' \emph{ACM Transactions on Software Engineering and Methodology}, vol.~32, no.~3, pp. 1--48, 2023.

\bibitem{kim2023evaluating}
J.~Kim, R.~Feldt, and S.~Yoo, ``Evaluating surprise adequacy for deep learning system testing,'' \emph{ACM Transactions on Software Engineering and Methodology}, vol.~32, no.~2, pp. 1--29, 2023.

\bibitem{kim2019guiding}
------, ``Guiding deep learning system testing using surprise adequacy,'' in \emph{2019 IEEE/ACM 41st International Conference on Software Engineering (ICSE)}.\hskip 1em plus 0.5em minus 0.4em\relax IEEE, 2019, pp. 1039--1049.

\bibitem{pei2017deepxplore}
K.~Pei, Y.~Cao, J.~Yang, and S.~Jana, ``Deepxplore: Automated whitebox testing of deep learning systems,'' in \emph{proceedings of the 26th Symposium on Operating Systems Principles}, 2017, pp. 1--18.

\bibitem{Ma2018DeepGaugeMT}
L.~Ma, F.~Juefei-Xu, F.~Zhang, J.~Sun, M.~Xue, B.~Li, C.~Chen, T.~Su, L.~Li, Y.~Liu, J.~Zhao, and Y.~Wang, ``Deepgauge: Multi-granularity testing criteria for deep learning systems,'' \emph{2018 33rd IEEE/ACM International Conference on Automated Software Engineering (ASE)}, pp. 120--131, 2018.

\bibitem{tian2018deeptest}
Y.~Tian, K.~Pei, S.~Jana, and B.~Ray, ``Deeptest: Automated testing of deep-neural-network-driven autonomous cars,'' in \emph{Proceedings of the 40th international conference on software engineering}, 2018, pp. 303--314.

\bibitem{sun2019structural}
Y.~Sun, X.~Huang, D.~Kroening, J.~Sharp, M.~Hill, and R.~Ashmore, ``Structural test coverage criteria for deep neural networks,'' \emph{ACM Transactions on Embedded Computing Systems (TECS)}, vol.~18, no.~5s, pp. 1--23, 2019.

\bibitem{papadakis2018mutation}
M.~Papadakis, D.~Shin, S.~Yoo, and D.-H. Bae, ``Are mutation scores correlated with real fault detection? a large scale empirical study on the relationship between mutants and real faults,'' in \emph{Proceedings of the 40th International Conference on Software Engineering}, 2018, pp. 537--548.

\bibitem{papadakis2019mutation}
M.~Papadakis, M.~Kintis, J.~Zhang, Y.~Jia, Y.~Le~Traon, and M.~Harman, ``Mutation testing advances: an analysis and survey,'' in \emph{Advances in Computers}.\hskip 1em plus 0.5em minus 0.4em\relax Elsevier, 2019, vol. 112, pp. 275--378.

\bibitem{jahangirova2020empirical}
G.~Jahangirova and P.~Tonella, ``An empirical evaluation of mutation operators for deep learning systems,'' in \emph{2020 IEEE 13th International Conference on Software Testing, Validation and Verification (ICST)}.\hskip 1em plus 0.5em minus 0.4em\relax IEEE, 2020, pp. 74--84.

\bibitem{panichella2021we}
A.~Panichella and C.~C. Liem, ``What are we really testing in mutation testing for machine learning? a critical reflection,'' in \emph{2021 IEEE/ACM 43rd International Conference on Software Engineering: New Ideas and Emerging Results (ICSE-NIER)}.\hskip 1em plus 0.5em minus 0.4em\relax IEEE, 2021, pp. 66--70.

\bibitem{andrews2006using}
J.~H. Andrews, L.~C. Briand, Y.~Labiche, and A.~S. Namin, ``Using mutation analysis for assessing and comparing testing coverage criteria,'' \emph{IEEE Transactions on Software Engineering}, vol.~32, no.~8, pp. 608--624, 2006.

\bibitem{just2014mutants}
R.~Just, D.~Jalali, L.~Inozemtseva, M.~D. Ernst, R.~Holmes, and G.~Fraser, ``Are mutants a valid substitute for real faults in software testing?'' in \emph{Proceedings of the 22nd ACM SIGSOFT International Symposium on Foundations of Software Engineering}, 2014, pp. 654--665.

\bibitem{aghababaeyan2021black}
Z.~Aghababaeyan, M.~Abdellatif, L.~Briand, M.~Bagherzadeh, and R.~S., ``Black-box testing of deep neural networks through test case diversity,'' \emph{IEEE Transactions on Software Engineering, arXiv preprint arXiv:2112.12591}, 2023.

\bibitem{li2019structural}
Z.~Li, X.~Ma, C.~Xu, and C.~Cao, ``Structural coverage criteria for neural networks could be misleading,'' in \emph{2019 IEEE/ACM 41st International Conference on Software Engineering: New Ideas and Emerging Results (ICSE-NIER)}.\hskip 1em plus 0.5em minus 0.4em\relax IEEE, 2019, pp. 89--92.

\bibitem{yang2022revisiting}
Z.~Yang, J.~Shi, M.~H. Asyrofi, and D.~Lo, ``Revisiting neuron coverage metrics and quality of deep neural networks,'' in \emph{2022 IEEE International Conference on Software Analysis, Evolution and Reengineering (SANER)}.\hskip 1em plus 0.5em minus 0.4em\relax IEEE, 2022, pp. 408--419.

\bibitem{ILSVRC15}
O.~Russakovsky, J.~Deng, H.~Su, J.~Krause, S.~Satheesh, S.~Ma, Z.~Huang, A.~Karpathy, A.~Khosla, M.~Bernstein, A.~C. Berg, and L.~Fei-Fei, ``{ImageNet Large Scale Visual Recognition Challenge},'' \emph{International Journal of Computer Vision (IJCV)}, vol. 115, no.~3, pp. 211--252, 2015.

\bibitem{humbatova2020taxonomy}
N.~Humbatova, G.~Jahangirova, G.~Bavota, V.~Riccio, A.~Stocco, and P.~Tonella, ``Taxonomy of real faults in deep learning systems,'' in \emph{Proceedings of the ACM/IEEE 42nd International Conference on Software Engineering}, 2020, pp. 1110--1121.

\bibitem{islam2019comprehensive}
M.~J. Islam, G.~Nguyen, R.~Pan, and H.~Rajan, ``A comprehensive study on deep learning bug characteristics,'' in \emph{Proceedings of the 2019 27th ACM Joint Meeting on European Software Engineering Conference and Symposium on the Foundations of Software Engineering}, 2019, pp. 510--520.

\bibitem{zhang2018empirical}
Y.~Zhang, Y.~Chen, S.-C. Cheung, Y.~Xiong, and L.~Zhang, ``An empirical study on tensorflow program bugs,'' in \emph{Proceedings of the 27th ACM SIGSOFT International Symposium on Software Testing and Analysis}, 2018, pp. 129--140.

\bibitem{wand1994kernel}
M.~P. Wand and M.~C. Jones, \emph{Kernel smoothing}.\hskip 1em plus 0.5em minus 0.4em\relax CRC press, 1994.

\bibitem{dola2021distribution}
S.~Dola, M.~B. Dwyer, and M.~L. Soffa, ``Distribution-aware testing of neural networks using generative models. in 2021 ieee/acm 43rd international conference on software engineering (icse),'' \emph{IEEE, 226{\'s}237. https://doi. org/10.1109/ICSE43902}, 2021.

\bibitem{feng2020deepgini}
Y.~Feng, Q.~Shi, X.~Gao, J.~Wan, C.~Fang, and Z.~Chen, ``Deepgini: prioritizing massive tests to enhance the robustness of deep neural networks,'' in \emph{Proceedings of the 29th ACM SIGSOFT International Symposium on Software Testing and Analysis}, 2020, pp. 177--188.

\bibitem{fahmy2021supporting}
H.~Fahmy, F.~Pastore, M.~Bagherzadeh, and L.~Briand, ``Supporting deep neural network safety analysis and retraining through heatmap-based unsupervised learning,'' \emph{IEEE Transactions on Reliability}, 2021.

\bibitem{simonyan2014very}
K.~Simonyan and A.~Zisserman, ``Very deep convolutional networks for large-scale image recognition,'' \emph{arXiv preprint arXiv:1409.1556}, 2014.

\bibitem{campello2013density}
R.~J. Campello, D.~Moulavi, and J.~Sander, ``Density-based clustering based on hierarchical density estimates,'' in \emph{Pacific-Asia conference on knowledge discovery and data mining}.\hskip 1em plus 0.5em minus 0.4em\relax Springer, 2013, pp. 160--172.

\bibitem{rousseeuw1987silhouettes}
P.~J. Rousseeuw, ``Silhouettes: a graphical aid to the interpretation and validation of cluster analysis,'' \emph{Journal of computational and applied mathematics}, vol.~20, pp. 53--65, 1987.

\bibitem{moulavi2014density}
D.~Moulavi, P.~A. Jaskowiak, R.~J. Campello, A.~Zimek, and J.~Sander, ``Density-based clustering validation,'' in \emph{Proceedings of the 2014 SIAM international conference on data mining}.\hskip 1em plus 0.5em minus 0.4em\relax SIAM, 2014, pp. 839--847.

\bibitem{biagiola2024testing}
M.~Biagiola and P.~Tonella, ``Testing of deep reinforcement learning agents with surrogate models,'' \emph{ACM Transactions on Software Engineering and Methodology}, vol.~33, no.~3, pp. 1--33, 2024.

\bibitem{attaoui2023black}
M.~Attaoui, H.~Fahmy, F.~Pastore, and L.~Briand, ``Black-box safety analysis and retraining of dnns based on feature extraction and clustering,'' \emph{ACM Transactions on Software Engineering and Methodology}, vol.~32, no.~3, pp. 1--40, 2023.

\bibitem{niu2020decade}
S.~Niu, Y.~Liu, J.~Wang, and H.~Song, ``A decade survey of transfer learning (2010--2020),'' \emph{IEEE Transactions on Artificial Intelligence}, vol.~1, no.~2, pp. 151--166, 2020.

\bibitem{montgomery2021introduction}
D.~C. Montgomery, E.~A. Peck, and G.~G. Vining, \emph{Introduction to linear regression analysis}.\hskip 1em plus 0.5em minus 0.4em\relax John Wiley \& Sons, 2021.

\bibitem{kumar2012bootstrap}
S.~Kumar and A.~N. Srivistava, ``Bootstrap prediction intervals in non-parametric regression with applications to anomaly detection,'' in \emph{The 18th ACM SIGKDD Conference on Knowledge Discovery and Data Mining}, no. ARC-E-DAA-TN6188, 2012.

\bibitem{deng2012mnist}
L.~Deng, ``The mnist database of handwritten digit images for machine learning research [best of the web],'' \emph{IEEE Signal Processing Magazine}, vol.~29, no.~6, pp. 141--142, 2012.

\bibitem{Cifar10}
\BIBentryALTinterwordspacing
K.~Alex, N.~Vinod, and H.~Geoffrey. The cifar-10 dataset. [Online]. Available: \url{http://www.cs.toronto.edu/$\sim$kriz/cifar.html}
\BIBentrySTDinterwordspacing

\bibitem{netzer2011reading}
Y.~Netzer, T.~Wang, A.~Coates, A.~Bissacco, B.~Wu, and A.~Y. Ng, ``Reading digits in natural images with unsupervised feature learning,'' 2011.

\bibitem{krizhevsky2009learning}
A.~Krizhevsky, G.~Hinton \emph{et~al.}, ``Learning multiple layers of features from tiny images,'' 2009.

\bibitem{lecun1998gradient}
Y.~LeCun, L.~Bottou, Y.~Bengio, and P.~Haffner, ``Gradient-based learning applied to document recognition,'' \emph{Proceedings of the IEEE}, vol.~86, no.~11, pp. 2278--2324, 1998.

\bibitem{he2016deep}
K.~He, X.~Zhang, S.~Ren, and J.~Sun, ``Deep residual learning for image recognition,'' in \emph{Proceedings of the IEEE conference on computer vision and pattern recognition}, 2016, pp. 770--778.

\bibitem{szegedy2016rethinking}
C.~Szegedy, V.~Vanhoucke, S.~Ioffe, J.~Shlens, and Z.~Wojna, ``Rethinking the inception architecture for computer vision,'' pp. 2818--2826, 2016.

\bibitem{kornblith2019better}
S.~Kornblith, J.~Shlens, and Q.~V. Le, ``Do better imagenet models transfer better?'' in \emph{Proceedings of the IEEE/CVF conference on computer vision and pattern recognition}, 2019, pp. 2661--2671.

\bibitem{chen2018isolating}
T.~Chen~Ricky, L.~Xuechen, G.~Roger, and D.~David, ``Isolating sources of disentanglement in vaes,'' in \emph{Proceedings of the 32nd International Conference on Neural Information Processing Systems}, 2018, pp. 2615--2625.

\bibitem{burgess2018understanding}
C.~P. Burgess, I.~Higgins, A.~Pal, L.~Matthey, N.~Watters, G.~Desjardins, and A.~Lerchner, ``Understanding disentangling in $\beta$-vae,'' \emph{arXiv preprint arXiv:1804.03599}, 2018.

\bibitem{computecanada}
``Digital research alliance of canada,'' \url{https://alliancecan.ca/}, 2016, accessed: July 31, 2023.

\bibitem{replicationpackage}
\BIBentryALTinterwordspacing
A.~Abbasishahkoo, M.~Dadkhah, L.~Briand, and D.~Lin. [Online]. Available: \url{https://figshare.com/s/fe3c9a593c110786027b}
\BIBentrySTDinterwordspacing

\bibitem{weiss2021review}
M.~Weiss, R.~Chakraborty, and P.~Tonella, ``A review and refinement of surprise adequacy,'' in \emph{2021 IEEE/ACM Third International Workshop on Deep Learning for Testing and Testing for Deep Learning (DeepTest)}.\hskip 1em plus 0.5em minus 0.4em\relax IEEE, 2021, pp. 17--24.

\bibitem{hu2024test}
Q.~Hu, Y.~Guo, X.~Xie, M.~Cordy, L.~Ma, M.~Papadakis, and Y.~Le~Traon, ``Test optimization in dnn testing: A survey,'' \emph{ACM Transactions on Software Engineering and Methodology}, 2018.

\bibitem{hu2022empirical}
Q.~Hu, Y.~Guo, M.~Cordy, X.~Xie, L.~Ma, M.~Papadakis, and Y.~Le~Traon, ``An empirical study on data distribution-aware test selection for deep learning enhancement,'' \emph{ACM Transactions on Software Engineering and Methodology (TOSEM)}, vol.~31, no.~4, pp. 1--30, 2022.

\bibitem{ma2021test}
W.~Ma, M.~Papadakis, A.~Tsakmalis, M.~Cordy, and Y.~L. Traon, ``Test selection for deep learning systems,'' \emph{ACM Transactions on Software Engineering and Methodology (TOSEM)}, vol.~30, no.~2, pp. 1--22, 2021.

\bibitem{byun2019input}
T.~Byun, V.~Sharma, A.~Vijayakumar, S.~Rayadurgam, and D.~Cofer, ``Input prioritization for testing neural networks,'' in \emph{2019 IEEE International Conference On Artificial Intelligence Testing (AITest)}.\hskip 1em plus 0.5em minus 0.4em\relax IEEE, 2019, pp. 63--70.

\bibitem{aghababaeyan2023deepgd}
Z.~Aghababaeyan, M.~Abdellatif, M.~Dadkhah, and L.~Briand, ``Deepgd: A multi-objective black-box test selection approach for deep neural networks,'' \emph{ACM Transactions on Software Engineering and Methodology}, 2023.

\bibitem{zohdinasab2023efficient}
T.~Zohdinasab, V.~Riccio, A.~Gambi, and P.~Tonella, ``Efficient and effective feature space exploration for testing deep learning systems,'' \emph{ACM Transactions on Software Engineering and Methodology}, vol.~32, no.~2, pp. 1--38, 2023.

\bibitem{jia2010analysis}
Y.~Jia and M.~Harman, ``An analysis and survey of the development of mutation testing,'' \emph{IEEE transactions on software engineering}, vol.~37, no.~5, pp. 649--678, 2010.

\bibitem{kochhar2015code}
P.~S. Kochhar, F.~Thung, and D.~Lo, ``Code coverage and test suite effectiveness: Empirical study with real bugs in large systems,'' in \emph{2015 IEEE 22nd international conference on software analysis, evolution, and reengineering (SANER)}.\hskip 1em plus 0.5em minus 0.4em\relax IEEE, 2015, pp. 560--564.

\bibitem{gerasimou2020importance}
S.~Gerasimou, H.~F. Eniser, A.~Sen, and A.~Cakan, ``Importance-driven deep learning system testing,'' in \emph{Proceedings of the ACM/IEEE 42nd International Conference on Software Engineering}, 2020, pp. 702--713.

\bibitem{chen2020deep}
J.~Chen, M.~Yan, Z.~Wang, Y.~Kang, and Z.~Wu, ``Deep neural network test coverage: How far are we?'' \emph{arXiv preprint arXiv:2010.04946}, 2020.

\bibitem{riccio2021deepmetis}
V.~Riccio, N.~Humbatova, G.~Jahangirova, and P.~Tonella, ``Deepmetis: Augmenting a deep learning test set to increase its mutation score,'' in \emph{2021 36th IEEE/ACM International Conference on Automated Software Engineering (ASE)}.\hskip 1em plus 0.5em minus 0.4em\relax IEEE, 2021, pp. 355--367.

\bibitem{wang2021prioritizing}
Z.~Wang, H.~You, J.~Chen, Y.~Zhang, X.~Dong, and W.~Zhang, ``Prioritizing test inputs for deep neural networks via mutation analysis,'' in \emph{2021 IEEE/ACM 43rd International Conference on Software Engineering (ICSE)}.\hskip 1em plus 0.5em minus 0.4em\relax IEEE, 2021, pp. 397--409.

\bibitem{wang2019adversarial}
J.~Wang, G.~Dong, J.~Sun, X.~Wang, and P.~Zhang, ``Adversarial sample detection for deep neural network through model mutation testing,'' in \emph{2019 IEEE/ACM 41st International Conference on Software Engineering (ICSE)}.\hskip 1em plus 0.5em minus 0.4em\relax IEEE, 2019, pp. 1245--1256.

\end{thebibliography}

%

\hfill

\begingroup
\setlength{\intextsep}{-1.5pt}
\begin{wrapfigure}{l}{36mm}     \includegraphics[width=2in,height=1.5in,clip,keepaspectratio]{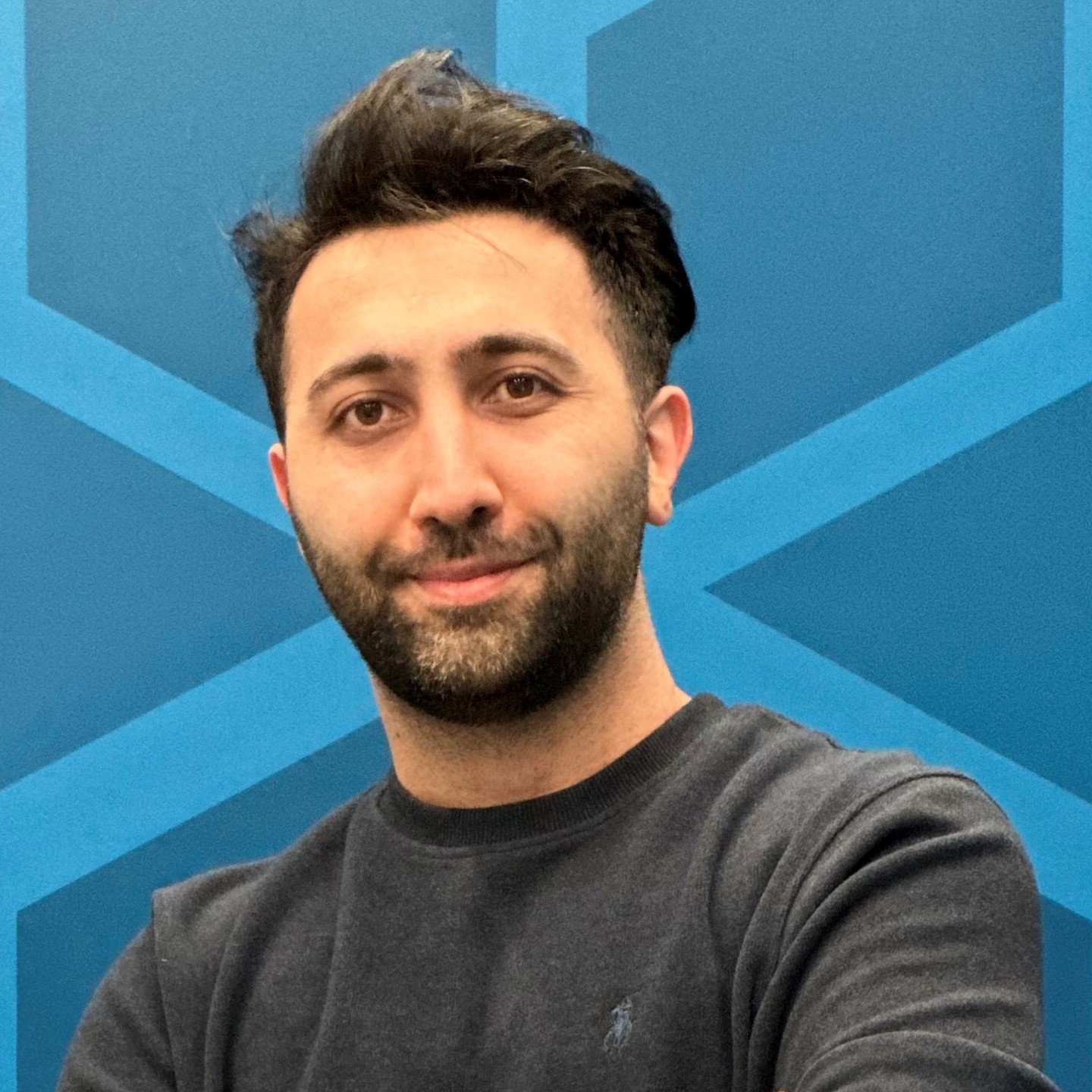}
  \end{wrapfigure}\par
  \noindent\textbf{Amin Abbasishahkoo} is a member of the Nanda Lab and is currently working toward the PhD degree in the School of EECS at the University of Ottawa. He has received several academic awards, including the PhD admission scholarship and an international doctoral scholarship from the University of Ottawa. He received his Master's degree in Artificial Intelligence at Shahid Beheshti University, where he was ranked 3rd among all peer students. His research interests focus on machine learning and automated testing and verification of AI-based systems.\par
\endgroup

\vfill

\begingroup
\setlength{\intextsep}{-1.5pt}
\begin{wrapfigure}{l}{36mm}     \includegraphics[width=2in,height=1.5in,clip,keepaspectratio]{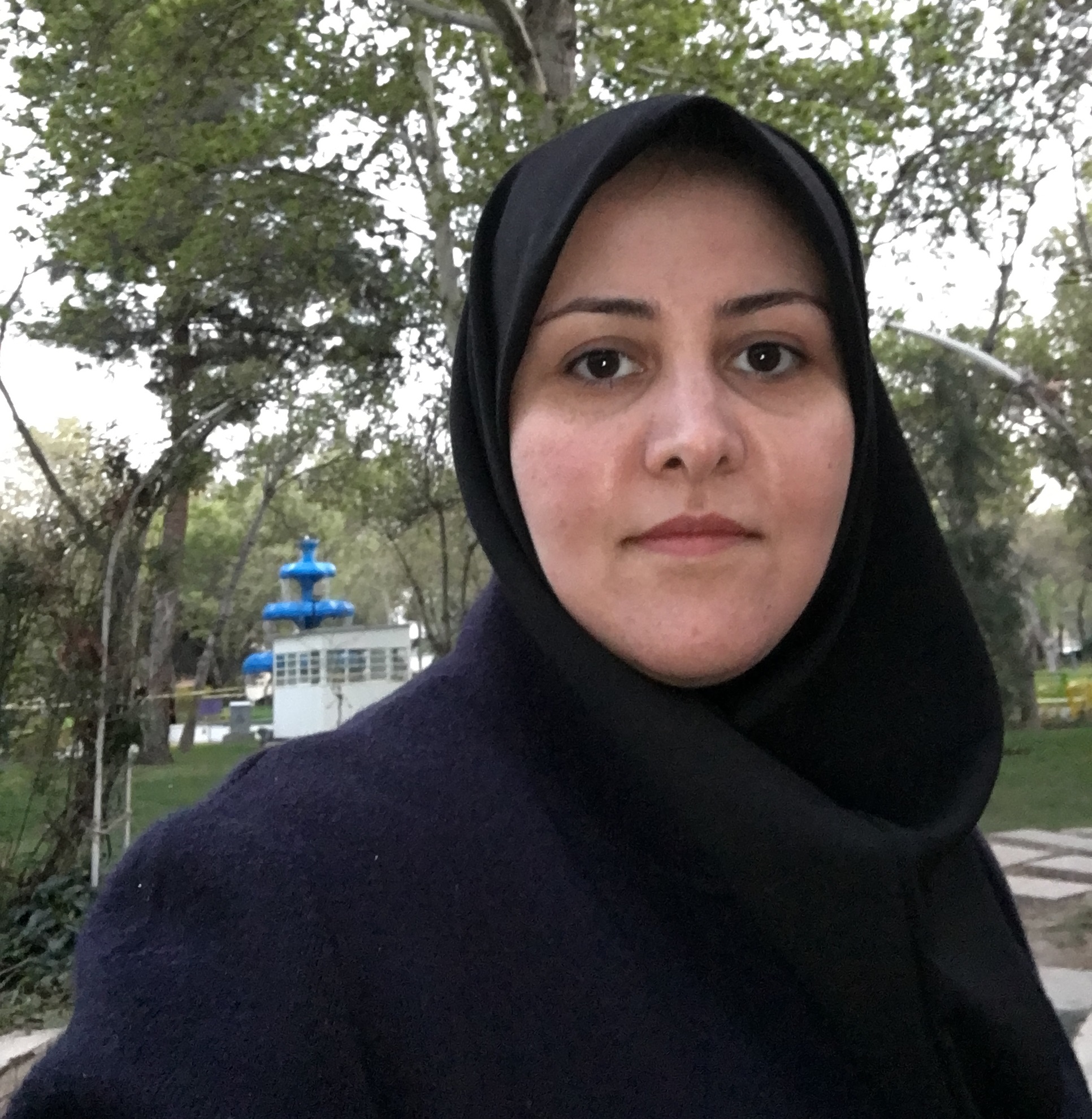}
  \end{wrapfigure}\par
  \noindent\textbf{Mahboubeh Dadkhah} received her Ph.D. in Computer Engineering from Ferdowsi University of Mashhad in 2021, where she was awarded a scholarship as a distinguished student. During her doctoral studies, she focused on semantic web-enabled techniques for testing large-scale systems. Currently, she is a postdoctoral research fellow at the School of EECS, University of Ottawa, working on testing deep neural networks. Her research interests include testing AI-based systems, automated software testing, and empirical software engineering.\par
\endgroup

\hfill 

\begingroup
\setlength{\intextsep}{-1.5pt}
\begin{wrapfigure}{l}{36mm}     \includegraphics[width=2in,height=1.5in,clip,keepaspectratio]{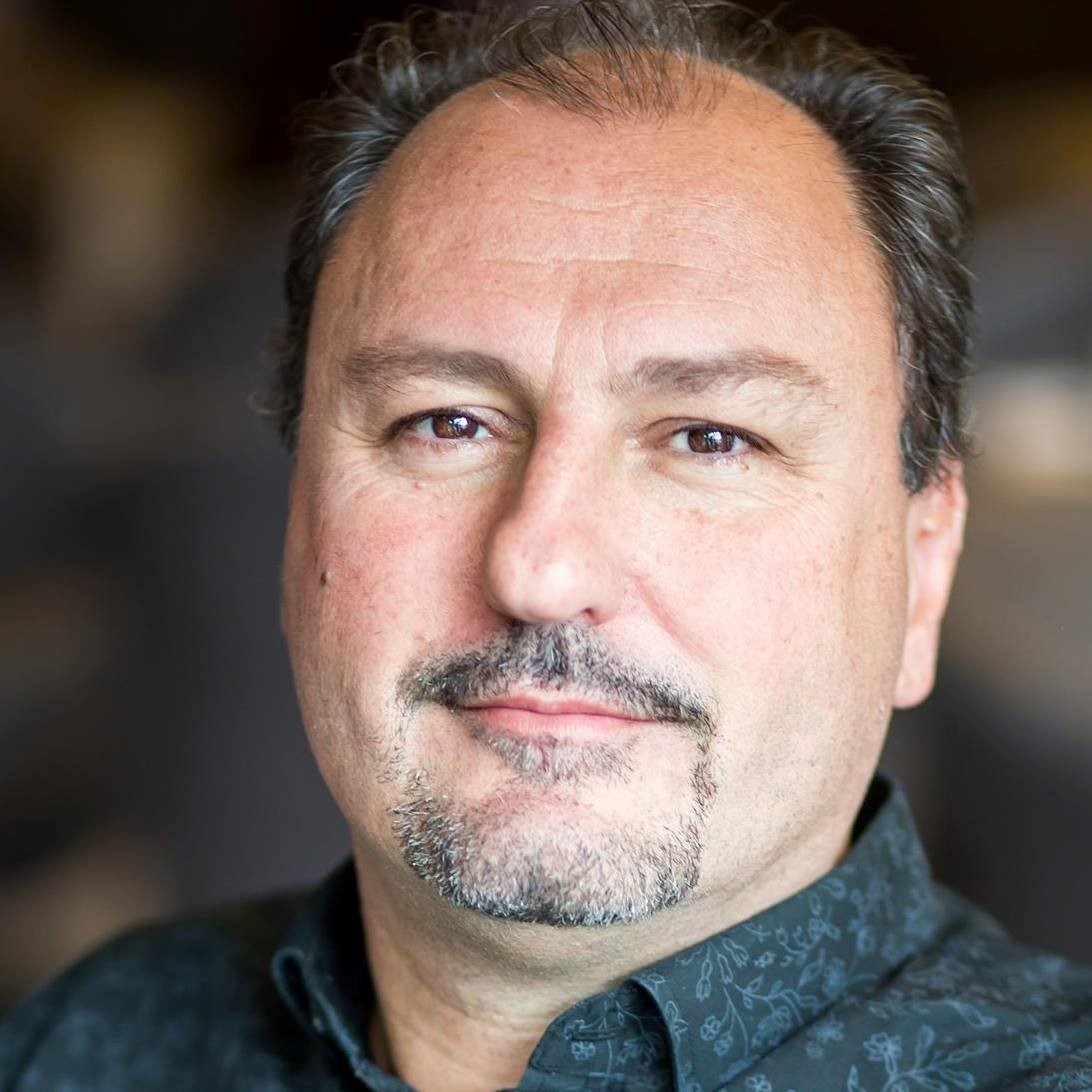}
  \end{wrapfigure}\par
  \noindent\textbf{Lionel C. Briand}  is professor of software engineering and has shared appointments between (1) The University of Ottawa, Canada, and (2) The Lero SFI Centre---the national Irish centre for software research---hosted by the University of Limerick, Ireland. In collaboration with colleagues, for over 30 years, he has run many collaborative research projects with companies in the automotive, satellite, aerospace, energy, financial, and legal domains. Lionel has held various engineering, academic, and leading positions in seven countries.  He currently holds a Canada Research Chair (Tier 1) on "Intelligent Software Dependability and Compliance" and is the director of Lero, the national Irish centre for software research. Lionel was elevated to the grades of IEEE Fellow and ACM Fellow for his work on software testing and verification. Further, he was granted the IEEE Computer Society Harlan Mills award, the ACM SIGSOFT outstanding research award, and the IEEE Reliability Society engineer-of-the-year award. He also received an ERC Advanced grant in 2016 on modelling and testing cyber-physical systems, the most prestigious individual research award in the European Union and was elected a fellow of the Academy of Science, Royal Society of Canada in 2023. More details can be found at: http://www.lbriand.info.\par
\endgroup

\hfill 

\begingroup
\setlength{\intextsep}{-1.5pt}
\begin{wrapfigure}{l}{36mm}     \includegraphics[width=2in,height=1.5in,clip,keepaspectratio]{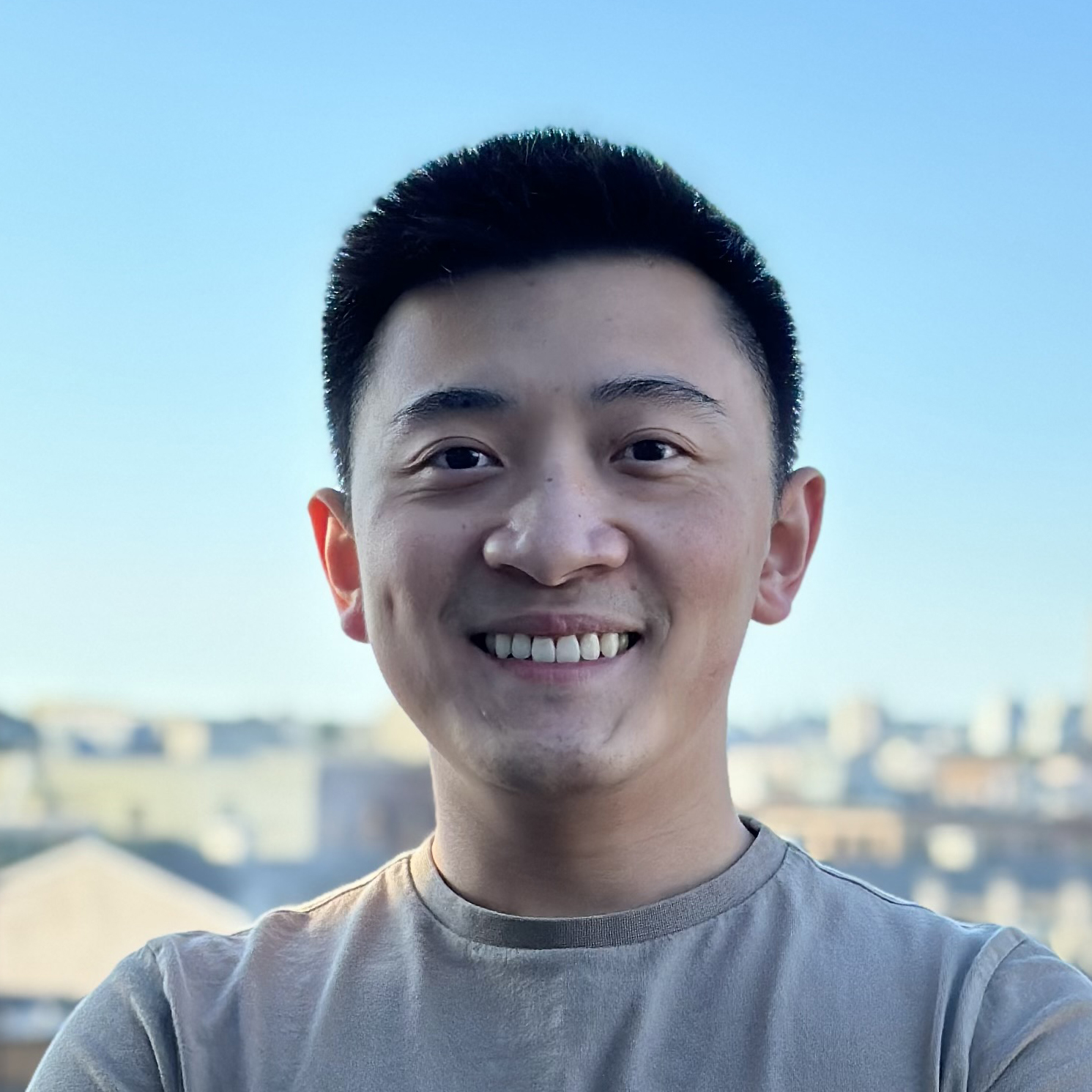}
  \end{wrapfigure}\par
  \noindent\textbf{Dayi Lin} is a Principal Researcher at the Centre for Software Excellence, Huawei Canada, working on the topics of SE for AI and foundation model applications. His research interests include SE for AI, AI for SE, game engineering, and mining software repositories. His work has been published at several top-tier software engineering venues, such as TSE, TOSEM, EMSE, and ICSE. His work on SE and games has also attracted wide media coverage including Kotaku, PC Gamers, Gamasutra, and national newspapers. He received a Ph.D. in Computer Science from Queen’s University, Canada. More about him can be found at https://lindayi.me\par
\endgroup

\end{document}